\documentclass[11pt,a4paper]{article}
\pdfoutput=1 

\usepackage{jheppub} 
\usepackage[T1]{fontenc}
\usepackage[utf8]{inputenc}
\usepackage{lmodern}      
\usepackage[dvipsnames, table]{xcolor} 
\usepackage{caption}
\usepackage{amsmath}
\usepackage{booktabs}
\usepackage{dsfont}
\usepackage{subfig} 
\usepackage{makecell}
\usepackage{subdepth}
\usepackage{bm}
\usepackage{xstring}
\usepackage{import}
\usepackage{multirow}
\usepackage{float}
\usepackage{rotating} 
\usepackage{lipsum}
\usepackage{xfrac}
\usepackage{romannum}
\usepackage{mathtools}

\newcommand{\etaoct}[0]{\eta_{\mathbf{8}}}
\newcommand{\etas}[0]{\ensuremath{\eta_{\mathbf{1}}}}
\newcommand{\omoct}[0]{\ensuremath{\omega_{\mathbf{8}}}}

\newcommand{\D}[0]{\ensuremath{D_{\bar{\mathbf{3}}}}}

\newcommand{\Dst}[0]{\ensuremath{D^*_{\bar{\mathbf{3}}}}}

\newcommand{\Dz}[0]{\ensuremath{D^*_{0,\bar{\mathbf{3}}}}}

\newcommand{\Do}[0]{\ensuremath{D_{1,\bar{\mathbf{3}}}}}

\newcommand{\Dopr}[0]{\ensuremath{D'_{1,\bar{\mathbf{3}}}}}

\newcommand{\Dt}[0]{\ensuremath{D^*_{2,\bar{\mathbf{3}}}}}

\newcommand{\SLJ}[3]{\ensuremath{{\:\!}^{{#1}\!}{#2}_{#3}}}
\newcommand{\chisq}[0]{\ensuremath{\chi^2/N_{\sf dof}}}
\newcommand{\grey}[1]{\textcolor{gray}{#1}}

\title{Exotic $T^*_{csJ}$  and $T^*_{c\bar{s}J}$ states and coupled-channel scattering at the $SU(3)$ flavour symmetric point from lattice QCD}

\author[a]{J. Daniel E. Yeo,}
\author[a]{Christopher E. Thomas,}
\author[a]{David J. Wilson}
\author{\\(for the Hadron Spectrum Collaboration)}

\affiliation[a]{DAMTP, University of Cambridge, Centre for Mathematical Sciences, Wilberforce Road, Cambridge, CB3 0WA, UK}

\emailAdd{jdey2@cam.ac.uk}
\emailAdd{c.e.thomas@damtp.cam.ac.uk}
\emailAdd{d.j.wilson@damtp.cam.ac.uk}

\abstract{Motivated by recent experimental observations of the flavour-exotic $T^*_{cs0}(2870)^0$ and $T^*_{c\bar{s}0}(2900)$, we present the first lattice QCD study of coupled-channel scattering of a charm meson with a light meson in the flavour-exotic sectors at the $SU(3)_f$ flavour symmetric point. 
Utilising five volumes with $m_\pi \approx 700$ MeV and employing large bases of meson-meson operators, finite-volume spectra are extracted and used to constrain infinite-volume scattering amplitudes with $J^P = \{0, 1, 2, 3, 4\}^+$ via the Lüscher formalism.
In the flavour $\mathbf{6}$ sector, each $S$-wave channel considered is found to be attractive with the scattering amplitudes having an associated pole singularity on an unphysical sheet below threshold, giving six flavour-exotic poles in the energy region constrained. 
In $J^P = 0^+$ there is a virtual bound state and a resonance. The latter is identified with the $T^*_{cs0}(2870)^0$ and $T^*_{c\bar{s}0}(2900)$, appearing as one state in the $SU(3)_f$ flavour symmetric limit, and suggests the existence of an isospin-$\frac{1}{2}$ partner. In $J^P =1^+$ there are three poles, one of which is identified as a $J^P =1^+$ partner of the $T^*_{cs0}(2870)^0$ and $T^*_{c\bar{s}0}(2900)$, and $J^P =2^+$ contains one pole which is identified as their $J^P =2^+$ partner.
Only mild interactions and no poles are seen in the $J^P = \{3, 4\}^+$ scattering amplitudes.
In the flavour $\overline{\mathbf{15}}$ sector, weak interactions are observed in $J^P = \{0, 1, 2, 3, 4\}^+$
with no well-determined poles in the energy region constrained.
}

\begin{document} 
\maketitle

\section{Introduction}
\label{section:intro}

Over the last few decades a vast array of exotic hadrons have been discovered in collider experiments, challenging our understanding of QCD.
Recently the LHCb experiment has observed the charmed-strange $T^*_{cs0}(2870)^0$,  $T^*_{c\bar{s}0}(2900)^0$,  $T^*_{c\bar{s}0}(2900)^{++}$  and $T^*_{cs1}(2900)$ \cite{LHCb:2020bls, LHCb:2020pxc, LHCb:2022sfr, LHCb:2022lzp}, which are unambiguously exotic owing to their combination of flavour quantum numbers. 
In particular, the  $T^*_{cs0}(2870)^0$ and $T^*_{cs1}(2900)$ both have a minimum quark content of $[c s\bar{u}\bar{d}]$,
whilst the $T^*_{c\bar{s}0}(2900)^0$ and $T^*_{c\bar{s}0}(2900)^{++}$ (collectively denoted as $T^*_{c\bar{s}0}(2900)$) have minimum quark content of $[c \bar{s}d\bar{u}]$ and $[c \bar{s}u\bar{d}]$, respectively. 
Analysis of the experimental data suggest that $T^*_{cs0}(2870)^0$,  $T^*_{c\bar{s}0}(2900)^0$ and  $T^*_{c\bar{s}0}(2900)^{++}$ are $J^P =0^+$, and 
$T^*_{cs1}(2900)$ is $J^P =1^-$, where $J$ and $P$ are the hadron's spin and parity, respectively.
Although a singly charged partner of the  $T^*_{c\bar{s}0}(2900)^0$ and $T^*_{c\bar{s}0}(2900)^{++}$ has currently not been observed, it is thought that the $T^*_{c\bar{s}0}(2900)^0$ and $T^*_{c\bar{s}0}(2900)^{++}$ are two components of an isospin-1 multiplet.
The $T^*_{cs0}(2870)^0$ is thought to be an isospin-0 state.

Evading a simple quark-model description, many ideas have been put forward regarding the nature of the $T^*_{cs0}(2870)$.
Refs.~\cite{Yue:2022mnf, Chen:2022svh, Duan:2023qsg, Lyu:2023ppb, Karliner:2020vsi, Agaev:2020nrc, Chen:2021erj, Yu:2023avh, Wang:2021lwy, Liu:2020nil, Huang:2020ptc, Mutuk:2020igv, Chen:2020aos, Chen:2021xlu, Hu:2020mxp, Kong:2021ohg}
argue that this state can be described as a $D^*\bar{K}^*$ molecule 
whilst Refs.~\cite{Zhang:2020oze, Lian:2023cgs, Jiang:2023rcn, Liu:2022hbk, Ortega:2023azl, Wang:2020xyc, Wang:2020prk, Agaev:2021knl, He:2020jna, Guo:2021mja} 
argue that it can be described as a compact tetraquark state.
Other ideas include that the experimental bump that is attributed to $T^*_{cs0}(2870)$ could arise from a singularity or kinematical cusp from triangle diagrams \cite{Liu:2020orv, Burns:2020epm}.
Although methods for distinguishing between the various models through different  experimental signatures have been discussed \cite{Burns:2020xne}, 
currently there is no agreement on the nature of this state.
Similarly, there is much speculation regarding the structure of the $T^*_{c\bar{s}0}(2900)$.
These include $D^*K^*$ molecule \cite{Agaev:2022duz, Agaev:2022eyk, Yue:2022mnf, Duan:2023qsg, Wang:2023hpp, Ding:2024dif}
and compact tetraquark \cite{Wei:2022wtr, Liu:2022hbk, Yang:2023evp, Ortega:2023azl, Mutuk:2025hql} descriptions, whilst 
others have suggested that the experimental peaks could arise from threshold effects~\cite{Ge:2022dsp, Molina:2022jcd}.

Whilst phenomenological models can be valuable, it is also important to study these states within QCD from a first-principles approach.
One such approach is lattice QCD, where the underlying spacetime is replaced by a finite-volume Euclidean lattice, allowing for the determination of non-perturbative phenomena directly from the QCD Lagrangian. 
A well-established method for investigating unstable hadrons from lattice QCD is the Lüscher method~\cite{Luscher:1990ux, Luscher:1986pf, Fu:2011xz, Luscher:1991cf, Kim:2005gf, Christ:2005gi, Leskovec:2012gb, Hansen:2012tf, Briceno:2014oea}
in which two-to-two hadron scattering amplitudes are constrained from the lattice spectrum and unstable hadrons are identified as poles in the complex energy plane.
So far, the  $T^*_{cs0}(2870)^0$ and $T^*_{c\bar{s}0}(2900)$ hadrons have not been investigated from a Lüscher scattering analysis as, even for fairly unphysically heavy pion masses, 
these states appear above three-(or more) hadron thresholds, thus requiring a three-(or more) hadron extension of the formalism.\footnote{There has been much theoretical development on the three-hadron formalism~\cite{Briceno:2012rv, Hansen:2014eka, Hansen:2015zga, Briceno:2017tce, Hammer:2017uqm, Hammer:2017kms, Mai:2017bge, Briceno:2018aml, Jackura:2019bmu, Blanton:2020gha, Blanton:2020gmf, Blanton:2021mih, Briceno:2019muc}, although it has not yet reached the maturity needed to study complex systems such as the ones studied here.}
In this paper, we present the first lattice QCD investigation of the $T^*_{cs0}(2870)^0$ and  $T^*_{c\bar{s}0}(2900)$, working at the flavour symmetric point (equal up, down and strange quark masses) with a pion mass of $m_{\pi} \approx 700$ MeV. 
Working with these unphysical-heavy light-quark masses  and $SU(3)_f$ flavour symmetry has three key advantages.
Firstly, with larger light-quark masses, the relevant three-(or more) hadron thresholds are pushed to higher energies, allowing for a two-hadron Lüscher analysis of these states.
Secondly,  when working at the flavour symmetric point there are fewer distinct scattering channels, which reduces the complexity of the scattering analysis. 
Finally, a common origin for these states can be investigated in terms of  $SU(3)_f$ multiplets. 

This work builds upon our study of elastic open-charm scattering at the $SU(3)_f$ flavour symmetric point~\cite{Yeo:2024chk} by investigating the coupled-channel region where these exotic states are expected to appear.
At the flavour symmetric point, the scattering of a $D-$meson and a light-meson yields the $SU(3)_f$ flavour decomposition $\bar{\mathbf{3}} \otimes \mathbf{8} = \bar{\mathbf{3}}\oplus \mathbf{6} \oplus \overline{\mathbf{15}}$ or $\bar{\mathbf{3}} \otimes \mathbf{1} = \bar{\mathbf{3}}$. 
Since the $T^*_{cs0}(2870)^0$ and  $T^*_{c\bar{s}0}(2900)$ states are manifestly flavour exotic, they appear in either the $\mathbf{6}$ or $\overline{\mathbf{15}}$ exotic sectors.
We therefore investigate coupled-channel scattering with $J^P =0^+$ in these flavour exotic sectors.
We also search for higher-spin partners of these states by exploring the pole structure of amplitudes with $J^P =\{1,2,3,4\}^+$ in both these flavour sectors.

The rest of the paper is organised as follows: a quick overview of our methodology for determining amplitudes from lattice QCD and the lattice ensembles used is given in Section~\ref{section:calculation_details}.
In Section~\ref{section:6f_analysis} the finite-volume spectra and the constrained  $J^P =\{0,1,2,3,4\}^+$ scattering amplitudes and poles are presented for the flavour $\mathbf{6}$ sector. 
Following this, we present the analogous results for the flavour  $\overline{\mathbf{15}}$ sector in Section~\ref{section:15bar_analysis}. 
A discussion of the determined poles is presented in Section~\ref{section:disc}.
In Section~\ref{section:concl_and_outlook} we finish with a summary and outlook.

\section{Details of calculation} 
\label{section:calculation_details}
\subsection{Methodology} \label{section:methodology}
We largely follow the same methodology as in Ref.~\cite{Yeo:2024chk}, with a few differences that we will highlight in this quick overview.
The finite-volume spectrum is extracted using the variational method~\cite{Michael:1985ne, Luscher:1990ck, Blossier:2009kd}, in which 
the following generalised eigenvalue problem (GEVP),
\begin{equation}
	C_{ij}(t) \, v^{(\mathbf{n})}_j = \lambda^{(\mathbf{n})}(t,t_0) \, C_{ij}(t_0) \, v^{(\mathbf{n})}_j ,
\end{equation}
is solved, where \smash{$C_{ij}(t) = \langle \Omega | \mathcal{O}_i(t)  \mathcal{O}^{\dagger}_j(0)| \Omega \rangle $} is a matrix of correlation functions.
The principal correlators (eigenvalues), $\lambda^{(\mathbf{n})}(t,t_0)$, can be shown to behave as $\lambda^{(\mathbf{n})}(t,t_0) \sim e^{-E_{\mathbf{n}}(t - t_0)}$ for large $t-t_0$,
where $E_{\mathbf{n}}$ is the $\mathbf{n}^\text{th}$ finite-volume energy level \cite{Luscher:1990ck}.
The principal correlators are fit to a single exponential or a double exponential to extract the corresponding energy, as described in Ref.~\cite{Dudek:2010wm}. 
For a handful of energy levels used in this study we employ model averaging, where we consider many time ranges for the principal correlator and weight them by a version of the Akaike information criterion \cite{Jay:2020jkz}, see Appendix~\ref{appendix:model_av} for more details.
The eigenvectors $v^{(\mathbf{n})}$ of the GEVP can be used to obtain operator overlaps $Z^{\mathbf{n}}_i =\langle \mathbf{n} |\mathcal{O}^{\dagger}_i | \Omega \rangle$,
which provide a useful metric for determining which operators were the most important for resolving each energy eigenstate.
For the correlation functions, we use a basis of meson-meson operators~\cite{Dudek:2012gj} consisting of optimised operators of an open-charm meson (in the flavour $\bar{\mathbf{3}}$ irrep) and a light meson (in the flavour $\mathbf{8}$ irrep).
These are projected into the $\mathbf{6}$ or $\overline{\mathbf{15}}$ irrep of $SU(3)_f$, 
the flavour sectors of interest in this study. 
No fermion bilinears carry the flavour quantum numbers of these flavour exotic sectors and, as motivated in our elastic study~\cite{Yeo:2024chk},  we do not include compact tetraquark-like operators in our basis.
Lists of operators used can be found in Appendix~\ref{appendix:interpolating_list}.

Once determined, the finite-volume spectra can be related to infinite-volume scattering amplitudes 
using the so-called Lüscher method~\cite{Luscher:1990ux, Luscher:1986pf, Fu:2011xz, Luscher:1991cf, Kim:2005gf, Christ:2005gi, Leskovec:2012gb, Hansen:2012tf, Briceno:2014oea}.
This relation is encoded in the quantisation condition (QC),
\begin{equation}
	\label{eq:luscher}
	\det \Big{[} \mathds{1} + i \rho (s) t(s)\big{(} \mathds{1} + i \overline{\mathcal{M}}^{\Lambda(\vec{P})}(s, L)\big{)}\Big{]}  = 0 \, ,
\end{equation}
where the determinant is over the space of hadron-hadron scattering channels and partial waves, 
$\rho(s)$ is a diagonal matrix of phase-space factors,
$t(s)$ is the (partial-wave projected) $t$-matrix,
and \smash{$\overline{\mathcal{M}}$} is a matrix of known functions depending on volume, energy and finite-volume irrep~\cite{Briceno:2014oea}.
For scattering involving hadrons with non-zero spin, a given hadron-hadron channel can have several partial waves of differing total intrinsic spin, $S$, and/or orbital angular momentum, $\ell$, with the same total angular momentum, $J$.
The QC is a function of Mandelstam $s (= E_{\sf cm}^2$) and its solutions are the finite-volume spectra for irrep $[\vec{P}]\Lambda$,
where $\vec{P}$ is the total three-momentum of the system.
As outlined in Ref.~\cite{Yeo:2024chk}, we parameterise the $t$-matrix and compare its predicted spectrum to the obtained lattice spectrum via means of a $\chi^2$ test. 
Fit parameters for a given parameterisation are obtained by minimising the $\chi^2$ as a function of these parameters~\cite{Dudek:2012xn, Woss:2020cmp}.
To reduce potential bias, we consider many functional forms of the $t$-matrix when fitting to the data.

Utilised throughout this study are the $K$-matrix parameterisations, 
\begin{equation}
	[t^{-1}]_{ij} = (2k_i)^{-\ell_i}[K^{-1}]_{ij}(2k_j)^{-\ell_j} + I_{ij}\, ,
\end{equation}
where the indices correspond to different hadron-hadron channels and partial waves.
$s$-channel unitarity is satisfied if $\text{Im}[I_{ij}(s)]= - \rho_i(s)\delta_{ij}$ above threshold and zero below.
For the remaining freedom in choice of $\text{Re}[I_{ij}(s)]$, we use either the \emph{simple phase-space prescription} \smash{$I_{ij}(s)= -i \rho_i(s)\delta_{ij}$}
or the \emph{Chew-Mandelstam prescription} \smash{$I_{ij}(s)= I_i(s)\delta_{ij}$}, where the real part of $I_i$ is related to its known imaginary part via a dispersive integral \cite{Wilson:2014cna}.\footnote{
When using the Chew-Mandelstam prescription, we set the subtraction point of $I_i$ equal to the $i^{\text{th}}$ channel's scattering threshold.}
The $K$-matrix must be real for real-valued $s$ and  must also be symmetric.
We consider $K$-matrix polynomial, inverse polynomial and ratio-of-polynomial parameterisations which take the form 
\begin{equation}
K_{ij}(s) = \sum_n \gamma^{(n)}_{ij} s^n, \quad 
[K^{-1}]_{ij}(s) = \sum_n c^{(n)}_{ij} s^n \quad \text{and} \quad
K_{ij}(s) = \frac{\sum_n c^{(n)}_{ij} s^n}{1 + \sum_n d^{(n)}_{ij} s^n},
\end{equation}
respectively, where $\gamma^{(n)}_{ij}$, $c^{(n)}_{ij}$ and $d^{(n)}_{ij}$ are real-valued, symmetric matrices.

Once reasonable parameterisations are determined, the scattering amplitudes can be analytically continued to the complex $s$-plane where hadrons are identified as poles.
Physical $s$-channel unitarity enforces a square-root singularity at each kinematic threshold, and so for an $n$ hadron-hadron channel scattering process the $s$-plane splits into $2^n$ Riemann sheets.  
Sheets are indexed by the sign of the imaginary part of the scattering momenta for each hadron-hadron channel (i.e. $\text{sign}[\text{Im}(k_a(s))]$ for $a \in \{ 1, \ldots, n \}$).
This is compactly written as an array of length $n$, $[\pm, \pm, \ldots, \pm]$, where channels are ordered by increasing threshold, $s^{(i)}_{\text{thr}}= (m_1^{(i)}+ m_2^{(i)})^2$.
The physical scattering amplitude is then defined as $t_{ij}(s +i\epsilon)$ on the $[+, +, +, \dots, +]$ sheet, also known as the \emph{physical sheet}. 
Any other sheet is known as an \emph{unphysical sheet}.  
For a given $s =E^2_{\sf cm}$, the  first unphysical sheet reached by moving down through the cut is referred to here as the \emph{proximal sheet}, while the rest are known as \emph{hidden sheets}.
On the physical sheet, poles can only be found on the real axis below the lowest kinematic threshold and these are identified as \emph{bound states} (asymptotic states of QCD).
On the unphysical sheets, poles can exist either on the real axis below  threshold or off the real axis in complex conjugate pairs.   
The former are \emph{virtual bound states} whilst the latter are \emph{resonances}. 
Near a pole, the $t$-matrix takes the factorised form 
\begin{equation}
	t_{ij} \sim \frac{c_ic_j}{s_{\text{pole}}-s},
\end{equation}
where $c_i$ gives the pole's coupling to the $i^\text{th}$ scattering channel and partial wave.
Note that experiments often measure production amplitudes rather than scattering amplitudes. 
These contain the $t$-matrix as well as a component describing the production process, so they feature the same poles as scattering amplitudes but potentially give different experimental signatures.

Our parameterisations satisfy unitarity requirements but are not guaranteed to satisfy requirements from analyticity. 
Parameterisations which violate these, such as those with a physical sheet pole off the real axis (close to the constrained energy region), are discarded.
As in our study of open-charm elastic scattering at the flavour symmetric point \cite{Yeo:2024chk}, the effect of left-hand cuts from $t$- and $u$-channel crossing symmetry or 
$t$- or $u$-channel one-particle exchange in the partial-wave projected amplitudes are not considered here and an investigation of their effect is left for future studies.
Recent developments on how to incorporate the latter of these can be found in Refs.~\cite{Raposo:2023oru, Raposo:2025dkb}.

\subsection{Lattice setup,  stable mesons and scattering channels}
\begin{table}
	\centering
		\begin{tabular}{cccc}
			\hline
			\rule{0pt}{12pt} $(L/a_s)^3\times (T/a_t)$   & $N_{\text{cfgs}}$   & $N_{\text{tsrcs}}$  & $N_{\text{vecs}}$   \\
			\hline
			\rule{0pt}{13pt} $14^3 \times 128$      	 & 397	& 8 & 64      \\
			$16^3 \times 128$      	     			     & 533	& 8 & 64      \\
			$18^3 \times 128$      	     				 & 357	& 8 & 96      \\
			$20^3 \times 128$      	    				 & 503	& 4 & 128     \\
			$24^3 \times 128$      	   				     & 607	& 4 & 160     \\
			\hline
		\end{tabular}
		\caption{Lattices used, with spatial and temporal extents, $L$ and $T$,
				 number of configurations $N_{\text{cfgs}}$, number of time sources $N_{\text{tsrcs}}$ and number of distillation vectors $N_{\text{vecs}}$.}
		\label{table:ensembles}
\end{table}

For this investigation, we use the same lattice ensembles as in our study of elastic open-charm scattering~\cite{Yeo:2024chk} with three degenerate dynamical ``light'' quarks and one quenched charm quark.
The ensembles are summarised in Table~\ref{table:ensembles}.
The anisotropy was determined to be $\xi=\frac{a_s}{a_t}= 3.471 \pm 0.028$~\cite{Yeo:2024chk},
where $a_s$ and $a_t$ are the spatial and temporal lattice spacing, respectively.
After scale setting using the $\Omega$-baryon mass, these ensembles are found to have  $a_t = (4655 \text{ MeV})^{-1}$ and a pion mass of  $m_\pi \approx 700$ MeV.
More details can be found in Section 3.1 of Ref.~\cite{Yeo:2024chk}.
We use several ensembles with these parameters, corresponding to different lattice sizes, $L^3\times T$. 
For each ensemble we use $N_{\text{cfg}}$ configurations and propagate uncertainties through jackknife resampling. 
Correlation functions are computed in the distillation framework \cite{HadronSpectrum:2009krc}.

\begin{table}[ht]
    \centering
	\begin{tabular}{cccccc}
		\hline
		\rule{0pt}{12pt}Meson      & $a_tm$        & $m/$MeV     & Flavour                & $J^{P(C)}$  \\
		\hline
		$\etaoct$      			   & 0.1478(1)	   & 688.0(5)    & $\boldsymbol{8}$       & $0^{-+}$  \\
		$\omoct$     	           & 0.2154(2)     & 1002.7(9)   & $\boldsymbol{8}$       & $1^{--}$  \\
		\hline
		\rule{0pt}{10pt}$\D$       & 0.42159(5)    & 1962.5(2)   & $\boldsymbol{\bar{3}}$ & $0^-$  \\
		$\Dst$			           & 0.44407(11)   & 2067.1(5)   & $\boldsymbol{\bar{3}}$ & $1^-$   \\
		$\Dz$		               & 0.52319(72)   & 2435(3)     & $\boldsymbol{\bar{3}}$ & $0^+$   \\
		$\Do$	             	   & 0.53977(44)   & 2513(2)     & $\boldsymbol{\bar{3}}$ & $1^+$   \\    
		$\Dopr$                    & 0.54280(44)   & 2527(2)     & $\boldsymbol{\bar{3}}$ & $1^+$   \\   
		$\Dt$    		           & 0.54757(44)   & 2549(2)     & $\boldsymbol{\bar{3}}$ & $2^+$   \\    
		\hline
	\end{tabular}
\caption{Mass, $SU(3)_f$ flavour irrep, and  $J^{P(C)}$ of relevant stable mesons.}
\label{table:mesons}
\end{table}

In this study we are interested in coupled-channel scattering in the open-charm positive-parity sectors.
At the flavour symmetric point, the hadron-hadron channels consist of a stable open-charm meson (in the flavour $\bar{\mathbf{3}}$ irrep) and a stable light meson (in the flavour $\mathbf{8}$ or $\mathbf{1}$ irrep).
Focusing on the flavour-exotic $\mathbf{6}$ and $\overline{\mathbf{15}}$ sectors, only channels with an open-charm meson and a light octet meson contribute.
The masses and quantum numbers of the relevant stable mesons are summarised in Table~\ref{table:mesons}, where the pseudoscalar and vector meson masses were determined in Refs.~\cite{Woss:2018irj,Yeo:2024chk} via dispersion-relation fits,
the $\Dz$ mass is determined from bound-state pole location in the $\D\etaoct(\SLJ{1}{S}{0})$ flavour $\bar{\mathbf{3}}$  scattering amplitude~\cite{Yeo:2024chk},
and the $\Do$, $\Dopr$ and $\Dt$ meson masses are determined by averaging their ground state energies from the relevant at-rest finite-volume irreps across volumes $L/a_s = \{ 12 ,14, 16 ,18,20, 24\}$.
The relevant open-charm scattering channels are summarised in Table~\ref{table:thresholds}.
We will be considering an energy region up to roughly the three-hadron threshold ($\D\etaoct\etaoct$), where all channels up to $\Dt\etaoct$ become kinematically open.
However, we note that for the positive-parity sectors, the $\Dz\etaoct$, $\Do\etaoct$, $\Dopr\etaoct$ and $\Dt\etaoct$ channels are expected to be suppressed since they contribute in a minimum of $P$-wave.
They will not be considered when constraining the scattering amplitudes and we justify neglecting these channels for each finite-volume irrep explored in the next section. 
Therefore, the focus of this study is on the four lowest-threshold channels in Table~\ref{table:thresholds}.
These channels consist of the lightest pseudoscalar or vector charmed meson, $\D \sim( D, D_s)$ or $\Dst \sim( D^*, D^*_s)$, and the lightest pseudoscalar or vector light meson, $\etaoct \sim(\pi, K, \bar{K}, \eta/\eta^{\prime}\text{-mixture})$ or $\omoct \sim(\rho, K^*, \bar{K}^*, \omega/\phi\text{-mixture})$.
The $\Dz$, $\Do$, $\Dopr$ and $\Dt$ mesons are not considered  further in our analysis. 

\begin{table}[ht]
    \centering
\begin{tabular}{cccccc}
    \hline
    \rule{0pt}{12pt}Channel      & $a_tE_{\text{thr}}$        & $E_{\text{thr}}/$MeV   \\
    \hline
    $\D\etaoct$      			   & 0.56939(11)	   & 2650.5(5)	    \\
    $\Dst\etaoct$      			   & 0.59187(15)	   & 2755.2(7)      \\
    $\D\omoct$      			   & 0.63699(21)	   & 2965(1)        \\
    $\Dst\omoct$      			   & 0.65947(23)	   & 3070(1)        \\
    $\Dz\etaoct$      			   & 0.67099(73)	   & 3123(3)        \\
    $\Do\etaoct$      			   & 0.68757(45)	   & 3201(2)        \\
    $\Dopr\etaoct$                 & 0.69060(45)       & 3215(2)        \\
    $\Dt\etaoct$      			   & 0.69537(45)	   & 3237(2)        \\
	$\D\etaoct\etaoct$             & 0.71719(21)	   & 3339(1)        \\
    \hline
\end{tabular}
\caption{Thresholds of open-charm scattering channels for the exotic flavour sectors (up to the first three-hadron channel).}
\label{table:thresholds}
\end{table}

\begin{figure}
  \centering
	\includegraphics[width=1\linewidth]{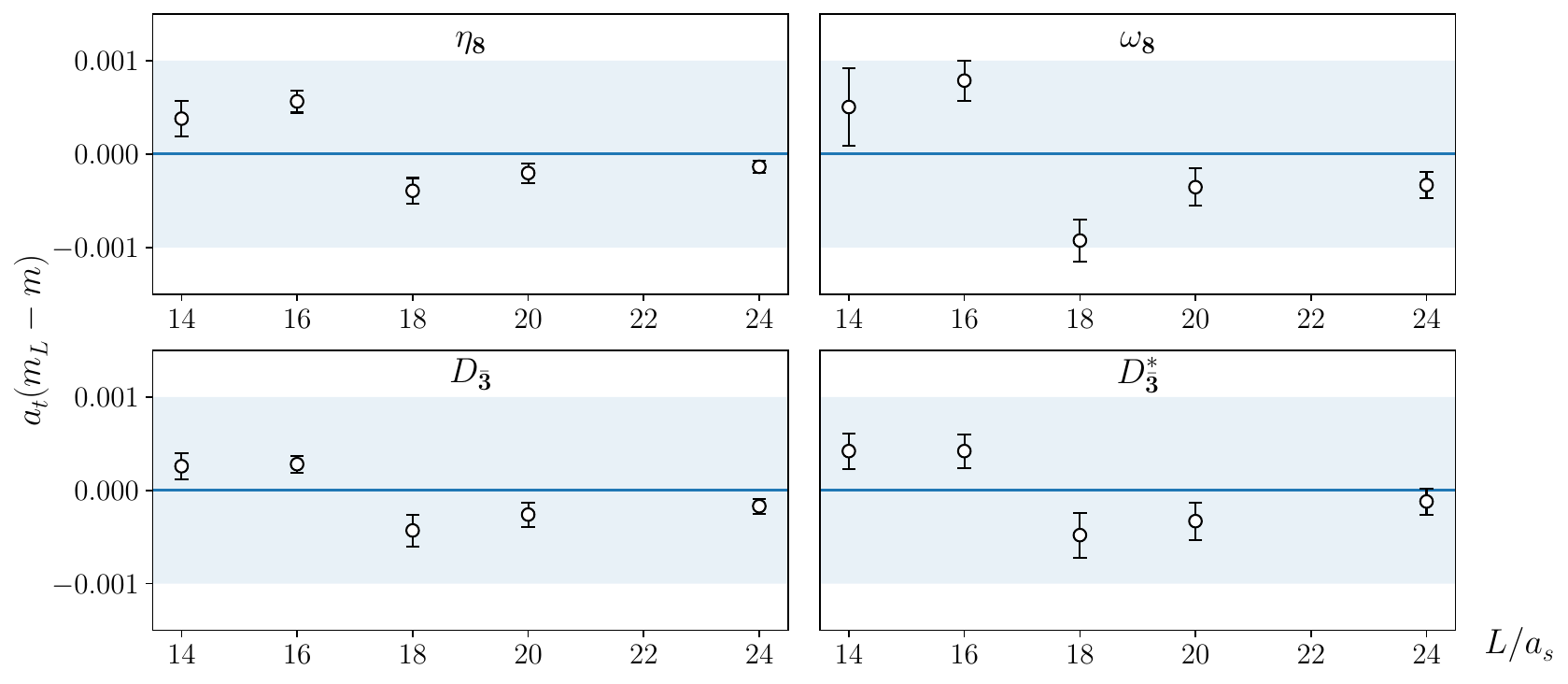}
	\caption{Masses of the $\etaoct$, $\omoct$, $\D$ and $\Dst$ mesons on each volume (points with errorbars), determined by fitting at-rest and in-flight levels from a single volume to the relativistic dispersion relation.
			 These are presented as $m_L-m$, where $m$ is the hadron's mass determined from a relativistic dispersion relation fit across all volumes (as quoted in Table~\ref{table:mesons}), and $m_L= m$ is indicated by the blue line.
			 For each hadron, the blue band indicates the $\delta E_{\text{add}}= 0.001$ envelope from its mass $m$.
			            }
  \label{fig:scattering_meson_masses}
\end{figure}

There are potential sources of uncertainty, such as discretisation effects, that are not included in the statistical analysis. 
One manifestation of these uncertainties is tension of the stable hadrons' masses between different volumes.
As done for the study of elastic $\D\etaoct$ scattering \cite{Yeo:2024chk}, an additional systematic uncertainty is added to the energy levels,
\begin{equation}
\delta E_{\mathbf{n}} \rightarrow \sqrt{\delta E_{\mathbf{n}}^2 + \delta E_{\text{add}}^2},
\end{equation}
to incorporate some of these unaccounted-for effects.
As a conservative approach, we estimate $\delta E_{\text{add}}$ by taking an envelope over the masses of the scattering hadrons across the different volumes (as shown in Fig.~\ref{fig:scattering_meson_masses}), yielding an additional systematic uncertainty of $\delta E_{\text{add}}= 0.001$.

\section{Flavour \texorpdfstring{$\mathbf{6}$}{} sector results} 
\label{section:6f_analysis}
We now discuss the results for the flavour $\mathbf{6}$ sector. We will start by presenting the finite-volume spectra before moving on to discuss amplitudes and  then poles found upon analytic continuation.

\subsection{Finite-volume spectra and partial waves} 
\label{section:6f_fv_spectra}
\begin{figure}
  \centering
	\includegraphics[width=0.9\linewidth]{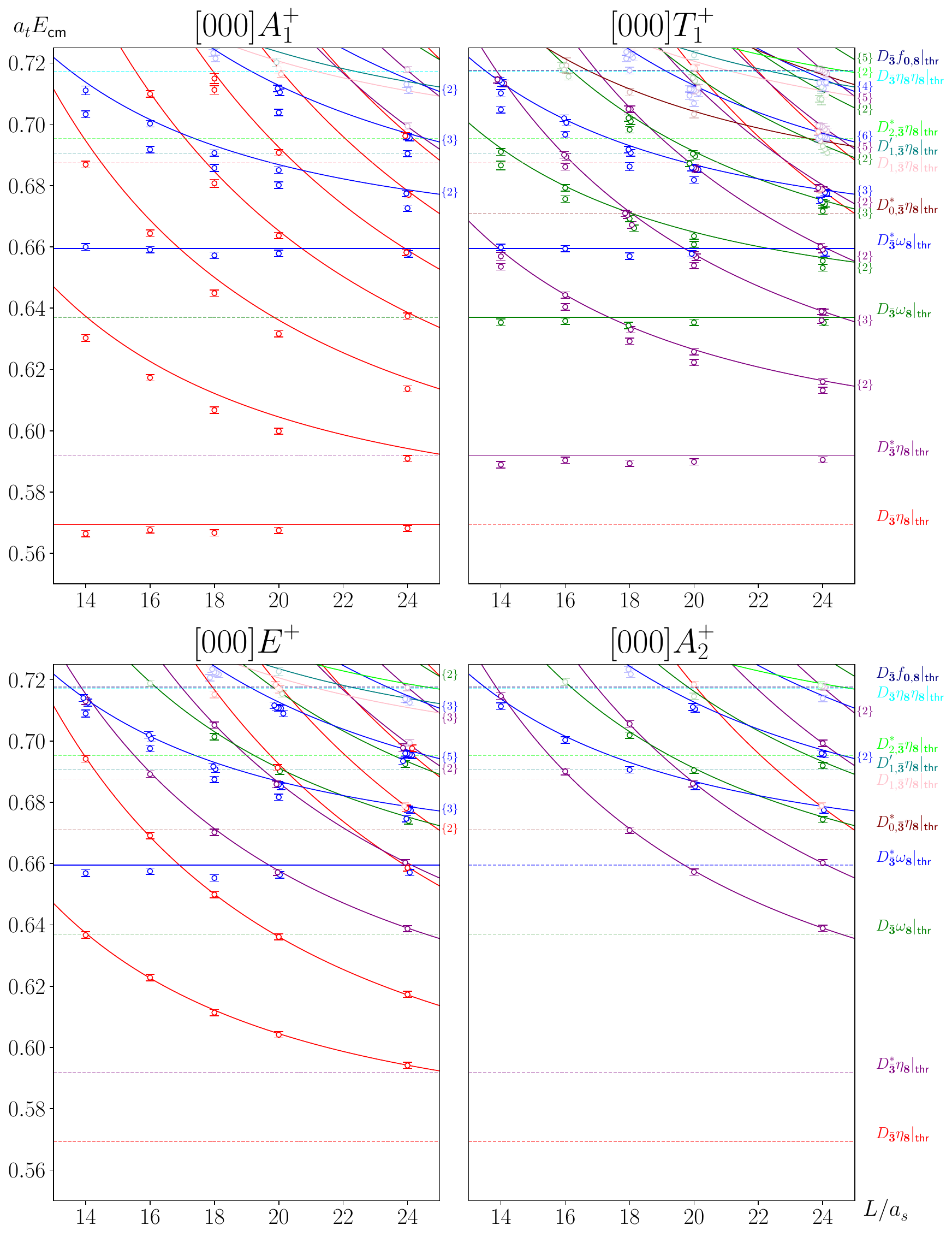}
	\caption{Finite-volume spectra in the $[000]A_1^+$, $[000]T_1^+$,  $[000]E^+$ and $[000]A_2^+$ irreps in the flavour $\mathbf{6}$ sector.
           For each irrep, energy levels (points with error bars) are displayed with their total uncertainty (statistical plus additional systematic uncertainty) 
           and coloured according to the hadron pair of their dominant operator overlap as described in the main text.
           Non-interacting meson-meson energies (curves) and two-hadron channel thresholds (dashed lines) are similarly coloured by hadron pair, with threshold labels (right) 
           indicating the colour key.
           The degeneracies of non-interacting energies are indicated in curly brackets whenever they are more than one.
           Levels not used in determining scattering amplitudes are faded out.}
  \label{fig:6f_spectra}
\end{figure}

\begin{figure}
  \centering
	\includegraphics[width=1\linewidth]{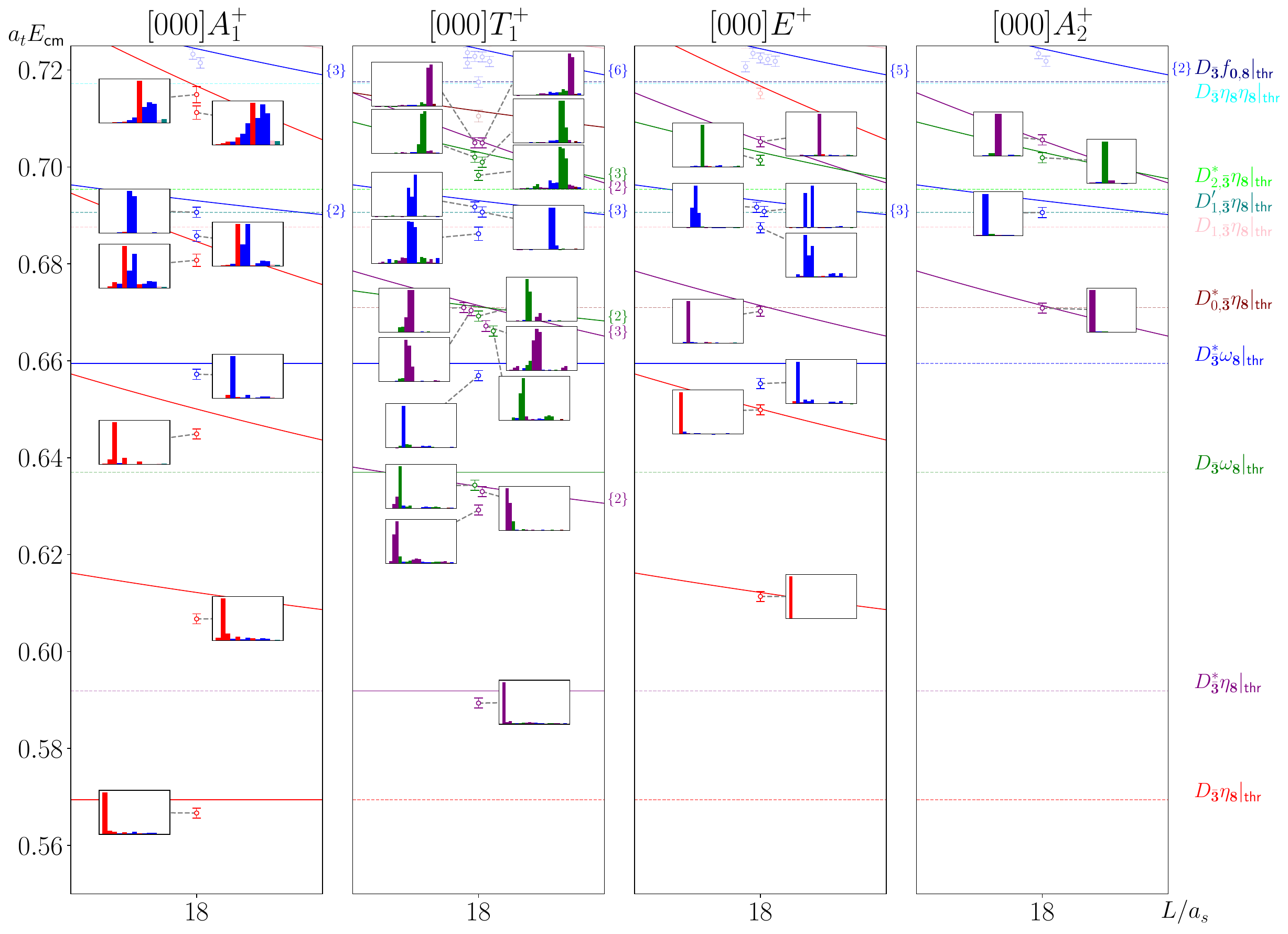}
	\caption{As Fig.~\ref{fig:6f_spectra} but only displaying $L/a_s = 18$ volume spectra. 
           Each energy level is displayed with a histogram of the magnitudes of its operator overlaps,  $|\tilde{Z}^{\mathbf{n}}_i|$, normalised as described in the text.
           Meson-meson operators are ordered by corresponding non-interacting energy (as in the tables in Appendix~\ref{appendix:interpolating_list}). 
           For the $[000]T_1^+$, only the first 20 operators are shown for visual clarity.
  }
  \label{fig:6f_withHist}
\end{figure}

As described in Section~\ref{section:methodology}, large bases of meson-meson interpolating operators were used to extract the finite-volume spectra via the variational method. 
This was done for the $[000]A_1^+$,  $[000]T_1^+$,  $[000]E^+$ and $[000]A_2^+$ finite-volume  irreps on the $L/a_s = \{ 14, 16 ,18, 20, 24\}$ volumes.
The full list of operators used can be found in Appendix~\ref{appendix:interpolating_list}. 
The resulting finite-volume spectra are presented in Fig.~\ref{fig:6f_spectra}, with each level colour coded according to the hadron pair of its dominant normalised operator overlap, $\tilde{Z}^{\mathbf{n}}_i =\langle \mathbf{n} |\mathcal{O}^{\dagger}_i | \Omega \rangle / \text{max}_{\mathbf{m}}\{|Z^{\mathbf{m}}_i|\}$.
Two-hadron decay thresholds and non-interacting meson-meson energy curves are also displayed and similarly coloured via hadron pair.\footnote{We also show the lowest three-hadron threshold, $\D \etaoct \etaoct$. 
All non-interacting three-meson energy curves and other three-hadron thresholds appear above the cutoff of the plot.}
In general, each energy level could be assigned to a single hadron pair via its dominant operator overlap(s),\footnote{We do not assume a correspondence between energy levels and hadron-hadron channels in the scattering analysis.} and in the energy region of interest (roughly up to $\D\etaoct\etaoct$ threshold) there were no ``extra levels'' beyond those expected based off of non-interacting meson-meson energies in that region.
Fig.~\ref{fig:6f_withHist} displays histograms of normalised operator overlaps for the $L/a_s = 18$ volume spectra in each irrep.
The pattern observed on this volume is a good representative of the overall pattern seen across all five volumes. 
The $J^P$ and partial-wave contributions to these irreps,
tabulating $J \leq 6$ and partial waves up to $G$-wave, can be found in Table~\ref{table:pw_sub}.

\begin{table}
  \centering
    \resizebox{\textwidth}{!}{%
    \begin{tabular}{ccccccccc}
    \hline
    Channel & $a_tE_{\rm{thr}}$ & $0^{+}$ & $1^{+}$ & $2^{+}$ & $3^{+}$ & $4^{+}$ & $5^{+}$ & $6^{+}$ \\
    \hline
    \rule{0pt}{12pt}$\D \etaoct$ & 0.569 & \SLJ{1}{S}{0} &  & \SLJ{1}{D}{2} &  & \grey{\SLJ{1}{G}{4}} &  &  \grey{\SLJ{1}{I}{6}} \\
    $\Dst \etaoct$ & 0.592 &  & \SLJ{3}{S}{1}, \SLJ{3}{D}{1} & \SLJ{3}{D}{2} & \SLJ{3}{D}{3}, \grey{\SLJ{3}{G}{3}} & \grey{\SLJ{3}{G}{4}} & \grey{\SLJ{3}{G}{5}} &  \\
    $\D \omoct$ & 0.637 &  & \SLJ{3}{S}{1}, \SLJ{3}{D}{1} & \SLJ{3}{D}{2} & \SLJ{3}{D}{3} \grey{\SLJ{3}{G}{3}} & \grey{\SLJ{3}{G}{4}} & \grey{\SLJ{3}{G}{5}} &  \\
    $\Dst \omoct$ & 0.659 & \SLJ{1}{S}{0}, \SLJ{5}{D}{0} & \SLJ{3}{S}{1}, \SLJ{3,5}{D}{1} & \SLJ{5}{S}{2}, \SLJ{1,3,5}{D}{2}, \grey{\SLJ{5}{G}{2}} & \SLJ{3,5}{D}{3}, \grey{\SLJ{3,5}{G}{3}} & \SLJ{5}{D}{4}, \grey{\SLJ{1,3,5}{G}{4}} & \grey{\SLJ{3,5}{G}{5}} & \grey{\SLJ{5}{G}{6}} \\
    $\Dz \etaoct$ & 0.671 &  & \grey{\SLJ{1}{P}{1}} &  & \grey{\SLJ{1}{F}{3}} &  &  &  \\
    $\Do \etaoct$ & 0.688 & \grey{\SLJ{3}{P}{0}} & \grey{\SLJ{3}{P}{1}} & \grey{\SLJ{3}{P}{2}, \SLJ{3}{F}{2}} & \grey{\SLJ{3}{F}{3}} & \grey{\SLJ{3}{F}{4}} &  &  \\
    $\Dopr \etaoct$ & 0.691 & \grey{\SLJ{3}{P}{0}} & \grey{\SLJ{3}{P}{1}} & \grey{\SLJ{3}{P}{2}, \SLJ{3}{F}{2}} & \grey{\SLJ{3}{F}{3}} & \grey{\SLJ{3}{F}{4}} &  &  \\
    $\Dt \etaoct$ & 0.695 &  & \grey{\SLJ{5}{P}{1}, \SLJ{5}{F}{1}} & \grey{\SLJ{5}{P}{2}, \SLJ{5}{F}{2}} & \grey{\SLJ{5}{P}{3}, \SLJ{5}{F}{3}} & \grey{\SLJ{5}{F}{4}} & \grey{\SLJ{5}{F}{5}} &  \\
    \hline
    \multicolumn{2}{c}{\rule{0pt}{12pt}  Finite-volume irrep} \\  
    \hline
    \multicolumn{2}{c}{$[000]A_1^+$} & $\square$  & &  &  & $\square$ & & \grey{$\triangle$}  \\
    \multicolumn{2}{c}{$[000]E^+$} &  & & $\square$ &  & $\square$ & \grey{$\triangle$} & \grey{$\triangle$} \\
    \multicolumn{2}{c}{$[000]A_2^+$} &   & &  & $\square$ & & & \grey{$\triangle$} \\
    \multicolumn{2}{c}{$[000]T_1^+$} &  & $\square$ &  & $\square$  & $\square$ & \grey{$\triangle$ $\triangle$} & \grey{$\triangle$}  \\[3pt]
    \hline
    \end{tabular}
    }
    \caption{\textbf{Top}: Partial-wave contributions to the positive-parity  sectors up to $J=6$, including channels up to the first three-hadron channel ($\D \etaoct \etaoct$) and partial waves up to $G$-wave,
             written in spectroscopic notation $\SLJ{2S+1}{\ell}{J}$.
             $\D \etaoct(\SLJ{1}{I}{6})$ is also tabulated (with no other $I$-waves tabulated) since this is the lowest-$\ell$ $\D \etaoct$ contribution to $[000]A_2^+$.
             Partial waves not included in the scattering analysis are in grey. 
             \textbf{Bottom}: Black squares and grey triangles indicate $J^P$ contributions to each finite-volume irrep \cite{Johnson:1982yq}, with grey triangles indicating $J^P$ sectors not considered in the scattering analysis.
             Note that $J^P =5^+$ contributes to $[000]T_1^+$ twice.}
    \label{table:pw_sub}
\end{table}

In the $[000]A_1^+$ irrep, downward shifts are seen in all five volumes for some energy levels whose dominant overlap is with either a $\D \etaoct$ (red) or a $\Dst \omoct$ (blue) operator. 
In particular, downward shifts are seen at both $\D\etaoct$ and $\Dst\omoct$ threshold.
For groups of levels with dominant $\Dst \omoct$ operator overlaps that would be degenerate in the absence of meson-meson interactions, only one level displays clear downward shifts whilst the rest lie close to their corresponding non-interacting energies.
These features suggest that the downward shifts are mainly driven by attractive $S$-wave interactions, with other partial waves displaying only weak interactions.
The left-most panel of Fig.~\ref{fig:6f_withHist} indicates energy levels above inelastic threshold significantly overlap onto both $\D \etaoct$ and $\Dst \omoct$ operators.
Such behaviour is typically an indication of cross-channel coupling.

These features are echoed in the other irreps.
In $[000]T_1^+$, downward shifts are seen in some levels with dominant $\Dst \etaoct$ (purple), $\D \omoct$ (green) or $\Dst \omoct$ (blue) operator overlaps,
again with one level significantly downward shifted whenever there is a group of levels would be degenerate in the absence of meson-meson interactions.
We note that the $\Dst \etaoct$, $\D \omoct$ and $\Dst \omoct$ channels each contribute in $S$-wave to this irrep.
In $[000]E^+$, downward shifts are observed in some $\Dst \omoct$ dominated levels -- where $\Dst \omoct$ is the only channel contributing in $S$-wave to this irrep in the energy region of interest -- with energy levels that dominantly overlap onto
the other hadron-hadron operators lying very close to non-interacting energy curves.
Finally, in the $[000]A_2^+$ irrep all energy levels show only small deviations from the non-interacting energy curves.
None of the channels of interest can contribute in $S$-wave in this irrep, supporting the suggestion that the attraction in the other irreps is driven by $S$-wave interactions.
Operator overlaps suggest small or negligible cross-channel coupling for the $[000]T_1^+$, $[000]E^+$ and $[000]A_2^+$ irreps (see Fig.~\ref{fig:6f_withHist}).

To go beyond this qualitative picture, we use the finite-volume spectra to constrain infinite-volume scattering amplitudes via the Lüscher formalism, as described in Section~\ref{section:methodology}.
We start with a discussion of the $[000]A_1^+$ irrep, where we justify neglecting $G$-wave contributions and therefore neglecting the $J^P=\{5, 6\}^+$ sectors, and present the constrained $J^P=\{0, 4\}^+$ amplitudes and poles.
After this, we investigate the $J^P =2^+$ amplitudes from the $[000]E^+$ irrep, with the $J^P=4^+$ contributions to this irrep fixed from the $[000]A_1^+$ analysis.
We then investigate the $J^P=3^+$ amplitudes from  $[000]A_2^+$. 
Finally, we constrain the $J^P = 1^+$ amplitudes from the $[000]T_1^+$ spectra, with contributions from the $J^P=3^+$ and  $J^P=4^+$ amplitudes fixed from the 
$[000]A_2^+$ and $[000]A_1^+$ analyses, respectively.

\subsection{\texorpdfstring{$J^P=\{0, 4\}^+$}{} amplitudes and poles}
\label{section:6f_A1p}

With the goal of constraining the coupled-channel region of $[000]A_1^+$ up to roughly the three-hadron threshold, we note the $\D\etaoct$, $\Dst \omoct$, $\Do \etaoct$, $\Dopr \etaoct$ and $\Dt \etaoct$ channels all become kinematically open and contribute to this irrep (the $\Dst \etaoct$ and $\D \omoct$ channels also contribute but in a minimum of $G$-wave, which we motivate ignoring below).
However, the $\Do \etaoct$ and $\Dopr \etaoct$ channels contribute in $P$-wave and  $\Dt \etaoct$ in $F$-wave, and these channels all have higher thresholds than the $\Dst \omoct$ channel.
By sticking to an energy region below where $\smash{D^{(\prime)}_{1,\bar{\mathbf{3}}} \etaoct }$ or $\Dt \etaoct$  dominated energy levels or non-interacting curves appear,
these channels can be ignored in the scattering analysis.
On the $L/a_s = 24$ volume, there is a faded-purple level at $a_tE_{\sf cm} \approx 0.70$ with dominant overlap on the $D^*_{\bar{\mathbf{3}}[210]} \eta_{\mathbf{8}[210]}$ operator and negligible contributions from other operators.
A $\Dst \etaoct$  partial-wave contribution in the quantisation condition is required to reproduce this level, the lowest\footnote{Lowest in terms of orbital-angular momentum.} of which is $\Dst \etaoct (\SLJ{3}{G}{4})$.
There is also a $G$-wave contribution from the lower channel, $\D \etaoct (\SLJ{1}{G}{4})$.
To test if these contributions were important, fits with and without the lowest two $G$-waves were compared (see Appendix~\ref{appendix:wGwave} for a more in-depth discussion).
The diagonal $G$-wave amplitudes were found to be small and their inclusion made negligible difference to the determination of the other amplitudes.
Therefore, throughout the rest of this study we ignore $G$-wave and higher contributions and levels that would appear as solutions to these ignored partial waves.
For the $[000]A^+_1$, this amounts to ignoring just this purple energy level.
Since  $J^P = \{5, 6\}^+$ only contribute to the $[000]A_1^+$, $[000]A_2^+$, $[000]E^+$ and $[000]T_1^+$ irreps in a minimum of $G$-wave, these $J^P$ are not explored any further in this work.  
Therefore, we use 46 energy levels from the $[000]A_1^+$ irrep to constrain the coupled-channel $\D \etaoct (\SLJ{1}{S}{0})- \Dst \omoct (\SLJ{1}{S}{0})- \Dst \omoct (\SLJ{5}{D}{0})$  and elastic $\Dst \omoct (\SLJ{5}{D}{4})$ amplitudes.
Partial waves considered in the scattering analysis are in black in Table~\ref{table:pw_sub} and  ignored partial waves are in grey. 
As done in our elastic study of $\D \etaoct (\SLJ{1}{S}{0})$ in the flavour $\mathbf{6}$ sector, we also use the three lowest levels from the $[100]A_1$ irrep and two lowest levels from $[110]A_1$ irrep on the $L/a_s =24$ volume to further constrain the near-threshold region of the $\D \etaoct (\SLJ{1}{S}{0}) \rightarrow \D \etaoct (\SLJ{1}{S}{0})$ amplitude.\footnote{In that study, we demonstrated these levels are dominated by the $\D \etaoct (\SLJ{1}{S}{0})$ channel and hence the
contributions from other partial waves to the in-flight irreps can be ignored.}
These five additional levels are presented in Fig.~7 of Ref.~\cite{Yeo:2024chk}.
In total, 51 levels were used to constrain the $J^P=\{0, 4\}^+$ amplitudes via the Lüscher quantisation condition.

\begin{figure}
  \centering
	\includegraphics[width=.9\linewidth]{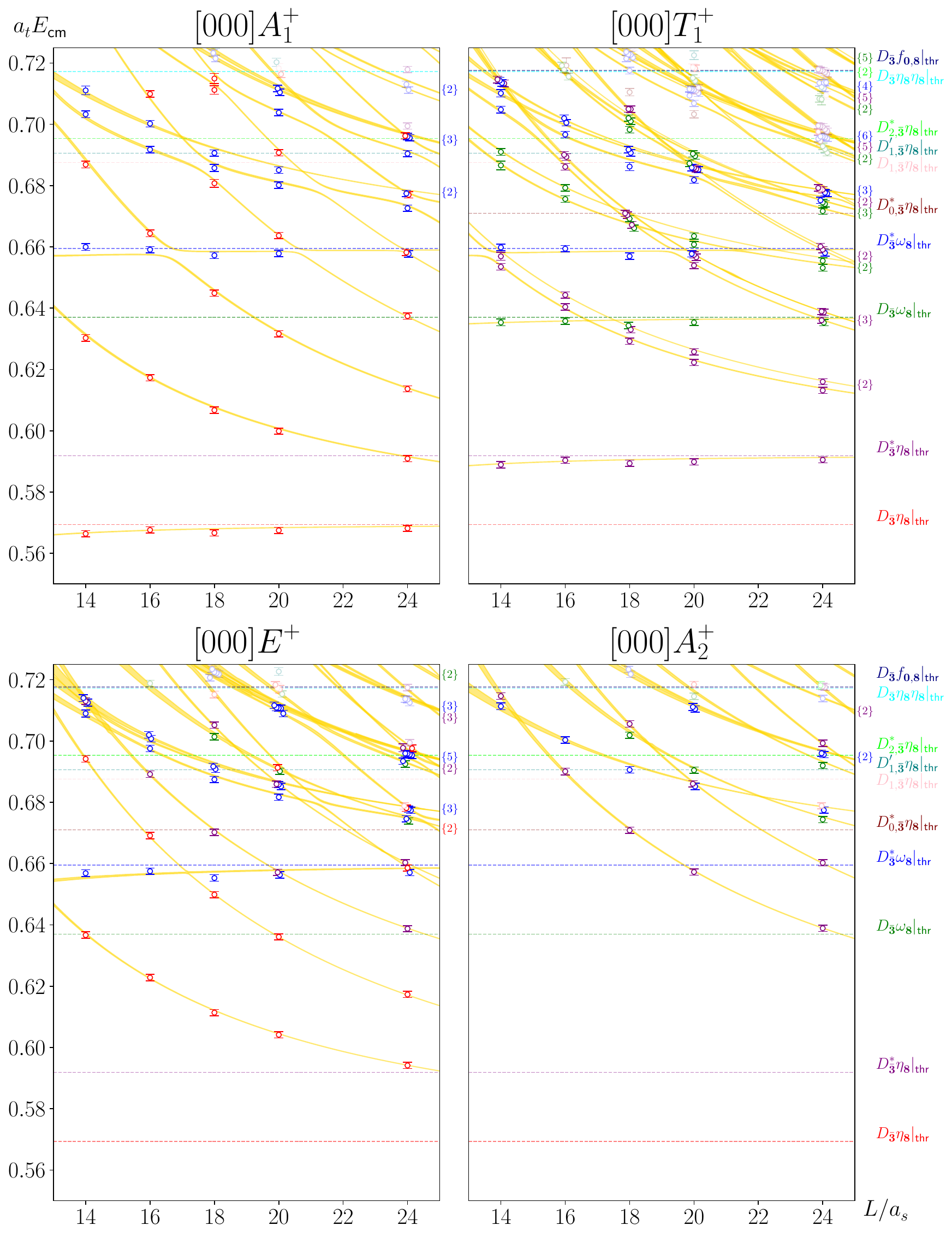}
	\caption{The finite-volume spectra and channel thresholds as in Fig.~\ref{fig:6f_spectra}, along with the finite-volume spectra from the reference parameterisation of each irrep displayed as a function of $L/a_s$ (golden bands). }
  \label{fig:6f_spectra_get_finite}
\end{figure}

For the $J^P=0^+$ amplitudes, $K$-matrix polynomial and inverse $K$-matrix polynomial parameterisations were considered with both phase space prescriptions,
whilst just a constant term with Chew-Mandelstam phase space was used for the diagonal $\Dst\omoct(\SLJ{5}{D}{4})$ amplitude.
We retain all parameterisations with a minimised $\chisq< 2$, which we refer to as ``reasonable parameterisations''.
A relatively simple parameterisation that gives a reasonable description of the data, which we call our \emph{reference parameterisation}, is an inverse $K$-matrix parameterisation with Chew-Mandelstam phase space. 
The fit parameters are
\begin{equation}
    \centering
    \resizebox{\textwidth}{!}{%
    \begin{tabular}{r@{ }ll}    
     $c^{(0)} \{\D \etaoct (\SLJ{1}{S}{0}) | \D \etaoct (\SLJ{1}{S}{0})\} =$ & $(2.21 \pm 0.13 \pm 7.14) $ &
     \resizebox{0.6\textwidth}{!}{
     \multirow{8}{*}{  $\left[ \begin{array}{rrrrrrrr}    
      1.00 & -0.56 &  0.19 &  0.20 &  0.10 &  0.11 &  0.06 & -0.12 \\
           &  1.00 & -0.03 &  0.42 &  0.22 &  0.11 &  0.10 & -0.12 \\
           &       &  1.00 &  0.82 &  0.43 &  0.14 &  0.19 & -0.19 \\
           &       &       &  1.00 &  0.58 &  0.17 &  0.25 & -0.29 \\
           &       &       &       &  1.00 & -0.69 &  0.17 & -0.20 \\
           &       &       &       &       &  1.00 &  0.05 & -0.07 \\
           &       &       &       &       &       &  1.00 &  0.09 \\
           &       &       &       &       &       &       &  1.00 \\
     \end{array} \right]$ } }
     \\ 
     $c^{(1)} \{\D \etaoct (\SLJ{1}{S}{0}) | \D \etaoct (\SLJ{1}{S}{0})\} =$ & $(-17.7 \pm 0.6 \pm 38.4)\cdot a_t^{2} $ & \\
     $c^{(2)} \{\D \etaoct (\SLJ{1}{S}{0}) | \D \etaoct (\SLJ{1}{S}{0})\} =$ & $(45.0 \pm 1.8 \pm 67.5)\cdot a_t^{4} $ & \\
     $c^{(0)} \{\D \etaoct (\SLJ{1}{S}{0}) | \Dst \omoct (\SLJ{1}{S}{0})\} =$ & $(1.2 \pm 0.2 \pm 1.2) $ & \\
     $c^{(0)} \{\Dst \omoct (\SLJ{1}{S}{0}) | \Dst \omoct (\SLJ{1}{S}{0})\} =$ & $(2.81 \pm 0.13 \pm 1.20) $ & \\
     $c^{(1)} \{\Dst \omoct (\SLJ{1}{S}{0}) | \Dst \omoct (\SLJ{1}{S}{0})\} =$ & $(-4.6 \pm 0.2 \pm 1.7)\cdot a_t^{2} $ & \\
     $c^{(0)} \{\Dst \omoct (\SLJ{5}{D}{0}) | \Dst \omoct (\SLJ{5}{D}{0})\} =$ & $(0.05 \pm 0.06 \pm 154.36)\cdot a_t^{-4} $ & \\
     $\gamma^{(0)} \{\Dst \omoct (\SLJ{5}{D}{4}) | \Dst \omoct (\SLJ{5}{D}{4})\} =$ & $(120 \pm 20 \pm 40)\cdot a_t^{4} $ & \\[10pt]
     &\multicolumn{2}{l}{$\chisq = \frac{50.80}{51-8} = 1.18$\,.}
    \end{tabular}
    }
\label{eq:A1p6f_ref}
\end{equation}

\begin{figure}
  \centering
	\includegraphics[width=1\linewidth]{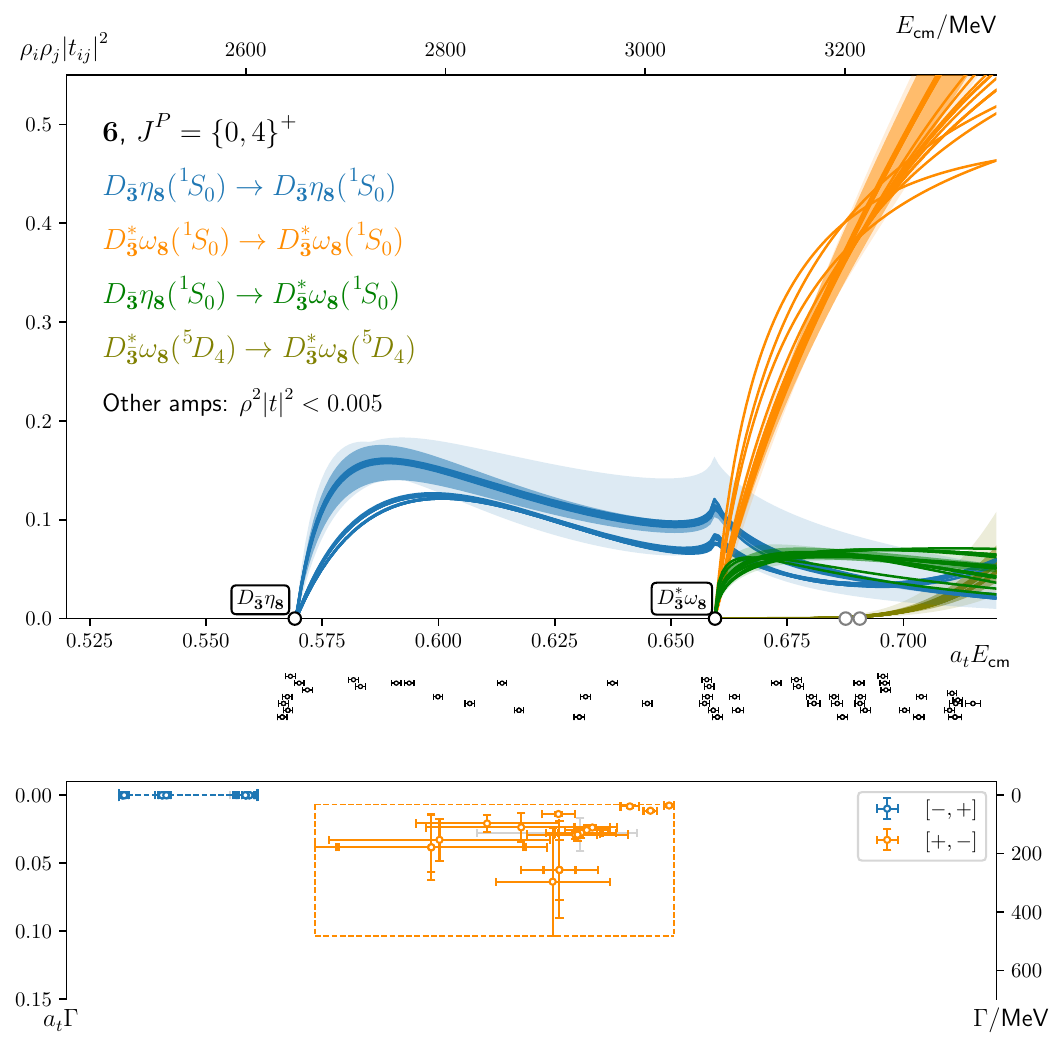}
	\caption{\textbf{Top panel}: $J^P = 0^+$ and $4^+$ amplitudes in the flavour $\mathbf{6}$ sector from reasonable fits to the $[000]A_1^+$ spectrum, displayed as $\rho_i \rho_j |t_{ij}|^2$. 
           Amplitudes from the reference parameterisation are shown with their central values (curves), statistical uncertainties (inner bands) and uncertainties from
           mass and anisotropy variations (outer bands).
           Central values of all other reasonable fits are also shown (curves).
           Circles on the horizontal axis indicate kinematic thresholds, with circles in grey corresponding to thresholds from ignored channels.
           \textbf{Middle panel}: Points indicate energies at which the $t$-matrix has been constrained. 
           \textbf{Bottom panel}: Nearest $J^P = 0^+$ pole locations in the complex energy plane and their statistical uncertainty for each reasonable parameterisation. 
           For the reference parameterisation, uncertainty from  mass and anisotropy variations are also shown (light grey error bars).
           In each parameterisation, a virtual bound state is found on the $[-,+]$ sheet (blue) and a resonance on the $[+,-]$ sheet (orange).
           Envelopes over the virtual bound state (blue dashed error bar) and resonance (orange dashed box) positions across the reasonable parameterisations are indicated.  
           }
  \label{fig:A1p6f_all_fits}
\end{figure}
\noindent shown with their central values, statistical uncertainty (first error), spread under mass-anisotropy variations (second error, where one sigma variations on the hadrons' masses and the anisotropy are considered and an envelope over their statistical uncertainties are taken) and corresponding correlation matrix.
The spectrum corresponding to this parameterisation, obtained by determining the solutions of the quantisation condition as a function of $L/a_s$, is plotted in Fig.~\ref{fig:6f_spectra_get_finite}.
There we see the finite-volume spectrum computed from lattice QCD is reproduced well from the parameterisation, with several avoided level crossings around and above $\Dst\omoct$ threshold emblematic of nonzero cross-channel coupling.
This parameterisation is plotted in Fig.~\ref{fig:A1p6f_all_fits} as $\rho_i \rho_j |t_{ij}|^2$.
Other reasonable parameterisations are also plotted in Fig.~\ref{fig:A1p6f_all_fits} and a summary of all of these can be found in Table~\ref{table:A1p6f_parameterisation_table} of Appendix~\ref{appendix:table_of_parameterisations}. 
When extrapolated away from the constrained energy region, some parameterisations can feature a bound-state pole  despite no corresponding energy level in the finite-volume spectrum.
The procedure we use~\cite{Yeo:2024chk, PitangaLachini:2025pxr} to push these ``spurious'' bound-state poles far enough below the constrained energy region as to not affect the physical scattering amplitude
 is discussed in Appendix~\ref{appendix:table_of_parameterisations}.
 Parameterisations for which this is invoked are appropriately labelled in the tables.

It was found that the diagonal $\D \etaoct (\SLJ{1}{S}{0})$,  $\Dst\omoct(\SLJ{1}{S}{0})$, and  $\Dst\omoct(\SLJ{5}{D}{0})$ amplitudes
required, respectively,  a second-order polynomial, a nonzero constant term (with higher-order terms improving the $\chi^2$), and a nonzero constant term to achieve reasonable fits to the data.
Furthermore, a nonzero term for the off-diagonal $\D\etaoct(\SLJ{1}{S}{0}) \rightarrow \Dst\omoct(\SLJ{1}{S}{0})$ amplitude was needed,
whilst other off-diagonal terms, when they were not set to zero, were found to be small and statistically consistent with zero.
In general, the inverse $K$-matrix parameterisations were found to provide a better description to the data, as evidenced by their smaller minimised \chisq\ (see Table~\ref{table:A1p6f_parameterisation_table}). 
Overall, we find strong interactions in the diagonal $\D \etaoct (\SLJ{1}{S}{0})$ and diagonal $\Dst \omoct (\SLJ{1}{S}{0})$ amplitudes with a nonzero coupling between these two partial waves.
This aligns with the qualitative discussion of the finite-volume spectra and operator overlaps in Section~\ref{section:6f_fv_spectra}.

\begin{figure}
  \centering
	\includegraphics[width=1\linewidth]{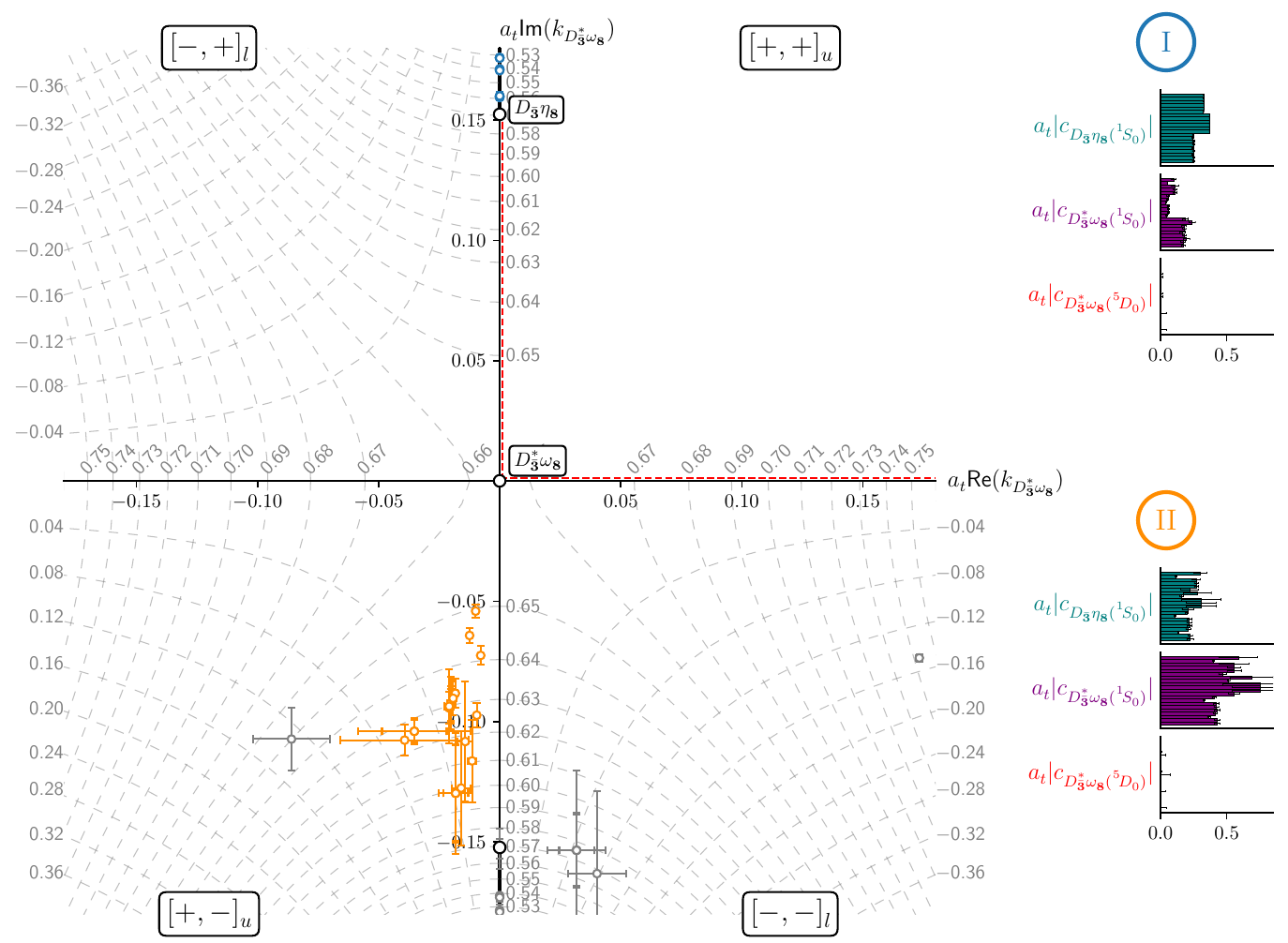}
	\caption{Poles in the $J^P=0^+$ amplitudes in the flavour $\mathbf{6}$ sector from each reasonable parameterisation plotted in the complex $\smash{k_{\Dst\omoct}}$ plane.
           Dashed red line denotes where the physical scattering amplitude is defined.
           Grey dashed lines show contours of constant $m= \text{Re}(\sqrt{s})$ (labelled on $x$- and $y$-axes)
           and $\Gamma= 2 \text{Im}(\sqrt{s})$ (labelled on left and right side of the plot).
           Virtual-bound-state locations on the $[-, +]$ sheet (blue points) and resonance locations on the $[+, -]$ sheet (orange points) are shown for each reasonable parameterisation, alongside additional poles (grey points).
           The magnitudes of partial-wave couplings for the virtual bound state and resonance are also displayed (histograms labelled ``$\mathrm{\Romannum{1}}$'' and ``$\mathrm{\Romannum{2}}$'', respectively), where each bar corresponds to the coupling value from a reasonable parameterisation.
           }
  \label{fig:A1p_poles_kcm}
\end{figure}

We next discuss the pole structure of reasonable parameterisations that resulted from analytic continuation to the complex energy plane, starting with the $J^P = 0^+$, $\D \etaoct (\SLJ{1}{S}{0})- \Dst \omoct (\SLJ{1}{S}{0})- \Dst \omoct (\SLJ{5}{D}{0})$ amplitudes.
With two distinct hadron-hadron channels in these amplitudes, there are four sheets where singularities could arise.  
For each reasonable parameterisation, a pole is found on the real axis below $\D \etaoct$ threshold on the $[-, +]$ sheet, corresponding to a virtual bound state.
The position in the complex $E_{\sf{cm}}$ plane is plotted in the bottom panel of Fig.~\ref{fig:A1p6f_all_fits} and in the complex $\smash{k_{\Dst\omoct}}$ plane in Fig.~\ref{fig:A1p_poles_kcm} (blue points).
Furthermore, in Fig.~\ref{fig:A1p_poles_kcm} the pole's couplings to each partial wave are presented (histograms labelled ``$\mathrm{\Romannum{1}}$''), where each bar corresponds to the coupling value from a different reasonable parameterisation.
Across these parameterisations, the pole was found to primarily couple to $\D \etaoct(\SLJ{1}{S}{0})$ with a smaller but nonzero coupling to $\Dst \omoct (\SLJ{1}{S}{0})$, and a negligible coupling  to $\Dst \omoct (\SLJ{5}{D}{0})$ (whenever it was not fixed to zero).
This is the same virtual bound state found in our elastic study of this sector \cite{Yeo:2024chk}.
We note, however, the spread of the pole's location across different parameterisations in this analysis is slightly larger than that found in the elastic study.
This can be understood due to the fact that now the parameterisations are being used to model the amplitudes over a larger energy region and a larger additional systematic uncertainty is being used.
Nevertheless, the picture of a virtual bound state below elastic $\D \etaoct$ threshold remains unchanged.
The pole location of this virtual bound state is
\begin{align}
a_t \sqrt{s_{\text{pole}}} &= 0.546 \pm 0.015, \\   
\sqrt{s_{\text{pole}}} &= 2540 \pm 70\, \text{MeV},
\end{align}
where we quote an envelope over the positions from different reasonable parameterisations.

In addition to the virtual bound state, a resonance  below $\Dst \omoct$ threshold is observed for each reasonable fit.
The location of this pole is similarly plotted for each parameterisation in  Fig.~\ref{fig:A1p6f_all_fits} and~\ref{fig:A1p_poles_kcm}  (orange points), with the latter also displaying the pole's partial-wave couplings (histograms labelled ``$\mathrm{\Romannum{2}}$'').
It is situated on the $[+,-]$ sheet, the hidden sheet reached from the proximal sheet by moving up through the cut between $\D\etaoct$ and $\Dst\omoct$ threshold.
Such hidden sheet poles will have a less direct impact on the physical scattering amplitude and will be less well constrained from the data if they are further from threshold.
Some parameterisations found this pole deep into the complex plane and with a large uncertainty.
However, all reasonable parameterisations we considered featured this pole.
It is found to predominantly couple to $\Dst \omoct (\SLJ{1}{S}{0})$ with a smaller but nonzero coupling to $\D \etaoct (\SLJ{1}{S}{0})$, and a coupling to $\Dst \omoct (\SLJ{5}{D}{0})$  that is statistically equivalent with zero (when not fixed to zero).
The pole's coupling to $\Dst \omoct (\SLJ{1}{S}{0})$ is large, resulting in a clear rapid turn on of the diagonal $\Dst \omoct (\SLJ{1}{S}{0})$ amplitude at threshold.
We give the location of this pole ($\sqrt{s_{\text{pole}}} = m_R + \frac{i}{2}\Gamma$) as 
\begin{align}
a_t \sqrt{s_{\text{pole}}}&= (0.61 \pm 0.04) + \tfrac{i}{2}(0.06 \pm 0.05 ) \\  
\sqrt{s_{\text{pole}}} &= (2850 \pm 180) + \tfrac{i}{2}(260 \pm 220 ) \, \text{MeV},                
\end{align}
similarly quoted as an envelope over the different reasonable parameterisations, providing a conservative estimate for the uncertainty on the mass and decay width.

For a select few parameterisations, there are additional poles and these are plotted in grey in Fig.~\ref{fig:A1p_poles_kcm}. 
The combination of these poles' couplings and distance from physical scattering, along with the fact that they don't appear in every parameterisation, suggests they are not needed to describe the physical scattering amplitude and are not robustly constrained by the data. 
In particular, these additional poles are not found in the inverse $K$-matrix parameterisations which have a better minimised \chisq.
Therefore, we conclude that there are two well-determined $J^P =0^+$ poles in the energy region considered.
In both cases, the pole is found not on the proximal sheet but on a hidden sheet.

To finish this section, we mention that no poles were found close to the constrained energy region in the $J^P=4^+$ amplitude.

\subsection{\texorpdfstring{$J^P = 2^+$}{} amplitudes and poles}

We constrain the $J^P=2^+$ amplitudes from the $[000]E^+$ finite-volume irrep (which also gets contributions from  $J^P = 4^+$, see Table~\ref{table:pw_sub}).
Analogously to the $[000]A_1^+$, there are levels on the $L/a_s = 24$ volume that require the inclusion of $G$-wave (or higher) contributions to be reproduced in the QC.
In particular, there are two near-degenerate $\D\etaoct$ dominated levels at $a_tE_{\sf cm} \approx 0.68$ (red and faded-red in Fig.~\ref{fig:6f_spectra}) which require 
the inclusion of the $\D\etaoct(\SLJ{1}{D}{2})$ and a second  $\D\etaoct$ partial wave, 
and two near-degenerate $\Dst\etaoct$ dominated levels at $a_tE_{\sf cm} \approx 0.70$ (purple and faded-purple levels in Fig.~\ref{fig:6f_spectra}) requiring $\Dst \etaoct(\SLJ{3}{D}{2})$  and a second $\Dst \etaoct$ partial wave. 
In each case, the next lowest contribution is $G$-wave.
Since we are neglecting these, we opt to ignore one of the two near-degenerate levels, which is faded out in the spectrum plot.
In practice, the choice of ignored level in both cases made small difference to the resulting amplitudes.\footnote{This was tested by comparing the reference parameterisation to fits where the choice of ignored level in each case was alternated.
It was found that there were small changes in  the \chisq\ and fit parameters, and the resulting poles were statistically consistent (within roughly one third of the standard deviation).}
Apart from the two aforementioned levels, we use all the energy levels up to approximately the three-hadron threshold or the nearest $\smash{D^{(\prime)}_{1,\bar{\mathbf{3}}} \etaoct }$ and $\Dt \etaoct$
 non-interacting energy curves or dominated levels.
These channels contribute in a minimum of $P$-wave and with this choice of cutoff on each volume, they can be ignored in the scattering analysis.
This leads to a total of 58 levels to constrain the  $J^P=2^+$, coupled-channel $\D \etaoct (\SLJ{1}{D}{2})- \Dst \etaoct (\SLJ{3}{D}{2})- \D \omoct (\SLJ{3}{D}{2})-\Dst \omoct (\SLJ{5}{S}{2})- \Dst \omoct (\SLJ{1}{D}{2}) - \Dst \omoct (\SLJ{3}{D}{2}) - \Dst \omoct (\SLJ{5}{D}{2})$
amplitudes.  
For the $J^P= 4^+$ amplitude, we opt to use a constant $K$-matrix with Chew-Mandelstam phase space, with its value fixed to the central value of the reference parameterisation in Eq.~\ref{eq:A1p6f_ref} (constrained from the $[000]A_1^+$ spectra) rather than to perform another independent fit of the amplitude.
As part of the parameterisation variations, we will consider fits with this fixed parameter shifted up and down by its one sigma statistical uncertainty.

\begin{figure}
  \centering
	\includegraphics[width=1\linewidth]{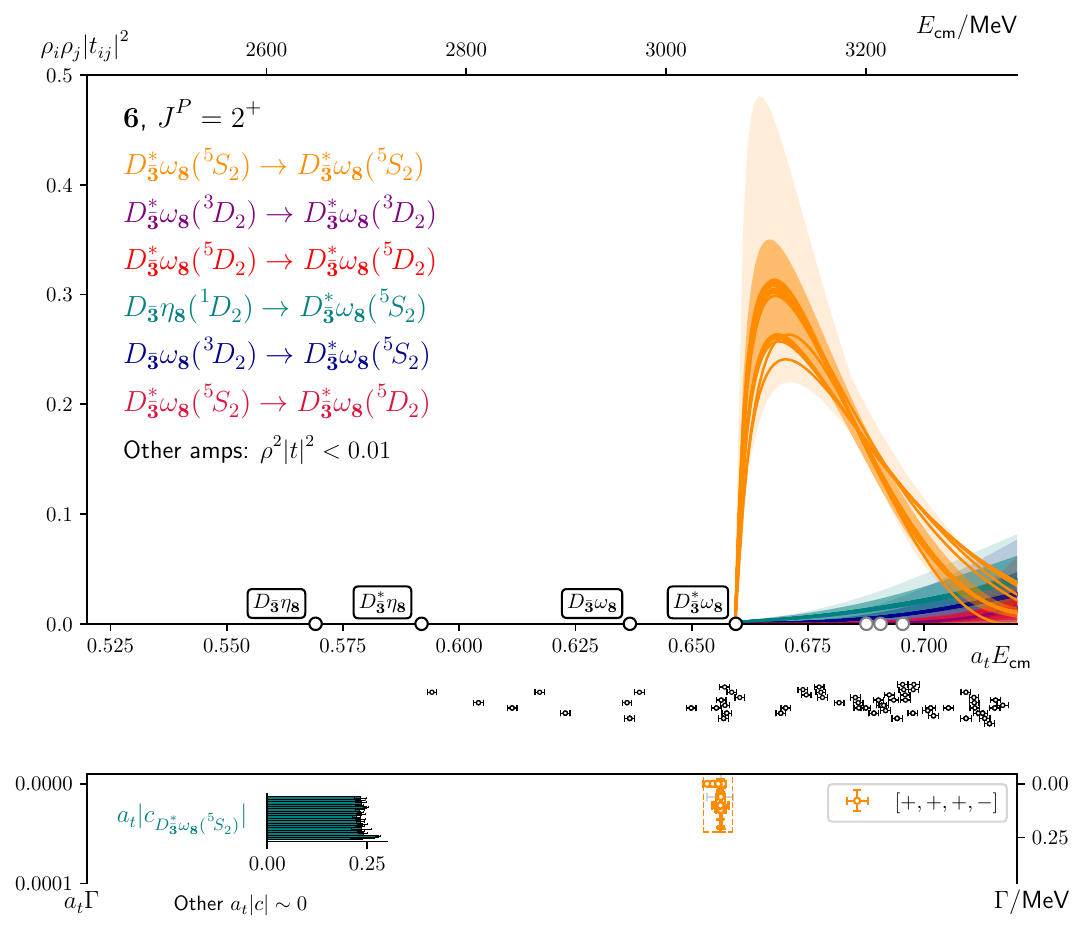}
	\caption{As Fig.~\ref{fig:A1p6f_all_fits} but for the $J^P = 2^+$ amplitudes in the flavour $\mathbf{6}$ sector from reasonable fits to the $[000]E^+$ spectrum.
           The fixed $J^P =4^+$ amplitude is not plotted.
           Bottom panel displays the nearest $J^P = 2^+$ poles for each reasonable parameterisation. 
           In each parameterisation, one pole is found just below $\Dst \omoct$ threshold.
           Histograms display the magnitudes of partial-wave couplings for the pole found on the $[+,+,+,-]$ sheet.}
  \label{fig:Ep6f_all_fits}
\end{figure}

For parameterising the $t$-matrix in the $J^P =2^+$ sector, we considered $K$-matrix polynomial, inverse polynomial and ratio-of-polynomials parameterisations with both phase-space prescriptions.
It was found only the latter could achieve reasonable descriptions to the data free of pathologies.
We take a parameterisation with Chew-Mandelstam phase space as our reference parameterisation.
Its fit parameters are 
\begin{equation}
  \centering
  \resizebox{\textwidth}{!}{%
  \begin{tabular}{r@{ }ll}    
$d^{(1)} \{\Dst\omoct(\SLJ{5}{S}{2}) |  \Dst\omoct(\SLJ{5}{S}{2})  \} =$  & $(-2.43 \pm 0.03 \pm 0.06) \cdot a_t^2 $ & \resizebox{1.19\textwidth}{!}{\multirow{15}{*}{ 
 $\left[ \begin{array}{rrrrrrrrrrrrrrr} 
1.00 &   0.20 &  -0.11 &   0.18 &  -0.05 &   0.43 &  -0.35 &   0.41 &   0.39 &   0.08 &   0.01 &   0.39 &   0.36 &   0.36 &  -0.23\\
&  1.00 &  -0.39 &   0.18 &   0.07 &   0.44 &  -0.10 &   0.25 &   0.17 &   0.23 &   0.00 &   0.18 &   0.04 &   0.01 &   0.00\\
&&  1.00 &  -0.13 &   0.23 &  -0.32 &   0.12 &  -0.05 &  -0.21 &  -0.24 &   0.01 &   0.01 &   0.14 &   0.05 &  -0.07\\
&&&  1.00 &  -0.21 &   0.24 &  -0.08 &   0.31 &   0.07 &   0.18 &  -0.00 &   0.23 &   0.13 &   0.00 &   0.02\\
&&&&  1.00 &  -0.03 &  -0.05 &   0.03 &   0.05 &  -0.10 &   0.00 &  -0.04 &   0.08 &   0.03 &  -0.03\\
&&&&&  1.00 &  -0.33 &   0.26 &   0.45 &   0.25 &   0.00 &   0.29 &   0.24 &  -0.03 &   0.08\\
&&&&&&  1.00 &  -0.24 &  -0.72 &  -0.05 &  -0.00 &  -0.04 &  -0.27 &   0.02 &  -0.08\\
&&&&&&&  1.00 &   0.18 &   0.19 &  -0.00 &   0.47 &   0.43 &   0.03 &   0.02\\
&&&&&&&&  1.00 &   0.04 &   0.01 &  -0.11 &   0.31 &  -0.02 &   0.10\\
&&&&&&&&&  1.00 &   0.05 &   0.31 &   0.09 &   0.04 &  -0.05\\
&&&&&&&&&&  1.00 &  -0.01 &  -0.02 &  -0.00 &   0.00\\
&&&&&&&&&&&  1.00 &   0.32 &   0.02 &   0.01\\
&&&&&&&&&&&&  1.00 &  -0.03 &   0.10\\
&&&&&&&&&&&&&  1.00 &  -0.99\\
&&&&&&&&&&&&&&  1.00
\end{array} \right]$ }
}\\
$c^{(0)} \{\D\etaoct(\SLJ{1}{D}{2}) |  \D\etaoct(\SLJ{1}{D}{2})  \} =$  & $(2 \pm 3 \pm 15)  \cdot a_t^4  $ & \\
$c^{(0)} \{\D\etaoct(\SLJ{1}{D}{2}) | \Dst\omoct(\SLJ{5}{S}{2}) \} =$ & $(-3.0 \pm 0.9 \pm 3.0)  \cdot a_t^2 $ & \\
$c^{(0)} \{\D\omoct(\SLJ{3}{D}{2}) |  \D\omoct(\SLJ{3}{D}{2})  \} =$ & $(22 \pm 12 \pm 12)  \cdot a_t^4 $ & \\
$c^{(0)} \{\D\omoct(\SLJ{3}{D}{2}) | \Dst\omoct(\SLJ{5}{S}{2}) \} =$  & $(-4.5 \pm 1.8 \pm 1.9)  \cdot a_t^2 $ & \\
$c^{(0)} \{\Dst\etaoct(\SLJ{3}{D}{2}) |  \Dst\etaoct(\SLJ{3}{D}{2})  \} =$ & $(8 \pm 3 \pm 9)  \cdot a_t^4 $ & \\
$c^{(0)} \{\Dst\etaoct(\SLJ{3}{D}{2}) |  \Dst\omoct(\SLJ{5}{S}{2}) \} =$ & $(0.2 \pm 0.8 \pm 2.0)  \cdot a_t^2 $ & \\
$c^{(0)} \{\Dst\omoct(\SLJ{1}{D}{2}) |  \Dst\omoct(\SLJ{1}{D}{2})  \} =$ & $(16 \pm 17 \pm 14)  \cdot a_t^4 $ & \\
$c^{(0)} \{\Dst\omoct(\SLJ{1}{D}{2}) |  \Dst\omoct(\SLJ{5}{S}{2})  \} =$ & $(1.6 \pm 1.5 \pm 5.2)  \cdot a_t^2 $ & \\
$c^{(0)} \{\Dst\omoct(\SLJ{3}{D}{2}) |  \Dst\omoct(\SLJ{3}{D}{2})  \} =$ & $(59 \pm 18 \pm 13) \cdot a_t^4  $ & \\
$c^{(0)} \{\Dst\omoct(\SLJ{3}{D}{2}) |  \Dst\omoct(\SLJ{5}{S}{2})  \} =$  & $(0 \pm 4 \pm 3)  \cdot a_t^2 $ & \\
$c^{(0)} \{\Dst\omoct(\SLJ{5}{D}{2}) |  \Dst\omoct(\SLJ{5}{D}{2})  \} =$ & $(60 \pm 30 \pm 50)  \cdot a_t^4 $ & \\
$c^{(0)} \{\Dst\omoct(\SLJ{5}{D}{2}) |  \Dst\omoct(\SLJ{5}{S}{2})  \} =$& $(4.4 \pm 1.4 \pm 0.2)  \cdot a_t^2 $ & \\
$c^{(0)} \{\Dst\omoct(\SLJ{5}{S}{2}) |  \Dst\omoct(\SLJ{5}{S}{2})  \} =$ & $(-2.46 \pm 0.17 \pm 0.48)$ & \\
$c^{(1)} \{\Dst\omoct(\SLJ{5}{S}{2}) |  \Dst\omoct(\SLJ{5}{S}{2})  \} =$ & $(4.6 \pm 0.3 \pm 1.0)  \cdot a_t^2 $ & \\[1.3ex]
&\multicolumn{2}{l}{ $ \chisq = \frac{38.03}{58-15} = 0.88$\,.}
\end{tabular}
  }
\label{eq:Ep6f_ref}
\end{equation}
\noindent The spectrum corresponding to the reference parameterisation is plotted in Fig.~\ref{fig:6f_spectra_get_finite},
where we see good agreement with the finite-volume spectrum computed using lattice QCD.
This parameterisation is plotted in  Fig.~\ref{fig:Ep6f_all_fits} with its statistical uncertainty and the uncertainty from
mass and anisotropy variations.
All other reasonable fits to the data are also displayed in Fig.~\ref{fig:Ep6f_all_fits} and are summarised in Table~\ref{table:Ep6f_parameterisation_table}.
Included are fits with the same form as the reference parameterisation but with the fixed parameter of the $\Dst \omoct (\SLJ{5}{D}{4})$ amplitude varied by plus or minus one sigma from its central value (from Eq.~\ref{eq:A1p6f_ref}). 
Whilst there were small differences in \chisq, it was found that variation of this parameter had an insignificant impact on the fit parameters of the $J^P=2^+$ amplitudes.

To describe the diagonal $\Dst \omoct (\SLJ{5}{S}{2})$ amplitude, a constant term plus a higher-order term was needed for both the numerator and the denominator, whilst only constants were needed for the diagonal $D$-wave amplitudes. 
Allowing for nonzero constants coupling the $\Dst \omoct (\SLJ{5}{S}{2})$ and $D$-wave partial waves yielded improved descriptions of the data, resulting in 
 a steeper growth at threshold and higher peak of the diagonal $\Dst \omoct (\SLJ{5}{S}{2})$ amplitude, but were not necessary to achieve reasonable parameterisations.
 Descriptions of the data were not improved when allowing for nonzero off-diagonal $D$-wave to $D$-wave amplitudes.
Overall, we find strong energy dependence in the  diagonal $\Dst \omoct (\SLJ{5}{S}{2})$ amplitude with a rapid growth at threshold but small diagonal $D$-wave amplitudes and only a slight preference for small nonzero couplings between $\Dst \omoct (\SLJ{5}{S}{2})$ and the other partial waves.

Moving to the pole structure of the resulting $J^P =2^+$ amplitudes, there are four  different scattering channels and hence the complex-energy plane splits into $16$ Riemann sheets.
Reasonable parameterisations of the data each had a pole on the $[+, +, +, -]$ sheet (where the channels are ordered $[\D \etaoct$, $\Dst \etaoct$, $\D \omoct$, $\Dst \omoct]$), with a very small or exactly zero (from the decoupled fits) imaginary component of its position, and a real component just below $\Dst \omoct$ threshold.
We note that similar to the $J^P =0^+$ poles, this pole is found on a hidden sheet.
The pole location for each reasonable parameterisation is plotted in the complex $E_{{\sf cm}}$ plane in the bottom panel of Fig.~\ref{fig:Ep6f_all_fits},
shown alongside its $\Dst \omoct (\SLJ{5}{S}{2})$ coupling (with all other partial-wave couplings found to be very small when not fixed to zero).
The pole location has a small spread across the different reasonable parameterisations and under mass-anisotropy variations.
Any other poles were found deep into the complex plane and far away from the constrained energy region.

Therefore, we conclude that only one pole is needed to describe the $J^P=2^+$ amplitudes in the energy region considered.
This pole is found on the $[+, +, +, -]$ sheet and is primarily coupled to $\Dst \omoct (\SLJ{5}{S}{2})$. 
We quote its location as
\begin{align}
a_t \sqrt{s_{\text{pole}}} &= 0.656 \pm 0.003 + \tfrac{i}{2}(0.00002 \pm 0.00003), \\   
\sqrt{s_{\text{pole}}} &= 3052 \pm 15 + \tfrac{i}{2}(0.08 \pm 0.15)\, \text{MeV}, 
\end{align}
obtained by taking an envelope over reasonable parameterisations.

\subsection{\texorpdfstring{$J^P = 3^+$}{} amplitudes}

We constrain the $J^P =3^+$ amplitudes from the $[000]A_2^+$ finite-volume spectrum.
Although our qualitative discussion in Section~\ref{section:6f_fv_spectra} of the $[000]A_2^+$ spectrum suggested small interactions, these amplitudes will be needed for the analysis of the $[000]T_1^+$ spectrum
when we determine the $J^P = 1^+$ amplitudes.

On the $L/a_s = 24$ volume, there is a $\D\etaoct$  dominated level  at $a_tE_{\sf cm} \approx 0.68$ (faded red in Fig.~\ref{fig:6f_spectra}), with negligible contributions from other operators.
The lowest $\D\etaoct$  partial-wave contribution to $[000]A_2^+$ is $\D\etaoct(\SLJ{1}{I}{6})$. 
Since we neglect $G$-wave and higher contributions, we chose to not include this level in the scattering analysis.
Apart from this level, we use all the energy levels up to approximately the three-hadron threshold.
The $\Dz \etaoct$ and $\smash{D^{(\prime)}_{1,\bar{\mathbf{3}}} \etaoct }$ contribute in a minimum of $F$-wave and $\Dt \etaoct$ in a minimum of $P$-wave,
and their lowest operator dominated energies or non-interacting curves for all four of these channels appear above the three-hadron threshold. 
We therefore do not include the $\Dz \etaoct$, $\smash{D^{(\prime)}_{1,\bar{\mathbf{3}}} \etaoct }$ and $\Dt \etaoct$ channels in the scattering analysis, and we use 22 energy levels to constrain the $\Dst\etaoct(\SLJ{3}{D}{3})- \D\omoct(\SLJ{3}{D}{3})- \Dst\omoct(\SLJ{3}{D}{3})- \Dst\omoct(\SLJ{5}{D}{3})$ amplitudes.

\begin{figure}
  \centering
	\includegraphics[width=\linewidth]{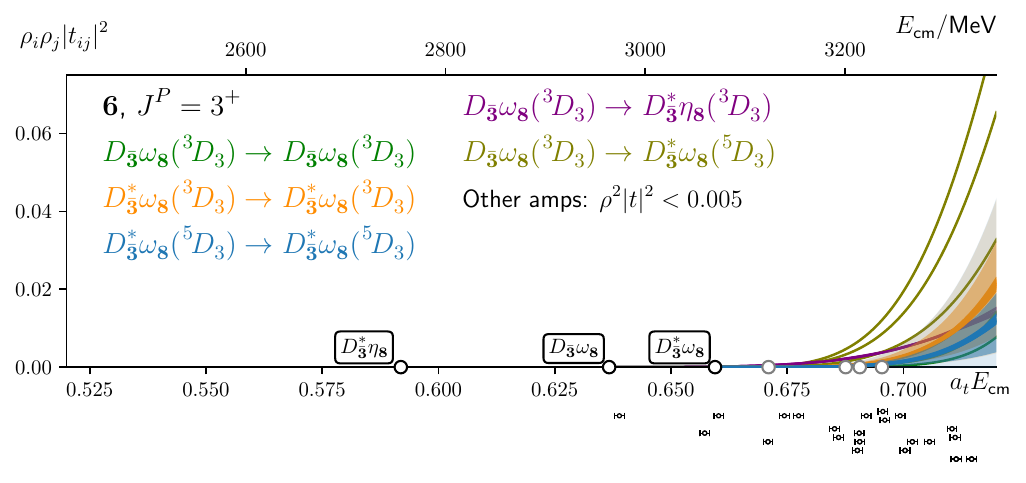}
	\caption{As Fig.~\ref{fig:A1p6f_all_fits} but for the $J^P = 3^+$ amplitudes in the flavour $\mathbf{6}$ sector from reasonable fits to the $[000]A_2^+$ spectrum.}
  \label{fig:A2p6f_all_fits}
\end{figure}

For the $J^P = 3^+$ amplitudes, we use $K$-matrix parameterisations with constant and linear terms and we consider both phase-space prescriptions.
For the reference parameterisation, we take a diagonal constant $K$-matrix with Chew-Mandelstam phase space. 
The resulting fit parameters are
\begin{equation}
    \centering
    \begin{tabular}{r@{ }ll} 
     $\gamma^{(0)} \{\D \omoct (\SLJ{3}{D}{3}) | \D \omoct (\SLJ{3}{D}{3})\} =$ & $(-2 \pm 10 \pm 18)\cdot a_t^{4} $ &
     \multirow{4}{*}{  $\left[ \begin{array}{rrrr}    
      1.00 &  0.39 &  0.19 &  0.41 \\
           &  1.00 &  0.34 &  0.71 \\
           &       &  1.00 &  0.30 \\
           &       &       &  1.00 \\
     \end{array} \right]$ }
     \\ 
     $\gamma^{(0)} \{\Dst \etaoct (\SLJ{3}{D}{3}) | \Dst \etaoct (\SLJ{3}{D}{3})\} =$ & $(1 \pm 3 \pm 11)\cdot a_t^{4} $ & \\
     $\gamma^{(0)} \{\Dst \omoct (\SLJ{3}{D}{3}) | \Dst \omoct (\SLJ{3}{D}{3})\} =$ & $(69 \pm 22 \pm 17)\cdot a_t^{4} $ & \\
     $\gamma^{(0)} \{\Dst \omoct (\SLJ{5}{D}{3}) | \Dst \omoct (\SLJ{5}{D}{3})\} =$ & $(52 \pm 19 \pm 37)\cdot a_t^{4} $ & \\[10pt]
     &\multicolumn{2}{l}{$\chisq = \frac{17.88}{22-4} = 0.99$\,.}
    \end{tabular}
\label{eq:A2p6f_ref}
\end{equation}
The spectrum corresponding to this parameterisation, which is plotted in Fig.~\ref{fig:6f_spectra_get_finite},  is in good agreement with the spectrum computed from lattice QCD.
This parameterisation is plotted in Fig.~\ref{fig:A2p6f_all_fits} with its statistical uncertainty and uncertainty from
mass and anisotropy variations.
Other reasonable parameterisations are also plotted in Fig.~\ref{fig:A2p6f_all_fits} and a summary of these can be found in Table~\ref{table:A2p6f_parameterisation_table}.
We considered diagonal parameterisations and parameterisations with nonzero off-diagonal amplitudes, both of which yielded reasonable descriptions of the data. 
For all parameterisations the amplitudes were found to be small, consistent with our qualitative discussion of the finite-volume spectra in Section~\ref{section:6f_fv_spectra}.

Upon analytic continuing the reasonable parameterisations, no poles were found close to the constrained energy region.
This matches our expectations given these amplitudes remain small over this region and have no signatures of nearby poles.

\subsection{\texorpdfstring{$J^P = 1^+$}{} single-channel amplitudes and poles}
\label{section:T1p6f_single_channel}
We constrain the $J^P = 1^+$ amplitudes from the $[000]T_1^+$ spectrum, which also gets contributions from  $J^P = \{3, 4\}^+$ -- see Table~\ref{table:pw_sub}. 
We begin by exploring the energy region where only the $\Dst \etaoct$ channel is open, using the first 10 $\Dst \etaoct$ dominated energy levels (purple in Fig.~\ref{fig:6f_spectra}) to constrain the amplitude up to $a_tE_{\sf cm} \approx 0.63$ .
To describe the near-degenerate levels between $\Dst\etaoct$ and $\D\omoct$ threshold, $\Dst \etaoct(\SLJ{3}{S}{1})$  and a second $\Dst\etaoct$ partial wave is needed in the QC, the next lowest of which is either $\Dst \etaoct(\SLJ{3}{D}{1})$ or $\Dst \etaoct(\SLJ{3}{D}{3})$.
With no relative threshold suppression between these two $D$-waves, we include both in the amplitude fits. 
For the $J^P =3^+$ contribution, we opt to use a constant with its value fixed to the central value of the
reference parameterisation in Eq.~\ref{eq:A2p6f_ref}, constrained from the $[000]A_2^+$ spectrum. 
We will test the sensitivity (if any) of the $J^P =1^+$ amplitudes to this choice of parameter by varying it within one sigma from its central value. 
In summary, we use the lowest 10 energies to constrain the $\Dst \etaoct(\SLJ{3}{S}{1}) -\Dst \etaoct(\SLJ{3}{D}{1})$ amplitudes.

\begin{figure}
	\minipage{1\textwidth}
	\includegraphics[width=\linewidth]{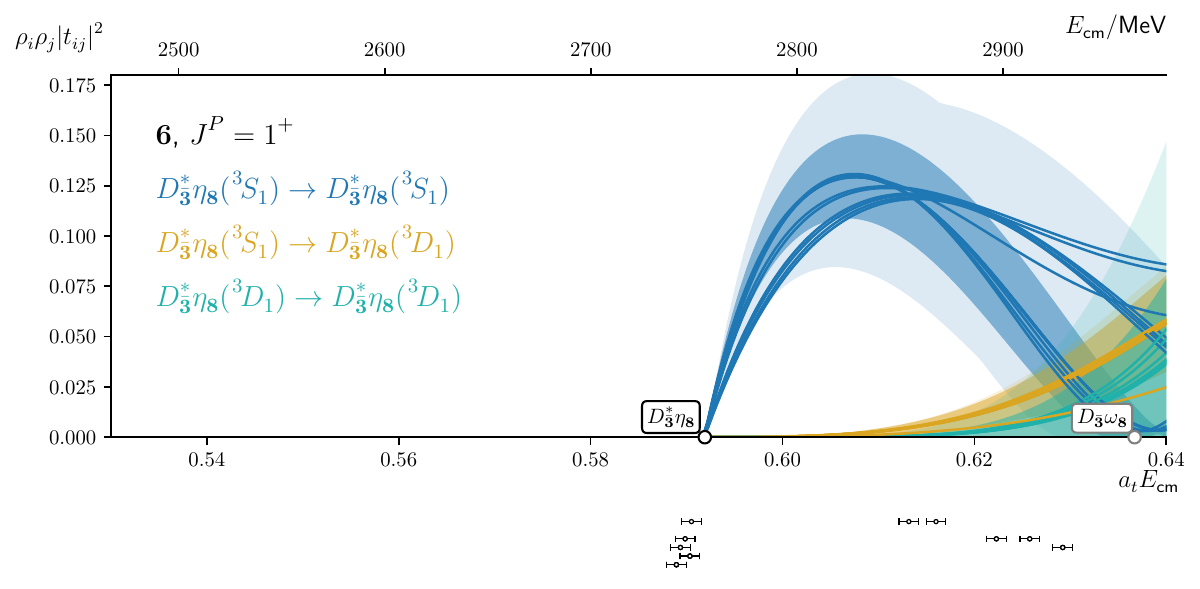}
	\caption{As top and middle panels of Fig.~\ref{fig:A1p6f_all_fits} but for the $J^P = 1^+$ $\Dst\etaoct$ partial-wave amplitudes in the flavour $\mathbf{6}$ sector from reasonable fits to the $[000]T_1^+$ spectrum, constrained up to $a_tE_{\sf cm} \approx 0.63$.
           }\label{fig:T1p6f_single_channel_all_fits}
	\endminipage
	\hfill
	\minipage{1\textwidth}
	\includegraphics[width=\linewidth]{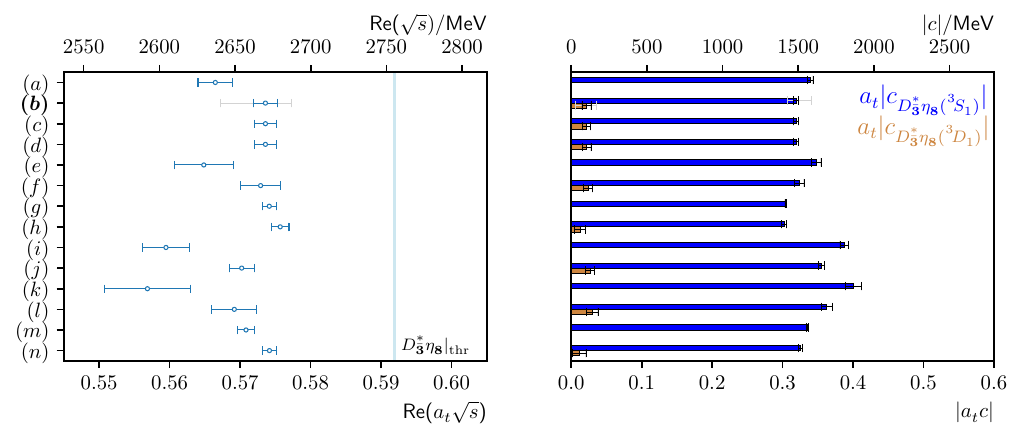}
	\caption{Virtual bound state location (left panel) and partial-wave couplings (right panel) from reasonable fits to the $[000]T_1^+$ spectrum, constrained up to $a_tE_{\sf cm} \approx 0.63$.
           Parameterisations are labelled according to Table~\ref{table:T1p6f_single_channel_parameterisation_table}.
           The reference parameterisation, $(b)$
           , is shown with its mass-anisotropy uncertainty (light grey error bars). }\label{fig:T1p6f_single_channel_poles}
\endminipage
\end{figure}

We use $K$-matrix polynomial parameterisations with both phase space prescriptions.
A parameterisation with Chew-Mandelstam phase space is taken to be our reference parameterisation.
Its fit parameters are 
\begin{equation}
    \centering
    \begin{tabular}{r@{ }ll}  
     $\gamma^{(0)} \{\Dst \etaoct (\SLJ{3}{D}{1}) | \Dst \etaoct (\SLJ{3}{D}{1})\} =$ & $(310 \pm 180 \pm 230)\cdot a_t^{4} $ &
     \multirow{4}{*}{  $\left[ \begin{array}{rrrr}    
      1.00 & -0.52 &  0.31 & -0.36 \\
           &  1.00 & -0.20 &  0.28 \\
           &       &  1.00 & -0.99 \\
           &       &       &  1.00 \\
     \end{array} \right]$ }
     \\ 
     $\gamma^{(0)} \{\Dst \etaoct (\SLJ{3}{D}{1}) | \Dst \etaoct (\SLJ{3}{S}{1})\} =$ & $(-17 \pm 4 \pm 10)\cdot a_t^{2} $ & \\
     $\gamma^{(0)} \{\Dst \etaoct (\SLJ{3}{S}{1}) | \Dst \etaoct (\SLJ{3}{S}{1})\} =$ & $(20.3 \pm 1.7 \pm 5.6) $ & \\
     $\gamma^{(1)} \{\Dst \etaoct (\SLJ{3}{S}{1}) | \Dst \etaoct (\SLJ{3}{S}{1})\} =$ & $(-49 \pm 5 \pm 15)\cdot a_t^{2} $ & \\[10pt]
     &\multicolumn{2}{l}{$\chisq = \frac{2.80}{10-4} = 0.47$\,.}
    \end{tabular}
\label{eq:T1p6f_elastic_ref}
\end{equation}
This is plotted in Fig.~\ref{fig:T1p6f_single_channel_all_fits} along with other reasonable parameterisations, including parameterisations with the same form as the reference one but with the fixed constant term describing the diagonal $\Dst \etaoct(\SLJ{3}{D}{3})$ amplitude varied by plus or minus one sigma. 
The $\Dst \etaoct(\SLJ{3}{S}{1}) -\Dst \etaoct(\SLJ{3}{D}{1})$ amplitudes were found to be insensitive to this choice of parameter. 
Reasonable parameterisations are summarised in Table~\ref{table:T1p6f_single_channel_parameterisation_table}.

To achieve reasonable descriptions of the data, it was found at least a constant plus a higher-order term was needed for the diagonal $\Dst \etaoct(\SLJ{3}{S}{1})$ amplitude, whilst only a constant term was needed for the diagonal $\Dst \etaoct(\SLJ{3}{D}{1})$ amplitude.
Coupling between these two waves was not needed to obtain reasonable parameterisations and when it was allowed for the off-diagonal amplitude was found to be small.

Moving to the pole structure of the reasonable parameterisations, we note that only one channel is open and so the complex energy plane splits into two Riemann sheets. 
A pole was found on the real axis of the unphysical sheet below threshold, corresponding to a virtual bound state, predominantly coupling to $\Dst \etaoct(\SLJ{3}{S}{1})$.
The pole is plotted in Fig.~\ref{fig:T1p6f_single_channel_poles} along with its partial-wave couplings for each reasonable parameterisation.
For parameterisations with the off-diagonal term not fixed to zero, the pole's coupling to $\Dst \etaoct(\SLJ{3}{D}{1})$ was found to be small.
No other poles were found near the constrained energy region.
We therefore find one well-determined pole in the $\Dst\etaoct$ amplitudes, constrained up to $a_tE_{\sf cm} \approx 0.63$.
In the next section, we consider the coupled-channel region of the $J^P =1^+$ sector and we refrain from quoting the pole location until then.

\subsection{\texorpdfstring{$J^P = 1^+$}{} coupled-channel amplitudes and poles}

\begin{figure}
  \centering
	\includegraphics[width=1\linewidth]{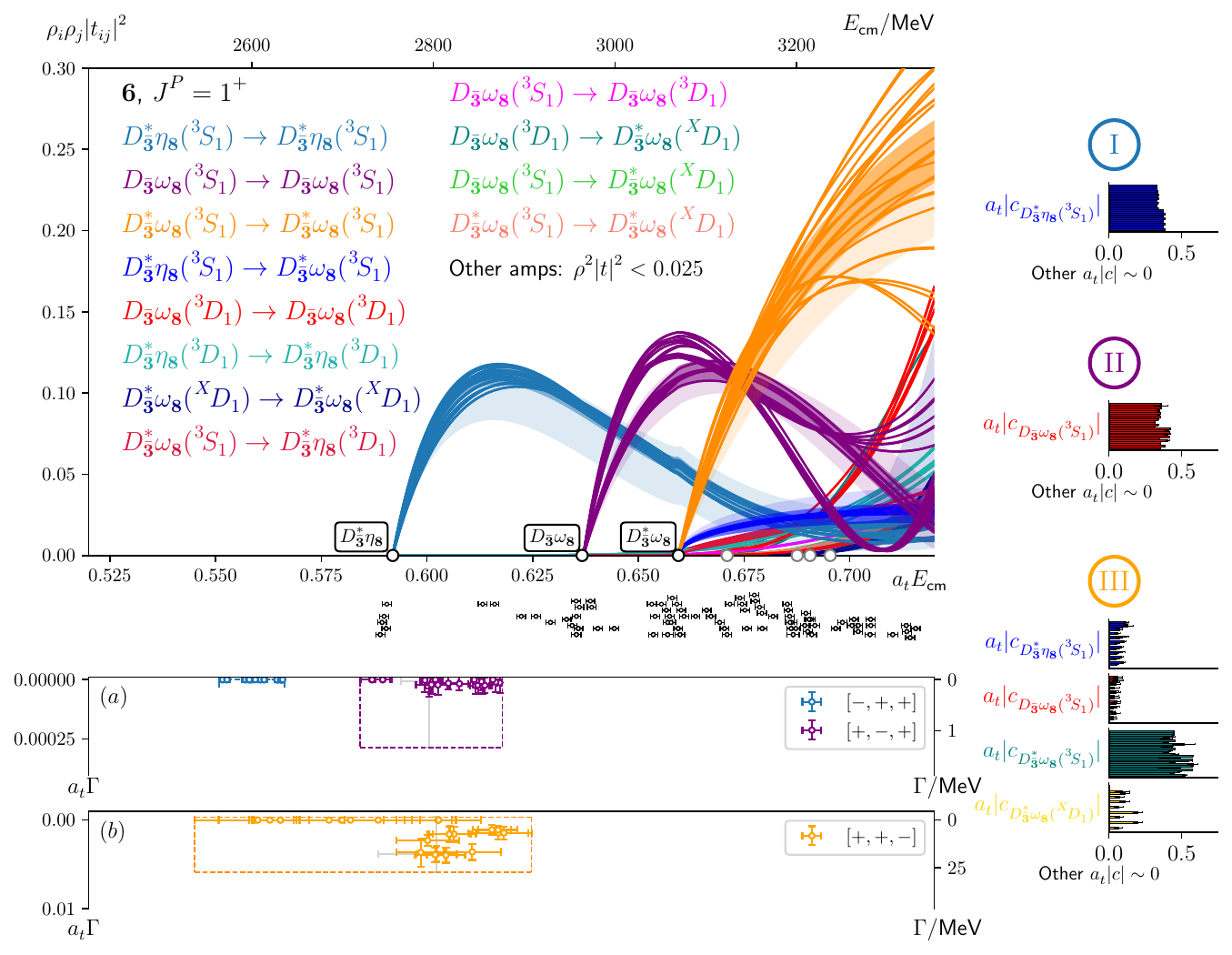}
	\caption{As Fig.~\ref{fig:A1p6f_all_fits} but for the $J^P = 1^+$ amplitudes in the flavour $\mathbf{6}$ sector from reasonable coupled-channel fits to the $[000]T_1^+$ spectrum.
           The fixed $J^P =\{3,4\}^+$ amplitudes are not plotted. 
           The poles are split into two panels -- $(a)$ and $(b)$ -- for visual clarity.}
  \label{fig:T1p6f_all_fits}
\end{figure}

To finish off our discussion of the flavour $\mathbf{6}$ sector, we constrain the $J^P =1^+$ amplitudes up to roughly the three-hadron threshold. 
On the $L/a_s = 24$ volume, there is a $D_{\bar{\mathbf{3}}[210]} \eta_{\mathbf{8}[210]}$ operator  dominated level at $a_tE_{\sf cm} \approx 0.68$ (faded red in Fig.~\ref{fig:6f_spectra}), which requires a $\D\etaoct$ partial wave in the QC to be reproduced.
The lowest such wave is $\D \etaoct (\SLJ{1}{G}{4})$ and neglecting $G$-wave (and higher) partial waves, we opt to ignore this level from the scattering analysis.
Apart from this energy level, we use all levels up to approximately the three-hadron threshold or nearest levels with dominant overlaps onto $\Dz \etaoct $,
 $\smash{D^{(\prime)}_{1,\bar{\mathbf{3}}} \etaoct }$  or  $\Dt \etaoct$ operators.
These channels contribute in a minimum of $P$-wave and with this choice of cutoff on each volume, they can be neglected from the scattering analysis. 
This leaves 82 energy levels to constrain the coupled-channel $\Dst \etaoct (\SLJ{3}{S}{1})- \Dst \etaoct (\SLJ{3}{D}{1})- \D \omoct (\SLJ{3}{S}{1})- \D \omoct (\SLJ{3}{D}{1})-\Dst \omoct (\SLJ{3}{S}{1})- \Dst \omoct (\SLJ{3}{D}{1}) - \Dst \omoct (\SLJ{5}{D}{1})$ amplitudes.
Note that since we are only using data from  an at-rest finite-volume irrep, the $\Dst \omoct (\SLJ{3}{D}{1})$ and $\Dst \omoct (\SLJ{5}{D}{1})$ partial waves cannot be distinguished in the quantisation condition.
 See Appendix~\ref{appendix:indist_Dwave} for more details. 
 To reflect this, we collectively refer to these waves as $\Dst \omoct(\SLJ{X}{D}{1})$ when displaying the results.\footnote{
This issue does not arise for other $J^P$ where multiple $\Dst \omoct$ $D$-waves occur or for other channels -- see Appendix~\ref{appendix:indist_Dwave}.} 
 For the $J^P = 3^+$ and $4^+$ contributions, we opt to use constant $K$-matrices with values fixed to the central values of the reference parameterisations of Eq.~\ref{eq:A2p6f_ref} and Eq.~\ref{eq:A1p6f_ref}, respectively.
We will test the sensitivity of the resultant $J^P =1^+$ coupled-channel amplitudes to this choice as part of the parameterisation variations. 

When parameterising the $J^P =1^+$ amplitudes, it was found that we could achieve sufficient variations considering only $K$-matrix polynomial parameterisations with both phase space prescriptions.
As a representative fit, we take a $K$-matrix with Chew-Mandelstam phase space as our reference parameterisation. 
Its fit parameters are
\begin{equation}
    \centering
    \resizebox{\textwidth}{!}{%
    \begin{tabular}{r@{ }ll}   
     $\gamma^{(0)} \{\D \omoct (\SLJ{3}{D}{1}) | \D \omoct (\SLJ{3}{D}{1})\} =$ & $(37 \pm 9 \pm 22)\cdot a_t^{4} $ &
     \resizebox{1.04\textwidth}{!}{
     \multirow{13}{*}{  $\left[ \begin{array}{rrrrrrrrrrrrr}    
      1.00 &  0.01 &  0.04 &  0.07 & -0.02 &  0.22 &  0.03 &  0.00 & -0.01 &  0.03 &  0.27 &  0.19 &  0.28 \\
           &  1.00 & -0.99 & -0.04 &  0.14 &  0.02 &  0.06 & -0.00 & -0.05 & -0.02 &  0.01 & -0.00 &  0.01 \\
           &       &  1.00 &  0.02 & -0.24 &  0.00 & -0.05 &  0.00 &  0.04 &  0.02 &  0.01 &  0.07 &  0.01 \\
           &       &       &  1.00 &  0.28 & -0.01 & -0.07 &  0.02 &  0.04 & -0.09 &  0.01 & -0.13 & -0.00 \\
           &       &       &       &  1.00 & -0.02 & -0.04 &  0.01 &  0.03 & -0.06 & -0.07 & -0.50 & -0.06 \\
           &       &       &       &       &  1.00 & -0.12 &  0.02 &  0.18 &  0.02 &  0.52 &  0.18 &  0.19 \\
           &       &       &       &       &       &  1.00 & -0.72 &  0.02 & -0.08 & -0.02 &  0.09 &  0.01 \\
           &       &       &       &       &       &       &  1.00 & -0.69 &  0.01 &  0.01 & -0.01 & -0.00 \\
           &       &       &       &       &       &       &       &  1.00 &  0.17 &  0.05 & -0.05 &  0.01 \\
           &       &       &       &       &       &       &       &       &  1.00 & -0.02 &  0.13 & -0.02 \\
           &       &       &       &       &       &       &       &       &       &  1.00 &  0.26 &  0.34 \\
           &       &       &       &       &       &       &       &       &       &       &  1.00 &  0.42 \\
           &       &       &       &       &       &       &       &       &       &       &       &  1.00 \\
     \end{array} \right]$ }
     } 
     \\ 
     $\gamma^{(0)} \{\D \omoct (\SLJ{3}{S}{1}) | \D \omoct (\SLJ{3}{S}{1})\} =$ & $(7.5 \pm 0.4 \pm 1.1) $ & \\
     $\gamma^{(1)} \{\D \omoct (\SLJ{3}{S}{1}) | \D \omoct (\SLJ{3}{S}{1})\} =$ & $(-13.7 \pm 0.8 \pm 2.3)\cdot a_t^{2} $ & \\
     $\gamma^{(0)} \{\D \omoct (\SLJ{3}{S}{1}) | \Dst \etaoct (\SLJ{3}{S}{1})\} =$ & $(-0.01 \pm 0.19 \pm 0.06) $ & \\
     $\gamma^{(0)} \{\D \omoct (\SLJ{3}{S}{1}) | \Dst \omoct (\SLJ{3}{S}{1})\} =$ & $(-0.42 \pm 0.24 \pm 0.19) $ & \\
     $\gamma^{(0)} \{\Dst \etaoct (\SLJ{3}{D}{1}) | \Dst \etaoct (\SLJ{3}{D}{1})\} =$ & $(8 \pm 2 \pm 11)\cdot a_t^{4} $ & \\
     $\gamma^{(0)} \{\Dst \etaoct (\SLJ{3}{S}{1}) | \Dst \etaoct (\SLJ{3}{S}{1})\} =$ & $(23.82 \pm 0.16 \pm 2.61) $ & \\
     $\gamma^{(1)} \{\Dst \etaoct (\SLJ{3}{S}{1}) | \Dst \etaoct (\SLJ{3}{S}{1})\} =$ & $(-95.2 \pm 0.5 \pm 10.8)\cdot a_t^{2} $ & \\
     $\gamma^{(2)} \{\Dst \etaoct (\SLJ{3}{S}{1}) | \Dst \etaoct (\SLJ{3}{S}{1})\} =$ & $(96.5 \pm 0.8 \pm 11.1)\cdot a_t^{4} $ & \\
     $\gamma^{(0)} \{\Dst \etaoct (\SLJ{3}{S}{1}) | \Dst \omoct (\SLJ{3}{S}{1})\} =$ & $(0.57 \pm 0.06 \pm 0.04) $ & \\
     $\gamma^{(0)} \{\Dst \omoct (\SLJ{3}{D}{1}) | \Dst \omoct (\SLJ{3}{D}{1})\} =$ & $(-84 \pm 15 \pm 70)\cdot a_t^{4} $ & \\
     $\gamma^{(0)} \{\Dst \omoct (\SLJ{3}{S}{1}) | \Dst \omoct (\SLJ{3}{S}{1})\} =$ & $(1.99 \pm 0.12 \pm 0.18) $ & \\
     $\gamma^{(0)} \{\Dst \omoct (\SLJ{5}{D}{1}) | \Dst \omoct (\SLJ{5}{D}{1})\} =$ & $(30 \pm 30 \pm 40)\cdot a_t^{4} $ & \\[10pt]
     &\multicolumn{2}{l}{$\chisq = \frac{69.31}{82-13} = 1.00$\,.}
    \end{tabular}
    }
  \label{eq:T1p6f_ref}
\end{equation}
The spectrum corresponding to this parameterisation is plotted in Fig.~\ref{fig:6f_spectra_get_finite} where we see the pattern of shifts in the energy levels obtained via the variational method are well reproduced.
This parameterisation is plotted in Fig.~\ref{fig:T1p6f_all_fits} along with other reasonable parameterisations, including parameterisations with the same form as the reference parameterisation in Eq.~\ref{eq:T1p6f_ref} but with the fixed $J^P = \{3,4\}^+$ amplitudes' parameters varied by plus or minus one sigma from their central values.
It was found the $J^P =1^+$ amplitudes were insensitive to the choice of these parameters. 
The reasonable parameterisations are summarised in Table~\ref{table:T1p6f_parameterisation_table}.

To achieve reasonable descriptions of the data, a second-order polynomial was needed for the diagonal 
$\Dst \etaoct(\SLJ{3}{S}{1})$ amplitude, at least a first-order polynomial was needed for the diagonal 
$\D \omoct(\SLJ{3}{S}{1})$ amplitude, and at least a constant term was needed for the diagonal $\Dst \omoct(\SLJ{3}{S}{1})$ amplitude. 
Only constants were needed for the diagonal $D$-wave amplitudes.
Parameterisations with and without constant terms coupling the different $S$-waves both yielded reasonable descriptions to the data.
However, parameterisations with these couplings had a smaller \chisq, with the $\Dst\etaoct(\SLJ{3}{S}{1}) \rightarrow \Dst\omoct(\SLJ{3}{S}{1})$ and $\D\omoct(\SLJ{3}{S}{1}) \rightarrow \Dst\omoct(\SLJ{3}{S}{1})$ amplitudes taking on small but statistically nonzero values.
In some parameterisations, large diagonal $D$-wave amplitudes were observed for $a_tE_{\sf cm} \gtrsim 0.69$.
We note, however, that this was not seen in every parameterisation and this likely arose due to too few constraints in this energy region.\footnote{We use a lower cutoff on the $L/a_s = \{20, 24\}$ volumes for $[000]T_1^+$ compared to other finite-volume irreps due to the presence of the $\Dz\etaoct$ channel.} 
Parameterisations with nonzero constants for the $S$-wave to $D$-wave amplitudes were considered but these amplitudes were found to be small.
Finally, we tested parameterisations with constant terms coupling $D$-wave to $D$-wave amplitudes (as well as $S$-wave to $D$-wave), however this did not yield an improved description of the data.

To summarise, strong interactions are observed in the diagonal $S$-wave amplitudes, but mostly weak interactions in the $D$-waves 
(with a couple of parameterisations displaying growing $D$-wave amplitudes at higher energy where there are fewer constraints, as discussed above). 
 When allowed for, the off-diagonal $\D\omoct(\SLJ{3}{S}{1}) \rightarrow \Dst\omoct(\SLJ{3}{S}{1})$
and  $\Dst\etaoct(\SLJ{3}{S}{1}) \rightarrow \Dst\omoct(\SLJ{3}{S}{1})$ amplitudes took on small but nonzero values, resulting in an improved \chisq.
These cross-channel couplings however were not necessary to obtain reasonable descriptions to the data.
All other off-diagonal amplitudes were found to be small when not fixed to zero.

Now we discuss the resulting pole structure from reasonable parameterisations. 
With three channels present in the $J^P =1^+$ amplitudes, there are eight Riemann sheets where poles could be situated.
For each reasonable parameterisation, a pole is found on the $[-, +, +]$ sheet (where the channels are ordered $[\Dst \etaoct$, $\D \omoct$, $\Dst \omoct]$) below $\Dst \etaoct$ threshold and is plotted in panel $(a)$ of Fig.~\ref{fig:T1p6f_all_fits} (blue points).
We further display the pole's $\Dst \etaoct(\SLJ{3}{S}{1})$ coupling for each parameterisation (histograms labelled ``$\mathrm{\Romannum{1}}$''),
with negligible coupling to other partial waves when they are not fixed to zero.
This is the same virtual bound state found in Section~\ref{section:T1p6f_single_channel}. 
The pole location in the coupled-channel fits is found to be at a similar energy to the pole location in the single-channel fits.
We quote the location as 
\begin{align}
a_t \sqrt{s_{\text{pole}}} &= 0.564 \pm 0.013 \\
\sqrt{s_{\text{pole}}} &= 2630 \pm 60 \, \text{MeV},
\end{align}
taking an envelope over both the single-channel fits of Section~\ref{section:T1p6f_single_channel} and the coupled-channel fits presented in this section.

In addition to the virtual bound state, each parameterisation contains a pole on the $[+,-,+]$ sheet, which is plotted in panel $(a)$ of Fig.~\ref{fig:T1p6f_all_fits} (purple points).
The pole's $\D \omoct(\SLJ{3}{S}{1})$ coupling for each parameterisations is similarly displayed in Fig.~\ref{fig:T1p6f_all_fits} (histograms labelled ``$\mathrm{\Romannum{2}}$''),
 with a negligible coupling to the other partial waves whenever coupling terms are not fixed to zero.
For parameterisations where cross-channel couplings were allowed, this pole had up a small decay width where the only kinematically allowed decay channel is $\Dst \etaoct$.
For decoupled parameterisations, the pole corresponds exactly to a $\D \omoct(\SLJ{3}{S}{1})$ virtual-bound state.
This pole has a relatively larger spread under mass-anisotropy variations (compared to parameterisation variations) than the other poles found in this study.
We quote its location as 
\begin{align}
a_t \sqrt{s_{\text{pole}}} &= 0.601 \pm 0.017 + \tfrac{i}{2}(0.00012\pm 0.00017) \\   
\sqrt{s_{\text{pole}}} &= 2800 \pm 80 + \tfrac{i}{2}(0.5\pm 0.8)\, \text{MeV}, 
\end{align}
where again we have taken an envelope across the reasonable parameterisations.

Furthermore, in each parameterisation a pole is found on the $[+,+,-]$ sheet below $\Dst \omoct$ threshold and is shown in the complex $E_{\sf cm}$ plane in panel $(b)$ of Fig.~\ref{fig:T1p6f_all_fits} (orange points) along with its partial-wave couplings (histograms labelled ``$\mathrm{\Romannum{3}}$'').
This pole is found to predominantly couple to $\Dst \omoct(\SLJ{3}{S}{1})$, with a smaller but nonzero coupling to $\Dst \etaoct(\SLJ{3}{S}{1})$ and $\D \omoct(\SLJ{3}{S}{1})$.
For parameterisations where the off-diagonal $D$-wave to $S$-wave amplitudes were not fixed to zero, this pole's coupling to $\Dst \omoct(\SLJ{X}{D}{1})$ was found to be small, and its coupling to all other $D$-waves were found to be very small and statistically consistent with zero.
We quote the location of this pole as
\begin{align}
a_t \sqrt{s_{\text{pole}}} &= 0.58 \pm 0.04 + \tfrac{i}{2}(0.003\pm 0.003) \\   
\sqrt{s_{\text{pole}}} &= 2720 \pm 190 + \tfrac{i}{2}(13\pm 14)\, \text{MeV}, 
\end{align}
where we have again taken an envelope across the reasonable parameterisations.

Finally, we note that any other poles were found far from the constrained energy region and were not consistently found in every parameterisation.

In summary, we find three well-determined $J^P =1^+$ poles in the energy region considered.
Each of these poles can be associated (via their couplings) with one of the three $S$-wave partial waves, and in each case is found on a hidden sheet.
Only the pole predominantly coupled to $\Dst\etaoct(\SLJ{3}{S}{1})$ is a virtual bound state for all parameterisations.
For parameterisations that have nonzero off-diagonal $S$-wave to $S$-wave amplitudes,
 the two poles predominantly coupled to $\D\omoct(\SLJ{3}{S}{1})$ and  $\Dst\omoct(\SLJ{3}{S}{1})$ are found to have small decay widths.
 However, since decoupled fits also yielded reasonable descriptions of the data, vanishing decay widths for these poles could not be ruled out. 
The pole primarily coupled to  $\Dst\omoct(\SLJ{3}{S}{1})$ leaves the largest imprint on the scattering amplitudes, resulting in the diagonal $\Dst\omoct(\SLJ{3}{S}{1})$  amplitude being roughly twice as large as in the other two diagonal $S$-wave amplitudes over the energy region of interest.

\section{Flavour \texorpdfstring{$\overline{\mathbf{15}}$}{} sector results}
\label{section:15bar_analysis}
We repeat the analysis of the previous section but now explore the finite-volume spectra and the $J^P = \{0, 1, 2, 3, 4\}^+$ scattering amplitudes in the flavour $\overline{\mathbf{15}}$ sector.
We will find that the spectra displays signs of either slight repulsion, slight attraction or weak interactions, and the $J^P = \{0, 1, 2, 3, 4\}^+$ amplitudes are found to be small with no robustly determined poles nearby in the complex energy plane (some but not all $J^P =0^+$ parameterisations find a single pole, see below).
To this end, we perform a less extensive study of this flavour sector, using fewer energy levels to constrain the amplitudes and we consider fewer parameterisation variations.

\subsection{Finite-volume spectra} 
\begin{figure}
    \centering
      \includegraphics[width=0.95\linewidth]{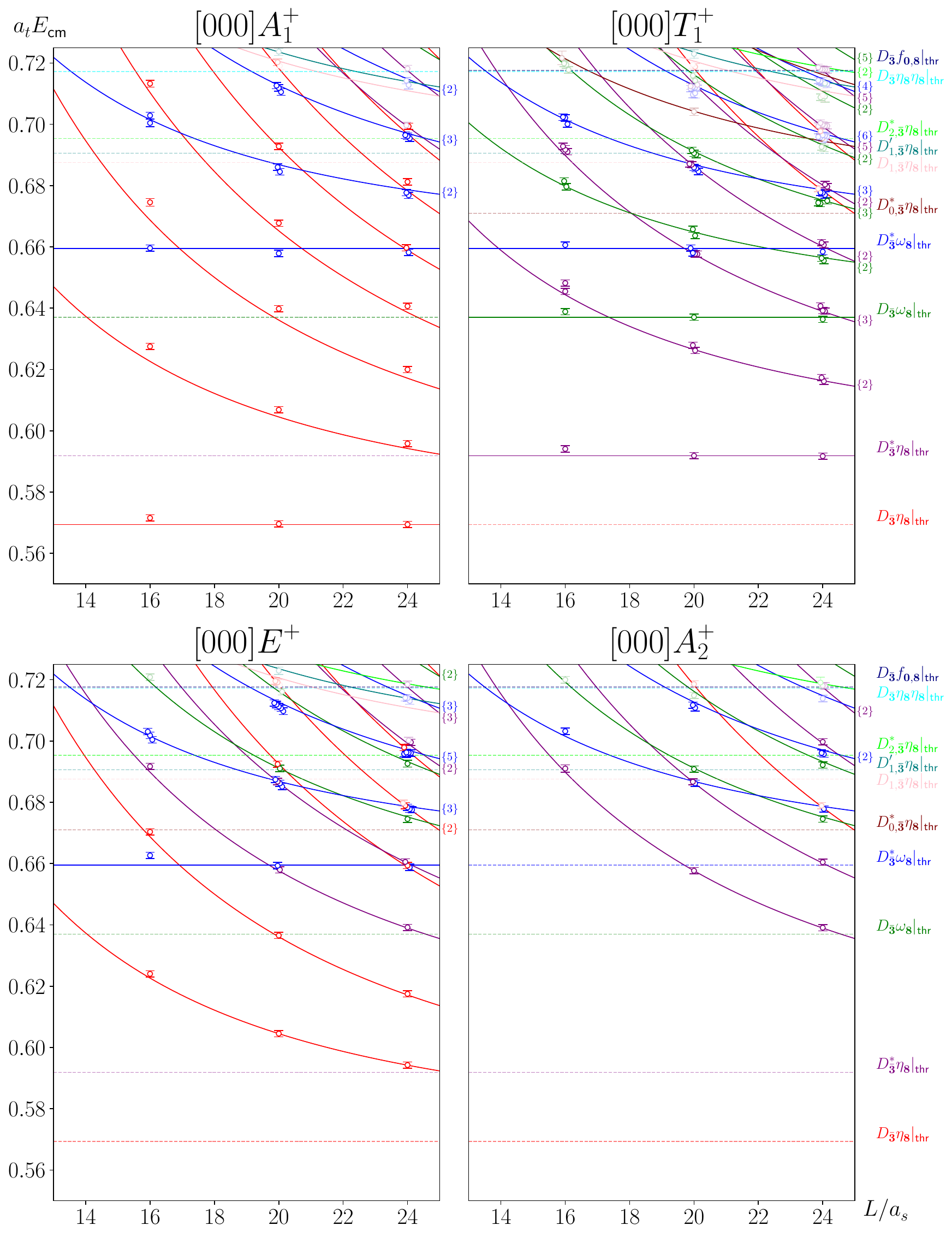}
      \caption{As for Fig.~\ref{fig:6f_spectra} but for the flavour $\overline{\mathbf{15}}$ sector.
              Note that the finite-volume spectrum is only computed on the $L/a_s = \{ 16, 20, 24\}$ volumes. }
    \label{fig:15bar_spectra}
  \end{figure}

The operator bases used to extract the spectra in flavour $\overline{\mathbf{15}}$ sector are the same as in the flavour $\mathbf{6}$ sector, but now we project the meson-meson operators onto the $\overline{\mathbf{15}}$ irrep.
Note, however, that the spectrum is only computed on the $L/a_s = \{16, 20, 24\}$ volumes for each finite-volume irrep.
The operator lists can be found in Appendix~\ref{appendix:interpolating_list}. 
The finite-volume spectra resulting from the variational analysis are shown in Fig.~\ref{fig:15bar_spectra}.

As for the flavour $\mathbf{6}$, most energy levels could be associated with a single operator via their dominant operator overlap and in the region of interest there are no ``extra levels'' beyond those expected from non-interacting meson-meson energies.
Across the irreps, nearly all energy levels are found close to non-interacting energy curves, with the exception of $\D\etaoct$ operator dominated levels in $[000]A_1^+$
which display small upward shifts. 
These features suggest some $\D\etaoct$ partial waves are repulsive and all other partial waves are at most weakly interacting. 
The operator overlaps across the irreps indicate at most small cross-channel coupling.

Again, to go beyond this qualitative picture, we use the finite-volume spectra to constrain scattering amplitudes via the Lüscher quantisation condition. 
We will first discuss the constraining of the $J^P = \{0,4\}^+$ amplitudes from the $[000]A_1^+$ spectrum.
After this, we briefly discuss the constraining of the  $J^P = \{1,2, 3\}^+$ amplitudes from the $[000]T_1^+$, $[000]E^+$ and $[000]A_2^+$ spectra.
In all cases, we will consider the same partial waves as in the flavour $\mathbf{6}$ analysis.

\subsection{\texorpdfstring{$J^P=\{0, 4\}^+$}{} amplitudes and poles}
\label{section:A1p_15bar}

Constraining the  $J^P=\{0, 4\}^+$ sectors from $[000]A_1^+$, we use the same energy cutoff on each volume as in the flavour $\mathbf{6}$ case, such that  contributions from $\smash{D_{1,\bar{\mathbf{3}}} \etaoct }$ and higher-threshold channels can be ignored.
We similarly ignore the energy level at $a_tE_{\sf cm} \approx 0.70$ on the  $L/a_s =24 $ volume (faded purple in Fig.~\ref{fig:15bar_spectra}) which would require a $G$-wave (or higher)  $\Dst \etaoct$ contribution in order to be reproduced.
This leaves 31 levels to constrain the $J^P=\{0, 4\}^+$ amplitudes.

\begin{figure}
    \centering
    \includegraphics[width=1\linewidth]{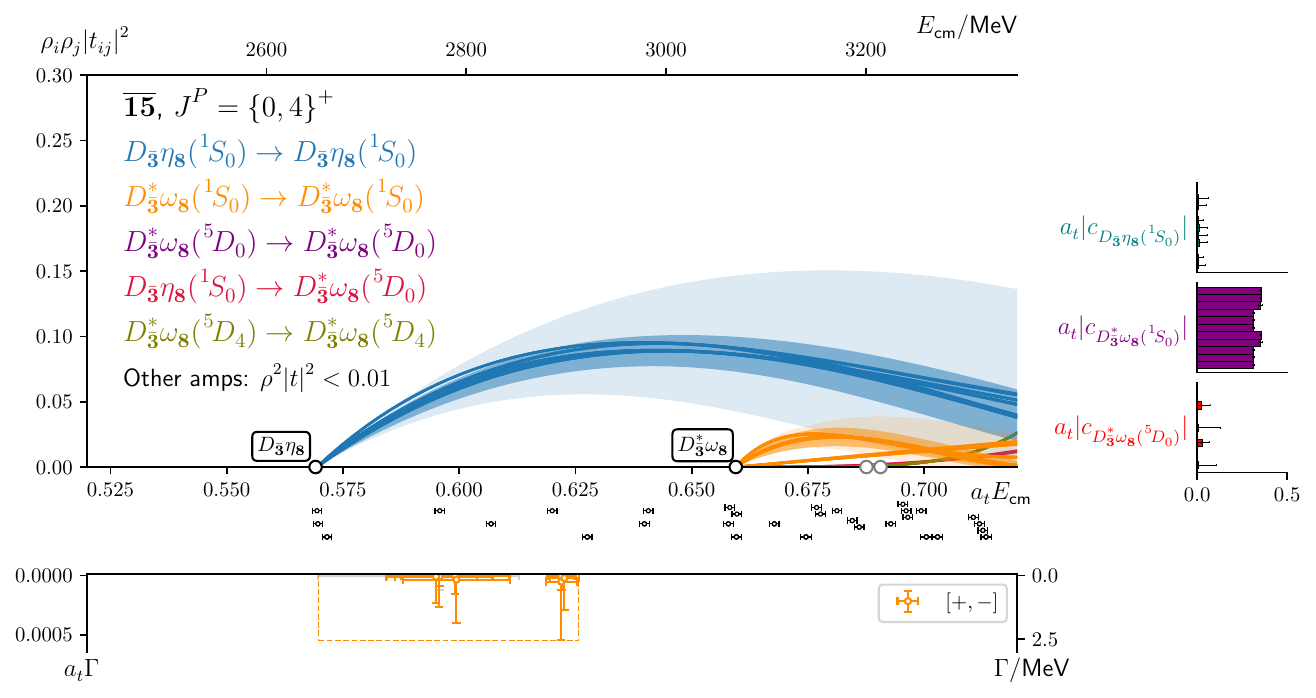}
    \caption{As Fig.~\ref{fig:A1p6f_all_fits} but for the $J^P = 0^+$ and $4^+$ amplitudes in the flavour $\overline{\mathbf{15}}$ sector.
             Some parameterisations find a pole on the $[+,-]$ sheet which are displayed (orange points) along with their couplings to the different channels and partial waves (histograms).
    }
    \label{fig:A1p15bar_all_fits}
\end{figure}

We parameterised the $J^P = 0^+$ amplitudes with $K$-matrix polynomials with Chew-Mandelstam phase space, whilst only a constant term with Chew-Mandelstam phase space was used for the $J^P = 4^+$ amplitude.
As an example parameterisation, the fit parameters of the reference parameterisation are
\begin{equation}
    \centering
    \resizebox{\textwidth}{!}{%
    \begin{tabular}{r@{ }ll}    
     $\gamma^{(0)} \{\D \etaoct (\SLJ{1}{S}{0}) | \D \etaoct (\SLJ{1}{S}{0})\} =$ & $(-2.1 \pm 0.3 \pm 0.3) $ &
     \multirow{7}{*}{  $\left[ \begin{array}{rrrrrrr}    
      1.00 & -0.98 & -0.00 &  0.04 & -0.04 & -0.00 &  0.04 \\
           &  1.00 &  0.01 & -0.04 &  0.04 &  0.03 &  0.01 \\
           &       &  1.00 & -0.00 & -0.00 & -0.00 & -0.01 \\
           &       &       &  1.00 & -1.00 &  0.08 &  0.17 \\
           &       &       &       &  1.00 & -0.06 & -0.14 \\
           &       &       &       &       &  1.00 &  0.31 \\
           &       &       &       &       &       &  1.00 \\
     \end{array} \right]$ }
     \\ 
     $\gamma^{(1)} \{\D \etaoct (\SLJ{1}{S}{0}) | \D \etaoct (\SLJ{1}{S}{0})\} =$ & $(3.4 \pm 0.6 \pm 0.9)\cdot a_t^{2} $ & \\
     $\gamma^{(0)} \{\D \etaoct (\SLJ{1}{S}{0}) | \Dst \omoct (\SLJ{1}{S}{0})\} =$ & $(0 \pm 0.28 \pm 0.07) $ & \\
     $\gamma^{(0)} \{\Dst \omoct (\SLJ{1}{S}{0}) | \Dst \omoct (\SLJ{1}{S}{0})\} =$ & $(5.5 \pm 1.3 \pm 1.4) $ & \\
     $\gamma^{(1)} \{\Dst \omoct (\SLJ{1}{S}{0}) | \Dst \omoct (\SLJ{1}{S}{0})\} =$ & $(-10 \pm 3 \pm 3)\cdot a_t^{2} $ & \\
     $\gamma^{(0)} \{\Dst \omoct (\SLJ{5}{D}{0}) | \Dst \omoct (\SLJ{5}{D}{0})\} =$ & $(-40 \pm 17 \pm 18)\cdot a_t^{4} $ & \\
     $\gamma^{(0)} \{\Dst \omoct (\SLJ{5}{D}{4}) | \Dst \omoct (\SLJ{5}{D}{4})\} =$ & $(0 \pm 20 \pm 30)\cdot a_t^{4} $ & \\[10pt]
     &\multicolumn{2}{l}{$\chisq = \frac{16.48}{31-7} = 0.69$\,.}
    \end{tabular}
    }
\label{eq:A1p15bar_ref}
\end{equation}
This is plotted in  Fig.~\ref{fig:A1p15bar_all_fits} along with the other reasonable parameterisations. 
A summary of these can be found in Table~\ref{table:A1p15bar_parameterisation_table}.
A polynomial with at least a constant and linear term for the diagonal $\D\etaoct(\SLJ{1}{S}{0})$ amplitude was needed to achieve reasonable descriptions of the data, whilst only constants were needed to describe the diagonal $\Dst \omoct$ amplitude.
Note, however, the  \chisq\ was significantly improved when higher-order terms were also included for the diagonal $\Dst \omoct(\SLJ{1}{S}{0})$ amplitude.
Parameterisations with nonzero constant terms for the off-diagonal amplitudes were considered, but these were found to be small and statistically consistent with zero.

\begin{figure}
  \centering
  \includegraphics[width=0.9\linewidth]{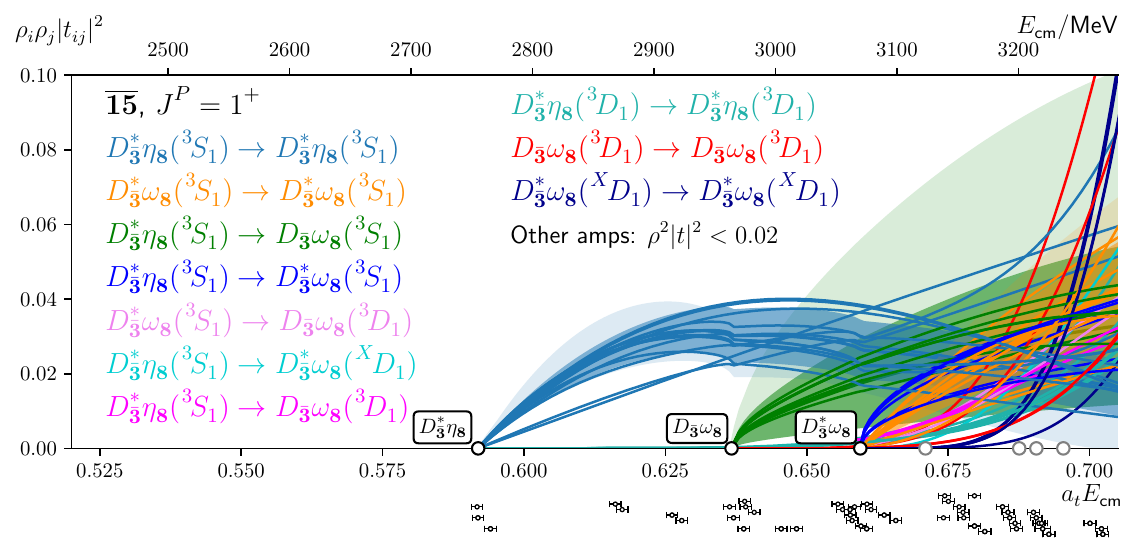}
  \includegraphics[width=0.9\linewidth]{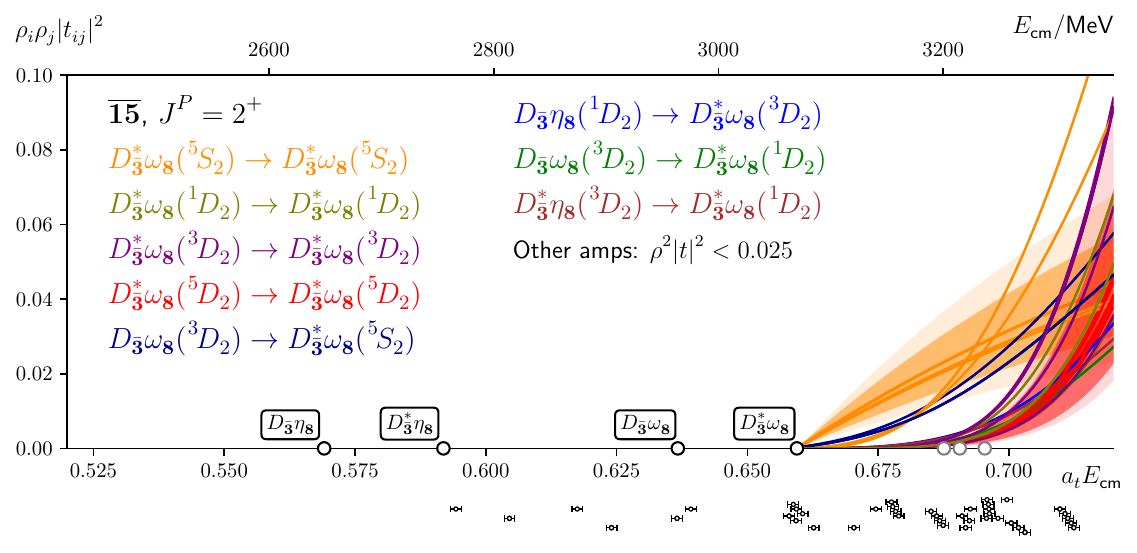}
  \includegraphics[width=0.9\linewidth]{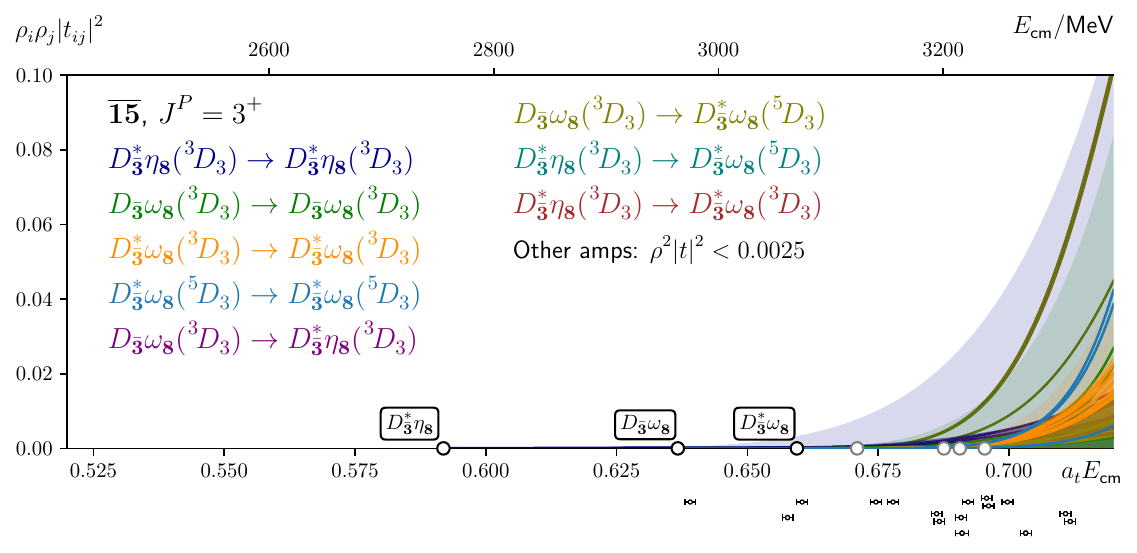}
  \caption{As Fig.~\ref{fig:A1p6f_all_fits} but for the $J^P = 1^+$ amplitudes (top panel), $J^P = 2^+$ amplitudes (middle panel) and $J^P = 3^+$ amplitudes (bottom panel) in the flavour $\overline{\mathbf{15}}$ sector.
  }
  \label{fig:15bar_other_all_fits}
\end{figure}

Upon analytic continuation, a pole predominantly coupled to $\Dst\omoct(\SLJ{1}{S}{0})$ was found in some but not all reasonable parameterisations.
In particular, parameterisations with only a constant term for the diagonal $\Dst\omoct(\SLJ{1}{S}{0})$ amplitude had no poles near the constrained energy region, whilst 
those with at least a constant plus linear term had a pole on the $[+,-]$ sheet.
These are plotted in the bottom panel of Fig.~\ref{fig:A1p15bar_all_fits}.
Although parameterisations with this pole had a better \chisq\ than those without, both achieved reasonable descriptions of the data. 
This pole is found at a similar location to the analogous flavour $\mathbf{6}$  resonance but with a smaller coupling to $\Dst \omoct(\SLJ{1}{S}{0})$, resulting in a small impact on the diagonal $\Dst \omoct(\SLJ{1}{S}{0})$ scattering amplitude. 
Therefore, we find small attraction in the diagonal $\Dst\omoct(\SLJ{1}{S}{0})$ amplitude but, since it is not found in all parameterisations, we cannot robustly confirm the existence of an associated pole.
We find no poles in the $J^P = 4^+$ amplitude near the energy region constrained.

\subsection{\texorpdfstring{$J^P = \{1, 2, 3\}^+$}{} amplitudes}
\label{section:other_15bar}

For constraining the $J^P = \{1, 2, 3\}^+$ amplitudes, we use the same workflow as in the flavour $\mathbf{6}$ sector.
  We constrain the $J^P =2^+$ from $[000]E^+$ with  $J^P=4^+$ contributions fixed to the reference parameterisation of the $[000]A_1^+$,
then constrain the $J^P =3^+$ amplitudes from $[000]A_2^+$, and finally constrain the $J^P =1^+$ amplitudes from the $[000]T_1^+$, with  $J^P =\{3,4\}^+$ contributions fixed to the reference parameterisation of the  
$[000]A_2^+$ and $[000]A_1^+$ analyses, respectively.
For each irrep, we consider the same energy cutoff on each volume as in the flavour $\mathbf{6}$ sector such that we can ignore $\smash{D^*_{0,\bar{\mathbf{3}}} \etaoct }$ and higher-threshold channels, 
and we neglect energy levels that require the inclusion of $G$-wave or higher contributions to be reproduced under the quantisation condition.
The considered partial waves are the same as in the flavour $\mathbf{6}$ sector and these can be found in see Table~\ref{table:pw_sub}.

We use $K$-matrix polynomial parameterisations for each of these $J^P$.
A more detailed discussion about the parameterisations, including the fit parameters of the reference parameterisations, can be found in Appendix~\ref{appendix:flavour_15bar_parameterisations}.
These are plotted as $\rho_i \rho_j |t_{ij}|^2$ in Fig.~\ref{fig:15bar_other_all_fits}, and for each $J^P$ are 
 found to be at most weakly attractive or repulsive, with amplitudes remaining small over the constrained energy region.
The most significant interactions were observed in the $S$-wave to $S$-wave amplitudes, but these are small in magnitude and smaller than in the $[000]A^+_1$ case, as expected from our qualitative discussion of the finite-volume spectrum.

Upon analytic continuation of the reasonable parameterisations, no poles are found in either of the  $J^P= \{1,2,3\}^+$ amplitudes near the constrained energy region.
This matches our expectations given that all amplitudes remain small over the energy region of interest and have no signatures of nearby poles.

\section{Discussion} 
\label{section:disc}
\label{section:pole_structure}

\begin{figure}
  \centering
	\includegraphics[width=1\linewidth]{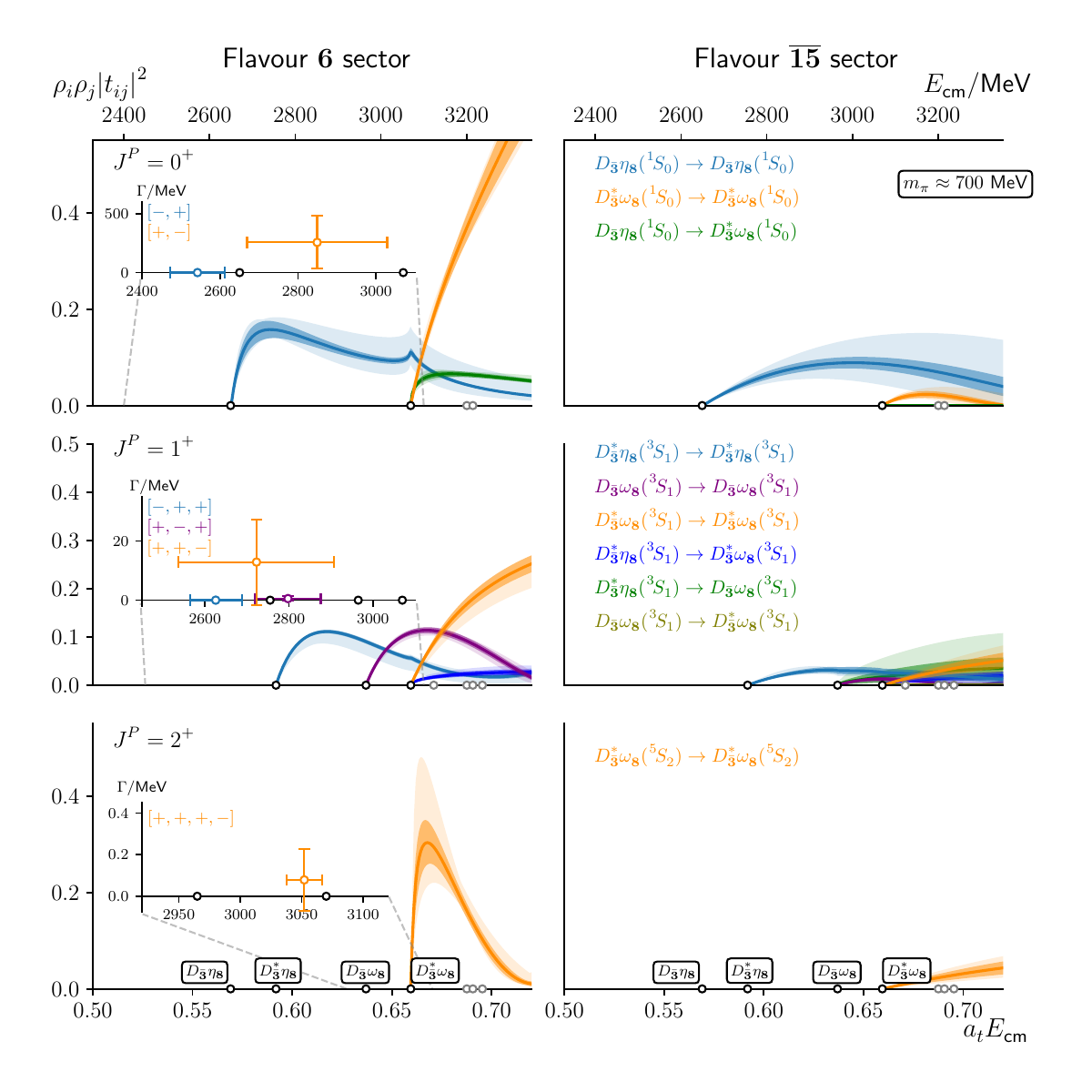}
	\caption{As Fig.~\ref{fig:A1p6f_all_fits} but for the reference parameterisations for the $J^P = 0^+$ (top panels), $1^+$ (middle panels) and $2^+$ (bottom panels) $S$-wave to $S$-wave amplitudes
          for both exotic flavour sectors.
           The behaviour of each reference parameterisation is a good representative of the reasonable parameterisations.
           Robustly determined poles are displayed in insets, shown as envelopes over all reasonable parameterisations.}
  \label{fig:summary_plot}
\end{figure}

Working with $SU(3)_f$ flavour symmetry and $m_\pi \approx 700$ MeV, this study finds attractive interactions in the diagonal $S$-wave to $S$-wave amplitudes in the flavour $\mathbf{6}$ sector, whilst the other flavour $\mathbf{6}$ amplitudes are small over the energy region considered. 
When multiple $S$-wave contributions could appear for a given $J^P$, at most small cross-channel coupling is observed, the largest of which is in $J^P = 0^+$. In the flavour $\overline{\mathbf{15}}$ sector, all amplitudes are found to be small in this energy region with at most slight repulsion or attraction. 
To visually summarise the results, the $S$-wave to $S$-wave amplitudes of the reference parameterisation for $J^P =\{0, 1,2 \}^+$ are displayed in Fig.~\ref{fig:summary_plot} as good representatives of the reasonable parameterisations.

The pattern of results approximately matches expectations from the heavy-quark limit~\cite{Isgur:1989vq, Isgur:1991wq}, as might be expected because the charm-quark mass lies well above the scale of QCD interactions ($\Lambda_{\mathrm{QCD}}$).\footnote{Similar observations were made in a lattice QCD study of the $D_0^\ast$, $D_1$ and $D_2^\ast$~\cite{Lang:2022elg}.}
In this limit the heavy quark spin decouples from the dynamics of the light and strange degrees of freedom, so near-threshold $\D\etaoct$ ($J^P=0^+$) and $\Dst\etaoct$ ($J^P=1^+$) are identical. 
Though $m_{\Dst} \ne m_{\D}$ implies the symmetry is broken, the qualitative behaviour across both the flavour $\mathbf{6}$ and $\mathbf{\overline{15}}$ is
consistent with it, including the near-threshold $0^+$ and $1^+$ poles discussed below. 
The $\D\omoct$ and $\Dst\omoct$ channels can also be related to each other in the heavy-quark limit, but the situation is more complicated because the $J^P=1^+$ $S$-wave
amplitudes get contributions from more than one configuration of light degrees of freedom.
In the flavour $\mathbf{6}$, there are clear similarities in the near-threshold behaviours of all of the relevant $S$-wave amplitudes.

We now discuss the pole singularities in these $SU(3)_f$ flavour symmetric amplitudes and then, in the following subsections, how these poles could appear when light (up and down) quark masses are lowered and flavour symmetry is broken.

\begin{table}
  \centering
    \resizebox{\textwidth}{!}{%
    \begin{tabular}{cccclll}
    \hline
    \rule{0pt}{13pt}$J^P $ & $m \pm \frac{i}{2}\Gamma/$ MeV & $SU(3)_f$ & $\SLJ{2S+1}{\ell}{J}$ &  $(-1,0)$ &  $(0,\frac{1}{2})$ & $(1,1)$  \\
    \hline    
    \rule{0pt}{11pt}$0^+$ & 2540(70)            & $\D \etaoct$   & $\SLJ{1}{S}{0}$ & $T^*_{cs0}$: $D\bar{K}$               & $D_0^{*\prime}$: $D\pi - D\eta - D_s\bar{K}- D\eta^{\prime}  $                                   & $T^*_{c\bar{s}0}$: $D_s\pi-DK$  \\
    $0^+$ & $2850(180) \pm \frac{i}{2}250(230)$  & $\Dst \omoct$  & $\SLJ{1}{S}{0}$ & $T^{*\prime}_{cs0}$: $D^*\bar{K}^*$  & $D_0^{*\prime\prime}$: $D^{*}\rho - D^{*}\omega - D_s^*\bar{K}^{*}- D^{*}\phi$                & $ T^{*\prime}_{c\bar{s}0}$: $D^*_s\rho-D^*K^*$  \\
    $1^+$ & 2630(60)                             & $\Dst \etaoct$ & $\SLJ{3}{S}{1}$ & $T_{cs1}$: $D^*\bar{K}$             & $D_1^{\prime\prime}$: $D^{*}\pi - D^{*}\eta - D_s^*\bar{K} - D^{*}\eta^{\prime} $               & $T_{c\bar{s}1}$: $D_s^{*}\pi-D^{*}K$  \\
    $1^+$ & $2800(80)\pm \frac{i}{2}0.5(0.8)$    & $\D \omoct$    & $\SLJ{3}{S}{1}$ & $T'_{cs1}$: $D\bar{K}^*$            & $D_1^{\prime\prime\prime}$: $D\rho - D\omega -  D_s\bar{K}^{*} - D\phi$                        & $T'_{c\bar{s}1}$: $D_s\rho-DK^{*}$  \\
    $1^+$ & $2720(190)\pm \frac{i}{2}13(14)$     & $\Dst \omoct$  & $\SLJ{3}{S}{1}$ & $T''_{cs1}$: $D^*\bar{K}^*$         & $D_1^{\prime \prime \prime \prime}$: $D^{*}\rho - D^{*}\omega - D_s^*\bar{K}^{*} - D^{*}\phi$  & $T''_{c\bar{s}1}$: $D_s^{*}\rho-D^{*}K^{*}$  \\
    $2^+$ & $3052(15)\pm \frac{i}{2}0.08(15)$    & $\Dst \omoct$  & $\SLJ{5}{S}{2}$ & $T^*_{cs2}$: $D^*\bar{K}^*$          & $D_2^{*\prime}$: $D^{*}\rho - D^{*}\omega - D_s^*\bar{K}^{*} - D^{*}\phi$                     & $T^*_{c\bar{s}2}$: $D_s^{*}\rho-D^{*}K^{*}$  \\
    \hline
    \end{tabular}
    }
    \caption{Summary of flavour $\mathbf{6}$ poles found in this study. 
            For each pole, the $J^P$, mass and decay width are given, 
            along with the channel and partial wave to which it is predominantly coupled
            (at the $SU(3)_f$ flavour symmetric point with $m_\pi \approx $ 700 MeV).
            Also shown are the pole’s ($S$, $I$) = (strangeness, isospin) components, each with a hadron label and the channels corresponding to their dominantly-coupled $SU(3)_f$ channel when $SU(3)_f$ flavour symmetry is broken. 
            Hadron labels follow the PDG naming convention~\cite{ParticleDataGroup:2024cfk}, with primes distinguishing poles of the same quantum numbers,
            and connections to experimentally-observed states are described in the main text.
            The $D_J^{(\ast)}$ states are all labelled with at least one prime to emphasise that they do not correspond to the lowest-energy state in these channels.
            }
             \label{table:pole_summary}
\end{table}

\subsection{Poles at the flavour symmetric point}
\label{sec:poles_su3}
In the flavour $\mathbf{6}$ sector, for each $S$-wave channel an associated pole is found below that channel's threshold predominantly coupled to that (kinematically-closed) $S$-wave channel.
This gives six well-determined poles: two with $J^P =0^+$, three with $J^P =1^+$ and one with $J^P =2^+$.
There are no $S$-wave contributions and no poles in $J^P = \{3, 4\}^+$ in the energy region considered.
Each pole is on a hidden sheet, so its effect on scattering and production amplitudes is less direct than for a more standard resonance, such as the $\rho$, which corresponds to a pole on a proximal sheet. Such a hidden-sheet pole may also be less well constrained by lattice data and show more dependence on the parameterisation used, particularly if it is not close to threshold.
In Fig.~\ref{fig:summary_plot} it can be seen that the main effect of each of these poles is rapid growth at threshold in the amplitude for the dominantly-coupled channel rather than a bump-like enhancement.
The mass and decay width of each well-determined pole (taking an envelope over all reasonable parameterisations) are displayed as insets in Fig.~\ref{fig:summary_plot} and summarised in Table~\ref{table:pole_summary}.

In $J^P =0^+$, the first flavour $\mathbf{6}$ pole, shown in blue in the top-left panel of Fig.~\ref{fig:summary_plot}, is a virtual bound state predominantly coupled to the pseudoscalar-pseudoscalar channel $\D\etaoct(\SLJ{1}{S}{0})$. This was found, along with a bound state in the flavour $\bar{\mathbf{3}}$ sector, in our study of elastic $\D\etaoct$ $S$-wave scattering~\cite{Yeo:2024chk}. A recent lattice study with $SU(4)_f$ flavour symmetry finds attraction and an analogous virtual bound state~\cite{Baeza-Ballesteros:2025iee}. The second $0^+$ pole, shown in orange in the figure, is a resonance predominantly coupled to the closed vector-vector channel $\Dst\omoct(\SLJ{1}{S}{0})$, but it would decay to its only open channel, $\D\etaoct(\SLJ{1}{S}{0})$.

In $J^P =1^+$, the lowest pole is a virtual bound state predominantly coupled to the vector-pseudoscalar channel $\Dst\etaoct(\SLJ{3}{S}{1})$ (blue in the middle-left panel of Fig.~\ref{fig:summary_plot}).
It is found just below $\Dst \etaoct$ threshold at a similar distance to that of the $J^P = 0^+$ virtual bound state from $\D \etaoct$ threshold, in agreement with the expected approximate heavy-quark spin symmetry that relates these two sectors \cite{Isgur:1989vq,Isgur:1991wq}.
The observation of attraction and a virtual bound state here, but only weak interactions in the $\overline{\mathbf{15}}$, agrees with the lattice study of Ref.~\cite{Gregory:2025ium} which explored only the shifts of the ground-state finite-volume energies.
Ref.~\cite{Guo:2025imr} argues that the observation of attraction at threshold and a corresponding pole in $J^P = \{0,1\}^+$ for the flavour $\mathbf{6}$, and only repulsion/weak interactions in the analogous flavour $\overline{\mathbf{15}}$ channels, is not consistent with a diquark-antidiquark model of a tetraquark and supports the identification of these poles as molecular states.

The second $J^P=1^+$ pole is just below $\D\omoct$ threshold and couples predominantly to $\D\omoct(\SLJ{3}{S}{1})$ (purple in the middle-left panel of Fig.~\ref{fig:summary_plot}).
Its properties closely match those of the first $J^P=1^+$ pole.
Despite having a mass above $\Dst \etaoct$ threshold, the vanishingly small coupling to this channel means it is effectively a virtual bound state in $\D\omoct$.
The third $J^P =1^+$ pole is predominantly coupled to the vector-vector channel, $\Dst\omoct(\SLJ{3}{S}{1})$ (orange in the figure).
Given this pole's similar mass and coupling pattern to the $J^P =0^+$ resonance, it can likely be identified as its $J^P =1^+$ partner.
This $1^+$ pole is found slighter further from $\Dst\omoct$ threshold than the corresponding $J^P=0^+$ pole, with less rapid growth in the diagonal $S$-wave $\Dst\omoct$ amplitude.
Although the parameterisations showed a preference for small nonzero couplings of this state to the lower-threshold channels, decoupled fits could not be ruled out.

\begin{table}
  \centering
    \begin{tabular}{ccccc}
    \hline
    $(S, I)$ & $SU(2)_I$ scattering channels  &  $SU(3)_f$ irrep   \\
    \hline    
    \rule{0pt}{12pt} $(-1, 0)$ & $D^{(*)}\bar{K}^{(*)}$ & $\mathbf{6} $  \\
    $(-1, 1)$                  & $D^{(*)}\bar{K}^{(*)}$ & $\overline{\mathbf{15}} $  \\
    $(0, \frac{1}{2})$         & $D^{(*)}\pi -D_s^{(*)}\bar{K}^{(*)} - D^{(*)}\rho-D^{(*)}\eta^{(\prime)} - D^{(*)}\omega -D^{(*)}\phi$ & $\bar{\mathbf{3}} \oplus \mathbf{6} \oplus \overline{\mathbf{15}}$  \\
    $(0, \frac{3}{2})$         & $D^{(*)}\pi-D^{(*)}\rho$ & $\overline{\mathbf{15}} $  \\
    $(1, 0)$                   & $D^{(*)}K^{(*)}-D_s^{(*)}\eta^{(\prime)} -D_s^{(*)}\omega -D_s^{(*)}\phi$ & $\bar{\mathbf{3}} \oplus \overline{\mathbf{15}}$  \\
    $(1, 1)$                   & $D^{(*)}K^{(*)}-D_s^{(*)}\pi -D_s^{(*)}\rho$ & $\mathbf{6} \oplus \overline{\mathbf{15}}$  \\
    $(2, \frac{1}{2})$        & $D_s^{(*)}K^{(*)}$ & $\overline{\mathbf{15}} $  \\
    \hline
    \end{tabular}
    \caption{$SU(3)_f$ irrep decomposition for each $(S, I)$ sector, shown alongside the lightest possible pseudoscalar-pseudoscalar, pseudoscalar-vector and vector-vector channels.}
    \label{table:SU3_contributions}
\end{table}

The pole in $J^P =2^+$ is just below $\Dst \omoct$ threshold (orange in the bottom-left panel of Fig.~\ref{fig:summary_plot}) and is predominantly coupled to $\Dst \omoct(\SLJ{5}{S}{2})$, the only $S$-wave channel in the energy region of interest. 
Given its mass and coupling pattern, it can be identified as a $J^P =2^+$ partner of the $J^P =0^+$ and $1^+$ resonances that are also predominantly coupled to $\Dst \omoct$.

In the flavour $\overline{\mathbf{15}}$ sector, no poles were reliably found across all parameterisations in $J^P = \{0, 1, 2,3,4\}^+$.
For a few parameterisations, a $J^P =0^+$ flavour $\overline{\mathbf{15}}$ pole was found predominantly coupled to the vector-vector channel, but it had a small influence on the physical scattering amplitude and it was not reliably determined. Future studies would be needed to confirm or refute its existence.

In the following subsections we discuss what might happen to the $(S, I)$ = (strangeness, isospin) components of the flavour $\mathbf{6}$ poles when the light quark masses are lowered and $SU(3)_f$ flavour symmetry is broken\footnote{but isospin symmetry is retained}.
The channels and $SU(3)_f$ flavour multiplets that contribute to the different $(S,I)$ sectors are shown in Table~\ref{table:SU3_contributions}.
We only make some qualitative comments because the trajectories of these hidden-sheet poles as the light-quark mass is reduced may be non-trivial, and the sheet structure is more complicated when $SU(3)_f$ flavour symmetry is broken. In addition, for physical light-quark masses all but the lightest hadrons in these channels become resonances, thus these channels become decays to three or more hadrons.
We also compare with experimental observations and other studies in the literature.

\subsection{Antistrange isoscalar sector}
The antistrange isoscalar sector, i.e.\ $(S,I)=(-1,0)$, has exotic flavour quantum numbers and only gets contributions from the $SU(3)_f$ flavour $\mathbf{6}$ (see Table~\ref{table:SU3_contributions}).
Our results suggest that there will be six states: two with $J^P =0^+$ ($T^*_{cs0}$, $T^{*\prime}_{cs0}$), three with $J^P =1^+$ ($T_{cs1}$, $T'_{cs1}$, $T''_{cs1}$) and one with $J^P =2^+$ ($T^*_{cs2}$).\footnote{Hadron labels assigned to each component do not correspond to experimental states unless specified otherwise.}
These are listed in Table~\ref{table:pole_summary} together with the channel to which they are predominantly coupled. 
While our calculations at the $SU(3)_f$ symmetric point suggest each dominantly couples to one channel in $S$-wave, the couplings may change when the light-quark masses are reduced towards their physical values, and so they may be seen in other final states.
If the poles remain on hidden sheets, they potentially have a less direct influence on physical scattering and production amplitudes than more standard resonance poles that appear on the proximal sheet.

The lightest $J^P = 0^+$ state, $T^{*}_{cs0}$, coupled dominantly to $D\bar{K}$, has been seen as a virtual bound state in previous lattice studies with $m_\pi \approx 239$ and $391$ MeV~\cite{Cheung:2020mql}. 
Because it has a relatively low mass, it does not appear to correspond to the experimentally-observed $T^{*}_{cs0}(2870)^0$ which we instead suggest is related to the other $0^+$ state (see below).
 
Under the assumption of approximate heavy-quark spin symmetry \cite{Isgur:1991wq, Isgur:1989vq}, the $D^*\bar{K}(\SLJ{3}{S}{1})$ and $D\bar{K}(\SLJ{1}{S}{0})$ systems are related. 
We therefore expect the $T_{cs1}$ (dominantly coupled to $D^*\bar{K}$) to have similar properties to the $T^{*}_{cs0}$ and to be a virtual bound state, at least for $m_\pi \approx 239$ and $391$ MeV.
The $T'_{cs1}$ (dominantly coupled to $D\bar{K}^*$) is very similar to the $T_{cs1}$ at the $SU(3)_f$ symmetric point -- for lighter light-quarks, 
the $T'_{cs1}$ may remain as an approximate virtual bound state or the different kinematics and couplings may mean it appears as a resonance.
The $T_{cs1}$ and $T'_{cs1}$ have so far not been studied in any other lattice QCD calculation or reported experimentally.

The $T^{*\prime}_{cs0}$, $T^{\prime \prime}_{cs1}$ and $T^{*}_{cs2}$ are all predominantly coupled to the vector-vector channel $D^*\bar{K}^*$.
Due to the proximity of the experimentally-observed $T^*_{cs0}(2870)^0$~\cite{LHCb:2020bls, LHCb:2020pxc} to $D^*\bar{K}^*$ threshold, we identify this with the $T^{*\prime}_{cs0}$.
With this identification, our work supports the $T^*_{cs0}(2870)^0$ having isospin 0 and suggests it will couple most strongly to $D^*\bar{K}^*(\SLJ{1}{S}{0})$,
although its couplings to the lower-threshold channels may change for lighter light-quarks. 
Since the $T''_{cs1}$ and $T^*_{cs2}$ also couple predominantly to $D^*\bar{K}^*$, we identify them as $J^P=1^+$ and $J^P=2^+$ partners of the $T^*_{cs0}(2870)^0$.
These were first predicted in Ref.~\cite{Molina:2010tx} within the hidden gauge formalism, but currently there are no other lattice QCD calculations of them.
Whilst a $J^P =1^+$ state would not appear in the decays investigated in Refs.~\cite{LHCb:2020bls, LHCb:2020pxc}, a $J^P =2^+$ state could.
However, given the statistics available, that study was unable to determine if the data could be described by a $J^P =0^+$ resonance and a $J^P =2^+$ resonance
(instead of the $T^*_{cs0}(2870)^0$ $J^P =0^+$ resonance and the $T^*_{cs1}(2900)$ $J^P =1^-$ resonance).
Refs.~\cite{Bayar:2022wbx, Song:2025ueo} have put forward methods to disentangle different resonant contributions from that data on $B^+ \rightarrow D^+D^-K^+$ decays~\cite{LHCb:2020bls, LHCb:2020pxc} and more recent data on $B^- \rightarrow D^-D^0K_S^0$ decays~\cite{LHCb:2024xyx} using the moments of the angular mass distribution, which would benefit in both cases from larger statistics. 
Whilst our findings suggest the existence of a $J^P = 2^+$ state, we make no comment on the $T^*_{cs1}(2900)$ since we have not investigated $J^P =1^-$ here.

\subsection{Isospin-$\frac{1}{2}$ sector}

The isospin-$\frac{1}{2}$, $(S,I)=(0,\frac{1}{2})$, sector is non-exotic in flavour and more complicated because it gets contributions from the flavour $\bar{\mathbf{3}}$, $\mathbf{6}$ and $\overline{\mathbf{15}}$ multiplets. 
Whilst our results suggest the $\overline{\mathbf{15}}$ is at most weakly interacting with no well-determined poles in this energy region, the flavour $\bar{\mathbf{3}}$ (which we have not done a scattering analysis of here) is expected to contain the ``conventional'' $D$ mesons. 
Above the stable ground states (listed in Table~\ref{table:mesons}), we find no indications of additional $\bar{\mathbf{3}}$ states below $a_tE_{\sf{cm}} \approx 0.65$ (3000 MeV) when the spectra on these lattices are computed using a basis of only fermion-bilinear operators -- the pattern resembles that found in Refs.~\cite{Moir:2013ub,Cheung:2016bym} where $m_\pi \approx 240, 391$ MeV.
Furthermore, the relevant $S$-wave $SU(3)_f$ channels each split into several channels, each with different thresholds, and some hadrons in these channels become admixtures of $SU(3)_f$ multiplets (for example, the  $\etas$ -- the lightest flavour-singlet pseudoscalar -- and the $I=0, I_z=0$ component of the $\etaoct$ mix to become $\eta$ and $\eta'$) making it more difficult to predict what will happen away from the $SU(3)_f$ symmetric point. 
However, the lightest two states, the virtual bound states in $J^P = \{0,1\}^+$, have been studied with lighter light quarks in lattice QCD and we can use this to guide our expectations.

In our study of elastic open-charm $J^P=0^+$ scattering at the flavour symmetric point~\cite{Yeo:2024chk}, as well as the virtual bound state in the flavour $\mathbf{6}$, we found a bound state in the flavour $\bar{\mathbf{3}}$ which we label $D_0^*$ here. We argued that the latter state is roughly in line with expectations for [$n\SLJ{2S+1}{\ell}{J} = 1\SLJ{3}{P}{0}$] quark-model $c\bar{q}$ configuration, where $n$ is the radial quantum number. 
The observation of an accompanying flavour $\mathbf{6}$ pole provided support for a two-pole structure describing $D\pi(\SLJ{1}{S}{0})$ scattering at the physical light-quark mass.
Note, however, a consensus on how the second pole, labelled $D_0^{*\prime}$ in Table~\ref{table:pole_summary}, manifests itself with light-quark masses close to their physical values is yet to be reached. Refs.~\cite{Moir:2016srx, Gayer:2021xzv, Yan:2024yuq} only robustly determined a single pole in the elastic-scattering region near the $D^*_0(2300)$. 
Ref.~\cite{Asokan:2022usm} argued a second pole is required if $SU(3)_f$ flavour constraints are placed on the $K$-matrix, with this pole appearing above $D\eta$ threshold and on a hidden sheet, coupled to $S$-wave $D\pi-D\eta-D_s\bar{K}$ (the coupling to the $D\eta^\prime$ was not determined).

Similar observations can be made in the axial-vector isospin-$\frac{1}{2}$ sector.
A recent coupled-channel study with $m_{\pi} \approx 391$ MeV found three $J^P =1^+$ poles~\cite{Lang:2025pjq}.
The two with the lowest masses, labelled $D_1$ and $D_1'$ and identified with the $D_1(2430)$ and $D_1(2420)$, roughly match the pattern expected for the quark model $[1\SLJ{1}{P}{1}]$ and $[1\SLJ{3}{P}{1}]$ configurations and are found on the physical and proximal sheets.
The third pole is found at higher energies on a hidden sheet which, analogously to the two-pole structure in $D\pi(\SLJ{1}{S}{0})$ scattering~\cite{Kolomeitsev:2003ac,Albaladejo:2016lbb,Du:2020pui,Meissner:2020khl}, can be identified with the axial-vector flavour $\mathbf{6}$ virtual bound state found here and labelled $D_1''$ in Table~\ref{table:pole_summary}.
Ref.~\cite{Lang:2025pjq} found this pole couples to $S$-wave $D^*\pi-D^*\eta-D_s^*\bar{K}$, with the largest coupling to $D^*\pi$, the channel with the largest available phase space (the coupling to $D^*\eta^\prime$ was not determined).

Taken together, these observations suggest that in the isospin-$\frac{1}{2}$ sector each flavour-exotic $\mathbf{6}$ pole manifests as a single pole, likely on a hidden sheet, coupled to the relevant $S$-wave channels,  appearing in addition to the quark-model-like poles from the flavour $\bar{\mathbf{3}}$ on the physical/proximal sheets.
Under these assumptions, we expect that the $D_1'''$, dominantly coupled to $\D\omoct(\SLJ{3}{S}{1})$ at the flavour symmetric point, will primarily couple to $S$-wave $D\rho - D\omega-D_s\bar{K}^{*} - D\phi$.
Note that, whilst the $(0, \frac{1}{2})$ component of $\D\omoct$ contains  $D\phi$, away from the flavour symmetric point $\phi$ is found to be mainly composed of strange quarks and with a significantly higher mass than the other light-vector mesons listed here. 
The $D_1'''$ may well couple differently to this channel than the others depending on the pattern of $SU(3)_f$ flavour symmetry breaking. 
Although the $D_1'''$ has very small coupling to the lower-threshold $\Dst\etaoct$ channel at the flavour symmetric point, this may change as the light-quark mass is reduced and so it might also couple to $D^*\pi-D^*\eta-D_s^*\bar{K}-D^*\eta'$.
This state has not been studied in any other lattice QCD calculations, but we expect it to lie above the three poles found in Ref.~\cite{Lang:2025pjq} (labelled here as $D_1$, $D_1'$ and  $D_1''$). 
Currently, no observations corresponding to the $D_1''$ and $D_1'''$ have been reported experimentally.

Finally, we have the $D_0^{*\prime \prime}$, $D_1''''$ and  $D^{*\prime}_2$ (see Table~\ref{table:pole_summary}).
In the flavour symmetric limit, these states are in same flavour $\mathbf{6}$ multiplets as the $T^{*\prime}_{cs0}$, $T^{\prime \prime}_{cs1}$ and $T^{*}_{cs2}$, respectively, which implies they are isospin-$\frac{1}{2}$ partners of the $T^*_{cs0}(2870)^0$. Our results suggest they will predominantly couple to $S$-wave $D^{*}\rho - D^{*}\omega - D_s^*\bar{K}^{*} - D^{*}\phi$.
Although the couplings to channels with lower thresholds are small at the flavour symmetric point, as the light-quark mass is reduced these couplings may become larger and so the $D_0^{*\prime \prime}$ and $D_1''''$ could couple to, respectively, pseudoscalar-pseudoscalar and vector-pseudoscalar channels.
Naïvely extrapolating to the physical pion mass by assuming the pole will be located near to its predominantly-coupled channels' thresholds
and using $[n\SLJ{2S+1}{\ell}{J}]$ notation to refer to quark-model-like states,
 we expect the $D_0^{*\prime \prime}$ to appear above the $D_0^{*}$ $[1\SLJ{3}{P}{0}]$ and the exotic $D_0^{*\prime}$, but below the first-excited $[2\SLJ{3}{P}{0}]$ state (which is expected at $\approx 3000$ MeV~\cite{Moir:2013ub,Cheung:2016bym}).
The  $D_1''''$ should appear above the $D_1$ $[1\SLJ{1}{P}{1}]$ and $D_1'$ $[1\SLJ{3}{P}{1}]$ and near in energy to the exotic $D_1''$ and $D_1'''$, but below the excited $[2\SLJ{1}{P}{1}]$ and $[2\SLJ{3}{P}{1}]$ states (which are both expected at $\approx 3000$ MeV~\cite{Moir:2013ub,Cheung:2016bym}).
The $D^{*\prime}_2$ should appear above the $D^*_2$ $[1\SLJ{3}{P}{2}]$ but below the first-excited $[2\SLJ{3}{P}{2}]$ state (expected at $\approx 3000$ MeV~\cite{Moir:2013ub,Cheung:2016bym}). 
Of these, only the $D_0^*$, $D_1$, $D_1'$ and $D_2^*$ are experimentally reported~\cite{ParticleDataGroup:2024cfk}.

In summary, under the assumption (motivated above) that each flavour $\mathbf{6}$ pole appears as a single pole in the isospin-$\frac{1}{2}$ sector, coupled to the relevant channels, and the flavour $\bar{\mathbf{3}}$ follows quark-model expectations, we expect four $J^P =0^+$ states, seven $J^P =1^+$ states and three $J^P=2^+$ states up to energies of $\approx 3000$ MeV. 
Six of these arise from the $\mathbf{6}$ and so are exotic in nature -- they may have a less direct influence on scattering and production amplitudes if they remain on hidden sheets for physical light-quark masses.
Studies exploring coupled-channel scattering in the flavour $\bar{\mathbf{3}}$ or in the isospin-$\frac{1}{2}$ sector away from the flavour symmetric point with $J^P =\{0,1,2,3,4\}^+$ would help confirm this picture.

\subsection{Strange isovector sector}

The strange isovector sector, $(S,I)=(1,1)$, gets contributions from the flavour $\mathbf{6}$ and $\overline{\mathbf{15}}$, with the latter observed to be weakly interacting with no well-determined poles in the energy region considered. 
Based on the observations made in the previous section, when breaking $SU(3)_f$ flavour symmetry and moving towards the physical pion mass, we expect the isospin-1 component of each of the six flavour $\mathbf{6}$ poles, labelled as $T^*_{c\bar{s}0}$,  $T^{*\prime}_{c\bar{s}0}$, $T_{c\bar{s}1}$, $T'_{c\bar{s}1}$, $T''_{c\bar{s}1}$ and $T^*_{c\bar{s}2}$ in Table~\ref{table:pole_summary}, to manifest as a single pole coupled to the relevant channels.
Therefore, our results suggest the $T^*_{c\bar{s}0}$ will predominantly couple to $D_s \pi - DK$, the $T_{c\bar{s}1}$ to $D^*_s \pi - D^*K$, and the $T'_{c\bar{s}1}$ to $D_s \rho - DK^*$, each in $S$-wave.
Again, we note the $T'_{c\bar{s}1}$ may couple more strongly to the lower-threshold channels when the light-quark mass is lowered. 
So far, these states have not been reported experimentally or studied in other lattice QCD calculations.
Since it has a relatively low mass, the $T^*_{c\bar{s}0}$ does not appear to correspond to the experimentally-observed $T^*_{c\bar{s}0}(2900)$.

The remaining three strange isospin-1 states are $T^{*\prime}_{c\bar{s}0}$, $T''_{c\bar{s}1}$ and $T^*_{c\bar{s}2}$.
We identify the $T^{*\prime}_{c\bar{s}0}$ with the experimentally-observed $T^*_{c\bar{s}0}(2900)$. 
Since the $T^{*\prime}_{cs0}$ and $T^{*\prime}_{c\bar{s}0}$ are components of the same flavour $\mathbf{6}$ pole, this suggests the $T^*_{cs0}(2870)^0$ and $T^*_{c\bar{s}0}(2900)$ correspond to a single state in the flavour symmetric limit, an idea first suggested in Ref.~\cite{Dmitrasinovic:2023eei}.
This identification supports the $T^*_{c\bar{s}0}(2900)$ being isospin-1 and hence implies the existence of a singly charged $T^*_{c\bar{s}0}(2900)^+$ (the $I_z=0$ component of the isospin-1 multiplet) which has not yet been observed experimentally.\footnote{The other $I_z$ components, $T^*_{c\bar{s}0}(2900)^0$ and $T^*_{c\bar{s}0}(2900)^{++}$, have been observed.}
Our results suggest the $T^*_{c\bar{s}0}(2900)$ is predominantly coupled to $S$-wave $D^*_s\rho-D^*K^*$.
Similarly, we expect the  $T''_{c\bar{s}1}$ and $T^*_{c\bar{s}2}$ to predominantly couple to $D^*_s\rho-D^*K^*$ in $S$-wave, 
which can be identified as $J^P = 1^+$ and $J^P = 2^+$ partners of the $T^*_{c\bar{s}0}(2900)$.
No candidates for the $T''_{c\bar{s}1}$ and $T^*_{c\bar{s}2}$ have been reported experimentally. 
Refs.~\cite{Lyu:2024zdo, Lyu:2025rsq} provide methods to identify the $T^*_{c\bar{s}2}$ from LHCb data on $B^+ \rightarrow D^{*-}D^+K^+$ decays~\cite{LHCb:2024vfz} and 
$B^+ \rightarrow D^{*-}D_s^+\pi^+$ decays~\cite{LHCb:2024vhs}, using the moments of the angular mass distribution.

\section{Summary and Outlook} 
\label{section:concl_and_outlook}
In summary, we have presented the first lattice QCD study of open-charm coupled-channel scattering with $SU(3)_f$ flavour symmetry and  $m_\pi \approx 700$ MeV, focused on  $J^P=\{0,1,2,3, 4\}^+$ in the flavour-exotic $\mathbf{6}$ and   $\overline{\mathbf{15}}$ sectors. 
A common pattern was evident.
In the flavour $\mathbf{6}$ sector, attractive interactions were observed and a corresponding pole was found whenever a scattering channel could contribute in $S$-wave.
This lead to six  flavour $\mathbf{6}$ poles: two with $J^P = 0^+$, three with $J^P = 1^+$ and one with $J^P = 2^+$.
Each of these were found on a hidden sheet and they are summarised in Table~\ref{table:pole_summary}. 
Only weak interactions and no poles were observed in $J^P =\{3,4\}^+$.
In the flavour $\overline{\mathbf{15}}$ sector, at most slight repulsion or slight attraction was observed, and no poles were consistently found with $J^P=\{0,1,2,3, 4\}^+$ in the energy region considered.
In $J^P = 0^+$, some parameterisations contained a flavour $\overline{\mathbf{15}}$ pole with a small coupling to the vector-vector channel, having only a small impact on the scattering amplitude.
The existence of this pole was not robustly determined in this study.

Of the poles in the flavour $\mathbf{6}$ sector, two correspond to virtual bound states, one with $J^P=0^+$ and one with $J^P=1^+$, predominantly coupled to $\D\etaoct(\SLJ{1}{S}{0})$ and $\Dst\etaoct(\SLJ{3}{S}{1})$, respectively.
The scalar pole was previously observed in our study of elastic $\D\etaoct$ scattering~\cite{Yeo:2024chk}.
By comparing to studies with lower pion masses~\cite{Moir:2016srx, Gayer:2021xzv, Yan:2024yuq, Asokan:2022usm, Lang:2025pjq},
we argued that when $SU(3)_f$ flavour symmetry is broken, each flavour $\mathbf{6}$ pole will appear as a single pole in each of the $(S,I)$ = (strangeness, isospin) = $(-1,0)$, $(0,\frac{1}{2})$ and $(1,1)$ sectors.
The results provide supporting evidence for a two-pole structure for $D\pi$ and $D^*\pi$ $S$-wave scattering~\cite{Kolomeitsev:2003ac, Albaladejo:2016lbb,Du:2020pui,Meissner:2020khl}.

The second $J^P=1^+$ pole predominantly couples to $\D\omoct(\SLJ{3}{S}{1})$.
Due to the small coupling observed between the two vector-pseudoscalar channels, this pole is effectively also a virtual bound state.
Our findings suggest a pole in each of the axial-vector $(S,I) = (-1,0)$, $(0,\frac{1}{2})$ and $(1,1)$  sectors (in addition to the $J^P =1^+$ virtual bound state mentioned above).
The pole in the isospin-$\frac{1}{2}$ sector will potentially appear at energies higher than that investigated in Ref.~\cite{Lang:2025pjq}.

The last three poles found in this study, one in each $J^P=\{0,1,2\}^+$,  are each coupled predominantly to a vector-vector channel, $\Dst\omoct(\SLJ{1}{S}{0})$, $\Dst\omoct(\SLJ{3}{S}{1})$ and $\Dst\omoct(\SLJ{5}{S}{2})$, respectively.
We argued that the $J^P=0^+$  pole could be identified with  the $T^*_{cs0}(2870)^0$ and $T^*_{c\bar{s}0}(2900)$ states, appearing as a single flavour $\mathbf{6}$ pole in the flavour symmetric limit.
This identification supports the isospin of these states being 0 and 1, respectively.
The analogous poles with $J^P=\{1,2\}^+$ were argued to be  $J^P=\{1,2\}^+$ partners of the $T^*_{cs0}(2870)^0$ and $T^*_{c\bar{s}0}(2900)$ states, which are currently unobserved.
To complete the picture of the pattern of open-charm hadrons at the flavour symmetric point, we suggest coupled-channel studies in the flavour $\bar{\mathbf{3}}$, which is expected to contain the ``conventional'' $D$ mesons.
Furthermore, to understand how the exotic poles found in this study appear away from the flavour symmetric point, we suggest coupled-channel studies in each of the 
$(S,I)$ = $(-1,0)$, $(0,\frac{1}{2})$ and $(1,1)$ sectors with lighter light-quark masses.
This would require  three-(or more) hadron extensions to the current Lüscher formalism for sufficiently small pion masses.

We also suggest similar studies to the one presented here but for the negative-parity sectors.
In particular, the $J^P =1^-$ sector, corresponding to the $T^*_{cs1}(2900)$, could be investigated. 
Other $J^-$ could be explored, including those where models predict additional exotic resonances~\cite{Wang:2025wpc}.
Furthermore, analogous studies with a bottom quark instead of a charm quark could be performed, probing the dependence of the phenomena on the heavy-quark mass and testing heavy-quark flavour symmetry.

\acknowledgments
We thank our colleagues within the Hadron Spectrum Collaboration (\url{www.hadspec.org}), in particular Jozef Dudek, for useful discussions.
JDEY, CET and DJW acknowledge support from the U.K. Science and Technology Facilities Council (STFC) [grant number ST/X000664/1].
DJW acknowledges support from a Royal Society University Research Fellowship.

The software codes {\tt Chroma}~\cite{Edwards:2004sx}, {\tt QUDA}~\cite{Clark:2009wm,Babich:2010mu} and {\tt Redstar}~\cite{Chen:2023zyy} were used.
Some software codes used in this project were developed with support from the U.S.\ Department of Energy, Office of Science, Office of Advanced Scientific Computing Research and Office of Nuclear Physics, Scientific Discovery through Advanced Computing (SciDAC) program; also acknowledged is support from the Exascale Computing Project (17-SC-20-SC), a collaborative effort of the U.S.\ Department of Energy Office of Science and the National Nuclear Security Administration.

This work used the Cambridge Service for Data Driven Discovery (CSD3), part of which is operated by the University of Cambridge Research Computing Service (\url{www.csd3.cam.ac.uk}) on behalf of the STFC DiRAC HPC Facility (\url{www.dirac.ac.uk}). The DiRAC component of CSD3 was funded by BEIS capital funding via STFC capital grants ST/P002307/1 and ST/R002452/1 and STFC operations grant ST/R00689X/1. Other components were provided by Dell EMC and Intel using Tier-2 funding from the Engineering and Physical Sciences Research Council (capital grant EP/P020259/1). This work also used the earlier DiRAC Data Analytic system at the University of Cambridge. This equipment was funded by BIS National E-infrastructure capital grant (ST/K001590/1), STFC capital grants ST/H008861/1 and ST/H00887X/1, and STFC DiRAC Operations grant ST/K00333X/1. DiRAC is part of the National E-Infrastructure.

Propagators used in this project were generated using DiRAC facilities, at Jefferson Lab under the USQCD Collaboration and the LQCD ARRA Project, under an ALCC award, using resources of the Oak Ridge Leadership Computing Facility at the Oak Ridge National Laboratory, which is supported by the Office of Science of the U.S. Department of Energy under Contract No. DE-AC05-00OR22725. and using resources of the National Energy Research Scientific Computing Center (NERSC), a DOE Office of Science User Facility supported by the Office of Science of the U.S. Department of Energy under Contract No. DE-AC02-05CH11231. The authors acknowledge the Texas Advanced Computing Center (TACC) at The University of Texas at Austin for providing HPC resources.
Gauge configurations were generated using resources awarded from the U.S. Department of Energy INCITE program at the Oak Ridge Leadership Computing Facility, the NERSC, the NSF Teragrid at the TACC and the Pittsburgh Supercomputer Center, using DiRAC facilities and at Jefferson Lab.

\vspace{0.5cm}

\noindent \textbf{Data Access Statement}
\vspace{0.2cm}

\noindent
Reasonable requests for data can be directed to the authors and will be considered in accordance with the Hadron Spectrum Collaboration's policies on sharing data.

\newpage
\appendix
\section{Model averaging of energy levels} 
\label{appendix:model_av}
\begin{figure}
	\includegraphics[width=\linewidth]{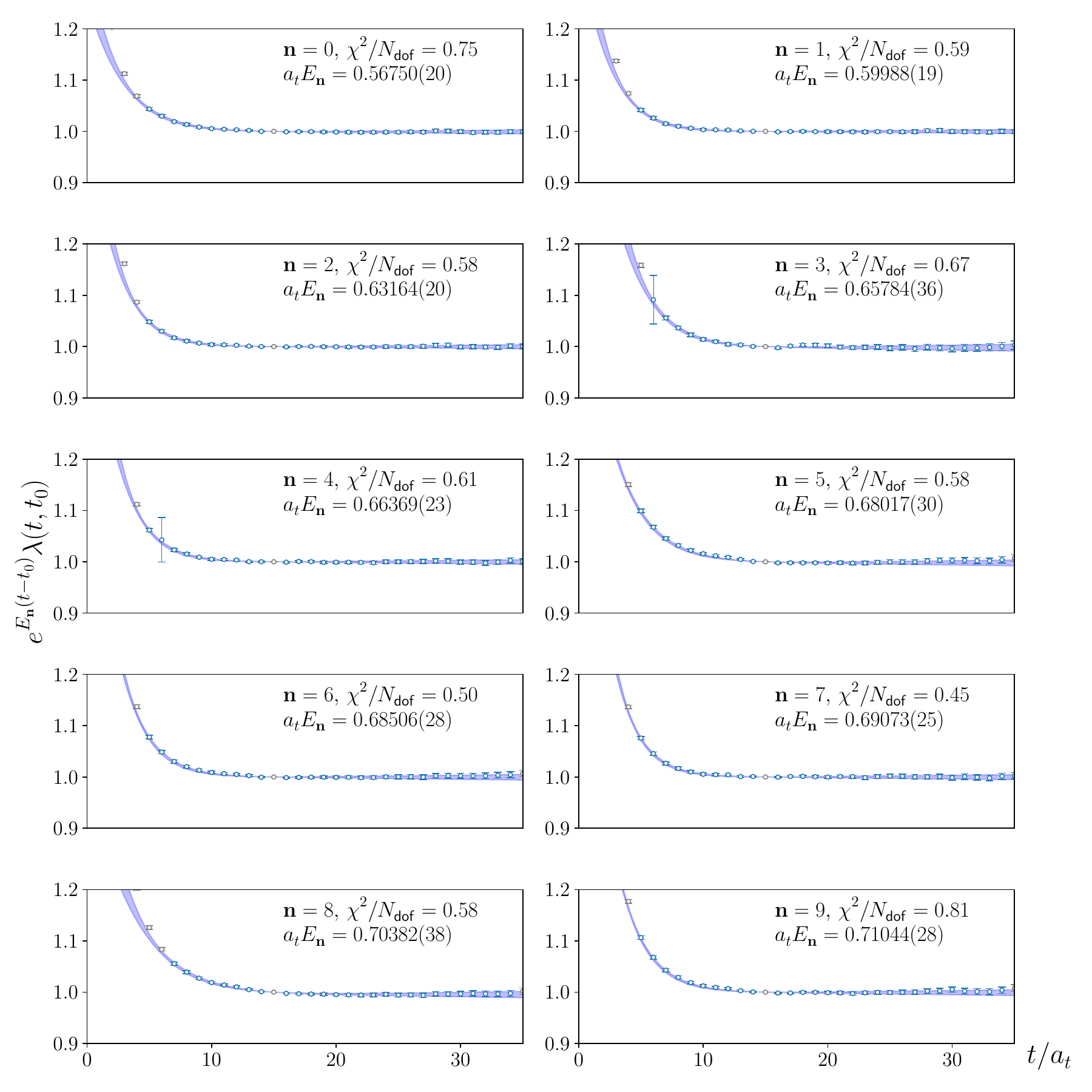}
	\caption{Principal correlators for the lowest ten levels on the $L/a_s = 20$ volume for $[000]A^+_1$ in the flavour $\mathbf{6}$ sector, ordered by extracted energy, displayed as $e^{E_{\mathbf{n}}(t-t_0)}\lambda^{(\mathbf{n})}(t,t_0)$ (points with error bars). 
             The fit with the best fit quality (lowest AIC) for each principal correlator  is similarly displayed (blue band) with its \chisq\ and the extracted energy.  
             Timeslices not used in the principal correlator fit are in grey.}
           \label{fig:prin_corrs_L20}
\end{figure}

As mentioned in Section~\ref{section:methodology}, the finite-volume energies are extracted by fitting principal correlators to either a single exponential
\begin{equation}
\lambda^{(\mathbf{n})}(t,t_0) =  e^{-E_{\mathbf{n}}(t - t_0)},
\end{equation}
or a double exponential of the form 
\begin{equation}
\lambda^{(\mathbf{n})}(t,t_0) = (1-A_{\mathbf{n}}) e^{-E_{\mathbf{n}}(t - t_0)} + A_{\mathbf{n}} e^{-E'_{\mathbf{n}}(t - t_0)},
\end{equation}
over a time range $[ t_{\text{min}}, t_{\text{max}}]$. 
The corresponding value for $E_{\mathbf{n}}$ is the $\mathbf{n}^{\text{th}}$ finite-volume energy.
In general, this method allowed for robust determination of the finite-volume spectrum.
Examples from the $L/a_s =20$ volume for the $[000]A^+_1$ irrep in the flavour $\mathbf{6}$ sector are shown in Fig.~\ref{fig:prin_corrs_L20}. 

However, for some principal correlators, fits with differing $t_{\text{min}}$ and $t_{\text{max}}$ gave similar fit qualities but different corresponding energies.
To account for this fit-range ambiguity, we consider a selection of different fit ranges and parameterisations, and weight the resulting energies via a version of Akaike information criterion (AIC)~\cite{Jay:2020jkz}, as described in Ref.~\cite{Radhakrishnan:2022ubg}.
Applying this technique in the context for principal correlator fitting, we consider single and double exponential fits over many time windows $t_{\text{min}}$ and $t_{\text{max}}$,
where each combination of parameterisation and time window are referred to as a model, and to each model assign an AIC, 
\begin{equation}
   \text{AIC}_i= \chi_i^2(\{\alpha^*\}) +2k_i +2N_{\text{cut},i},
\end{equation}
where $\chi_i^2(\{\alpha^*\})$ is the minimised chi-squared, $k_i$ is the number of parameters and $N_{\text{cut},i}$ is the number of data points excluded from model $i$,
and an  corresponding  (unnormalised) weight,
\begin{equation}
\begin{split}
  w_i =& \exp\Big\{-\frac{1}{2}\text{AIC}_i\Big\}.
  \end{split}
\end{equation}
Retaining the 25 models with the highest weights (or lowest AIC), the ``model averaged'' energy is given by a weighted average, 
\begin{equation}
  E^{\text{av}}_{\mathbf{n}} = \sum_{i=1}^{25}\tilde{w}_i E_{\mathbf{n},i}
\end{equation}
where $E_{\mathbf{n},i}$ is the extracted energy for model $i$, and the $\tilde{w}_i  = w_i/\sum_{i=1}^{25}w_i$ factor is interpreted as the probability that model $i$ is the best one out of the model set given the data. 
The corresponding uncertainty is given by 
\begin{equation}
  \sigma^2_{ E_{\mathbf{n}}} = \sum_{i=1}^{25} \tilde{w}_i  \sigma^2_{ E_{\mathbf{n},i}}  + \frac{1}{2}\sum_{i,j=1}^{25} \tilde{w}_i  \tilde{w}_j ( E_{\mathbf{n},i}- E_{\mathbf{n},j})^2
\end{equation}
 where $\sigma^2_{ E_{\mathbf{n},i}}$ is the statistical uncertainty on model $i$.

\begin{figure}
  \centering
	\includegraphics[width=1\linewidth]{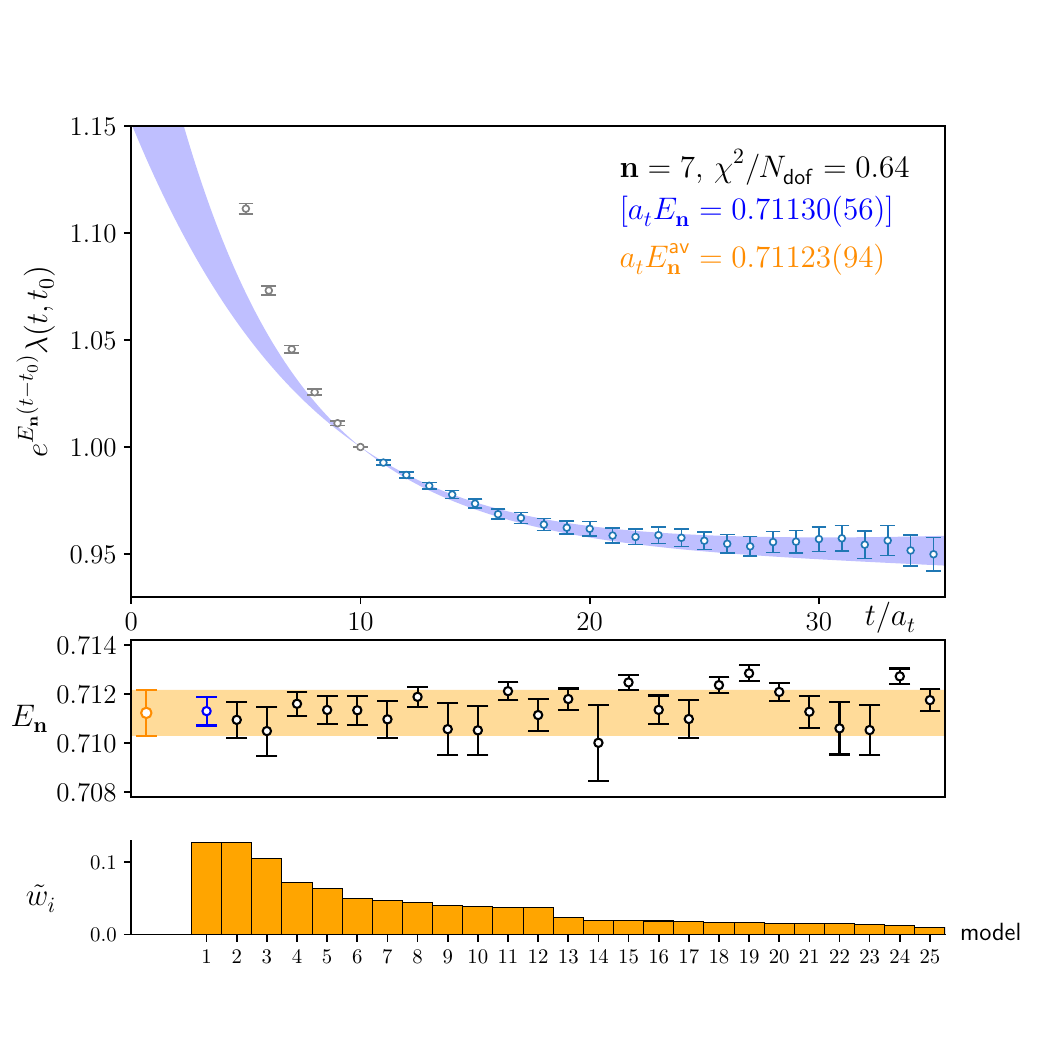}
	\caption{The lowest-AIC principal-correlator fit for $\mathbf{n} = 7$ on the $L/a_s = 18$ volume for $[000]A^+_1$ in the flavour $\mathbf{6}$ sector is displayed in the top panel, 
           along with its \chisq, its extracted energy (blue) and model-averaged value for the energy (orange).
           Model-averaged energy (orange point) is displayed in the middle panel, along with the extracted energies of the 25 lowest-AIC models (blue and black points in the middle panel), ordered by increasing AIC. Energy from the lowest-AIC principal-correlator fit is in blue. 
           Bottom panel displays the normalised AIC weights for each model. }
  \label{fig:prin_corrs_with_modeav}
\end{figure}

 We note that this method was not used extensively during this study. 
Model averaging was only used if there was a clear fit-range ambiguity and the model-averaged value and uncertainty differed from 
the best fit value.
An example of this scenario is shown in  Fig.~\ref{fig:prin_corrs_with_modeav}.
Overall, only 16 out of 352 energy levels utilised in the scattering analysis employed model averaging. 
None of the energy levels common to our elastic study \cite{Yeo:2024chk} and this study were model averaged.\footnote{These levels have slightly different numerical values in this work compared to Ref.~\cite{Yeo:2024chk} since here we use larger operator bases for the variational method and larger additional systematic uncertainty.}

\clearpage
\section{Interpolating operators} 
\label{appendix:interpolating_list}
In Tables~\ref{table:A1p_interpolating_list}-\ref{table:T1p_interpolating_list}, we list the meson-meson operators used in each irrep and volume.
For each table, all meson-meson operators up to roughly the three-hadron threshold are tabulated, ordered by increasing corresponding non-interacting energy,  and any operators not used in the basis are in grey.
We note that no energy levels near the non-interacting energies of absent operators  were used in the scattering analysis.
The meson-meson operator lists for the flavour $\mathbf{6}$  and $\overline{\mathbf{15}}$ sectors are the same but with different target $SU(3)_f$ irreps.
For the flavour $\mathbf{6}$ sector, the spectrum was computed on all five volumes, whereas for the $\overline{\mathbf{15}}$ sector the spectrum was only computed on the $L/a_s = \{16, 20, 24\}$ volumes.

\begin{table}
	\centering
        \resizebox{1\columnwidth}{!}{%
        \begin{tabular}{ccccc}
        \hline
        $L/a_s=14$ & $L/a_s=16$ & $L/a_s=18$ & $L/a_s=20$ & $L/a_s=24$\\
        \hline
        $D_{\bar{\mathbf{3}}[000]}\eta_{\mathbf{8}[000]}$            & $D_{\bar{\mathbf{3}}[000]}\eta_{\mathbf{8}[000]}$            & $D_{\bar{\mathbf{3}}[000]}\eta_{\mathbf{8}[000]}$            & $D_{\bar{\mathbf{3}}[000]}\eta_{\mathbf{8}[000]}$            & $D_{\bar{\mathbf{3}}[000]}\eta_{\mathbf{8}[000]}$            \\
        $D_{\bar{\mathbf{3}}[100]}\eta_{\mathbf{8}[100]}$            & $D_{\bar{\mathbf{3}}[100]}\eta_{\mathbf{8}[100]}$            & $D_{\bar{\mathbf{3}}[100]}\eta_{\mathbf{8}[100]}$            & $D_{\bar{\mathbf{3}}[100]}\eta_{\mathbf{8}[100]}$            & $D_{\bar{\mathbf{3}}[100]}\eta_{\mathbf{8}[100]}$            \\
        $D^*_{\bar{\mathbf{3}}[000]}\omega_{\mathbf{8}[000]}$        & $D^*_{\bar{\mathbf{3}}[000]}\omega_{\mathbf{8}[000]}$        & $D_{\bar{\mathbf{3}}[110]}\eta_{\mathbf{8}[110]}$            & $D_{\bar{\mathbf{3}}[110]}\eta_{\mathbf{8}[110]}$            & $D_{\bar{\mathbf{3}}[110]}\eta_{\mathbf{8}[110]}$            \\
        $D_{\bar{\mathbf{3}}[110]}\eta_{\mathbf{8}[110]}$            & $D_{\bar{\mathbf{3}}[110]}\eta_{\mathbf{8}[110]}$            & $D^*_{\bar{\mathbf{3}}[000]}\omega_{\mathbf{8}[000]}$        & $D^*_{\bar{\mathbf{3}}[000]}\omega_{\mathbf{8}[000]}$        & $D_{\bar{\mathbf{3}}[111]}\eta_{\mathbf{8}[111]}$            \\
        $D^*_{\bar{\mathbf{3}}[100]}\omega_{\mathbf{8}[100]}\{2\}$   & $D^*_{\bar{\mathbf{3}}[100]}\omega_{\mathbf{8}[100]}\{2\}$   & $D_{\bar{\mathbf{3}}[111]}\eta_{\mathbf{8}[111]}$            & $D_{\bar{\mathbf{3}}[111]}\eta_{\mathbf{8}[111]}$            & $D_{\bar{\mathbf{3}}[200]}\eta_{\mathbf{8}[200]}$            \\
        $D_{\bar{\mathbf{3}}[111]}\eta_{\mathbf{8}[111]}$            & $D_{\bar{\mathbf{3}}[111]}\eta_{\mathbf{8}[111]}$            & $D^*_{\bar{\mathbf{3}}[100]}\omega_{\mathbf{8}[100]}\{2\}$   & $D^*_{\bar{\mathbf{3}}[100]}\omega_{\mathbf{8}[100]}\{2\}$   & $D^*_{\bar{\mathbf{3}}[000]}\omega_{\mathbf{8}[000]}$        \\
        $D_{1,\mathbf{\bar{3}}[100]}\eta_{\mathbf{8}[100]}$          & $D_{1,\mathbf{\bar{3}}[100]}\eta_{\mathbf{8}[100]}$          & $D_{\bar{\mathbf{3}}[200]}\eta_{\mathbf{8}[200]}$            & $D_{\bar{\mathbf{3}}[200]}\eta_{\mathbf{8}[200]}$            & $D_{\bar{\mathbf{3}}[210]}\eta_{\mathbf{8}[210]}$            \\
        $D'_{1,\mathbf{\bar{3}}[100]}\eta_{\mathbf{8}[100]}$         & $D^*_{\bar{\mathbf{3}}[110]}\omega_{\mathbf{8}[110]}\{3\}$   & $D^*_{\bar{\mathbf{3}}[110]}\omega_{\mathbf{8}[110]}\{3\}$   & $D^*_{\bar{\mathbf{3}}[110]}\omega_{\mathbf{8}[110]}\{3\}$   & $D^*_{\bar{\mathbf{3}}[100]}\omega_{\mathbf{8}[100]}\{2\}$   \\
        $D^*_{\bar{\mathbf{3}}[110]}\omega_{\mathbf{8}[110]}\{3\}$   & $D'_{1,\mathbf{\bar{3}}[100]}\eta_{\mathbf{8}[100]}$         & $D_{1,\mathbf{\bar{3}}[100]}\eta_{\mathbf{8}[100]}$          & $D_{\bar{\mathbf{3}}[210]}\eta_{\mathbf{8}[210]}$            & $D_{\bar{\mathbf{3}}[211]}\eta_{\mathbf{8}[211]}$            \\
                                                                     & $D_{\bar{\mathbf{3}}[200]}\eta_{\mathbf{8}[200]}$            & $D'_{1,\mathbf{\bar{3}}[100]}\eta_{\mathbf{8}[100]}$         & $D_{1,\mathbf{\bar{3}}[100]}\eta_{\mathbf{8}[100]}$          & $D^*_{\bar{\mathbf{3}}[110]}\omega_{\mathbf{8}[110]}\{3\}$   \\
                                                                     &                                                              &                                                              & $D'_{1,\mathbf{\bar{3}}[100]}\eta_{\mathbf{8}[100]}$         & $D^*_{\bar{\mathbf{3}}[210]}\eta_{\mathbf{8}[210]}$          \\
                                                                     &                                                              &                                                              &                                                              & $\grey{D_{1,\mathbf{\bar{3}}[100]}\eta_{\mathbf{8}[100]}}$   \\
                                                                     &                                                              &                                                              &                                                              & $\grey{D'_{1,\mathbf{\bar{3}}[100]}\eta_{\mathbf{8}[100]}}$  \\
                                                                     &                                                              &                                                              &                                                              & $D^*_{\bar{\mathbf{3}}[111]}\omega_{\mathbf{8}[111]}\{2\}$   \\
                                                                     &                                                              &                                                              &                                                              & $D^*_{\bar{\mathbf{3}}[211]}\eta_{\mathbf{8}[211]}$          \\[3pt]
        \hline
        \end{tabular}
        }
		\caption{Meson-meson operators with corresponding non-interacting energies       
        up to roughly the three-hadron threshold on each volume for the $[000]A_1^+$ irrep, ordered by non-interacting energy.
             Operators not included in the operator basis are in grey. 
             Multiplicities are labelled in $\{\ldots \}$ whenever more than one.
                }
		\label{table:A1p_interpolating_list} 
\end{table}

\begin{table}
		\centering
		\resizebox{1\columnwidth}{!}{%
        \begin{tabular}{ccccc}
        \hline
        $L/a_s=14$ & $L/a_s=16$ & $L/a_s=18$ & $L/a_s=20$ & $L/a_s=24$\\
        \hline
        $D_{\bar{\mathbf{3}}[100]}\eta_{\mathbf{8}[100]}$            & $D_{\bar{\mathbf{3}}[100]}\eta_{\mathbf{8}[100]}$            & $D_{\bar{\mathbf{3}}[100]}\eta_{\mathbf{8}[100]}$            & $D_{\bar{\mathbf{3}}[100]}\eta_{\mathbf{8}[100]}$            & $D_{\bar{\mathbf{3}}[100]}\eta_{\mathbf{8}[100]}$            \\
        $D^*_{\bar{\mathbf{3}}[000]}\omega_{\mathbf{8}[000]}$        & $D^*_{\bar{\mathbf{3}}[000]}\omega_{\mathbf{8}[000]}$        & $D_{\bar{\mathbf{3}}[110]}\eta_{\mathbf{8}[110]}$            & $D_{\bar{\mathbf{3}}[110]}\eta_{\mathbf{8}[110]}$            & $D_{\bar{\mathbf{3}}[110]}\eta_{\mathbf{8}[110]}$            \\
        $D_{\bar{\mathbf{3}}[110]}\eta_{\mathbf{8}[110]}$            & $D_{\bar{\mathbf{3}}[110]}\eta_{\mathbf{8}[110]}$            & $D^*_{\bar{\mathbf{3}}[000]}\omega_{\mathbf{8}[000]}$        & $D^*_{\bar{\mathbf{3}}[110]}\eta_{\mathbf{8}[110]}$          & $D^*_{\bar{\mathbf{3}}[110]}\eta_{\mathbf{8}[110]}$          \\
        $D^*_{\bar{\mathbf{3}}[100]}\omega_{\mathbf{8}[100]}\{3\}$   & $D^*_{\bar{\mathbf{3}}[110]}\eta_{\mathbf{8}[110]}$          & $D^*_{\bar{\mathbf{3}}[110]}\eta_{\mathbf{8}[110]}$          & $D^*_{\bar{\mathbf{3}}[000]}\omega_{\mathbf{8}[000]}$        & $D_{\bar{\mathbf{3}}[200]}\eta_{\mathbf{8}[200]}$            \\
        $D^*_{\bar{\mathbf{3}}[110]}\eta_{\mathbf{8}[110]}$          & $D^*_{\bar{\mathbf{3}}[100]}\omega_{\mathbf{8}[100]}\{3\}$   & $D^*_{\bar{\mathbf{3}}[100]}\omega_{\mathbf{8}[100]}\{3\}$   & $D^*_{\bar{\mathbf{3}}[111]}\eta_{\mathbf{8}[111]}$          & $D^*_{\bar{\mathbf{3}}[000]}\omega_{\mathbf{8}[000]}$        \\
        $D_{\bar{\mathbf{3}}[110]}\omega_{\mathbf{8}[110]}$          & $D_{\bar{\mathbf{3}}[110]}\omega_{\mathbf{8}[110]}$          & $D_{\bar{\mathbf{3}}[110]}\omega_{\mathbf{8}[110]}$          & $D^*_{\bar{\mathbf{3}}[100]}\omega_{\mathbf{8}[100]}\{3\}$   & $D^*_{\bar{\mathbf{3}}[111]}\eta_{\mathbf{8}[111]}$          \\
                                                                     & $D^*_{\bar{\mathbf{3}}[111]}\eta_{\mathbf{8}[111]}$          & $D^*_{\bar{\mathbf{3}}[111]}\eta_{\mathbf{8}[111]}$          & $D_{\bar{\mathbf{3}}[110]}\omega_{\mathbf{8}[110]}$          & $D_{\bar{\mathbf{3}}[110]}\omega_{\mathbf{8}[110]}$          \\
                                                                     & $D_{1,\mathbf{\bar{3}}[100]}\eta_{\mathbf{8}[100]}$          & $D_{\bar{\mathbf{3}}[200]}\eta_{\mathbf{8}[200]}$            & $D_{\bar{\mathbf{3}}[200]}\eta_{\mathbf{8}[200]}$            & $D_{\bar{\mathbf{3}}[210]}\eta_{\mathbf{8}[210]}\{2\}$       \\
                                                                     & $D^*_{\bar{\mathbf{3}}[110]}\omega_{\mathbf{8}[110]}\{5\}$   & $D^*_{\bar{\mathbf{3}}[110]}\omega_{\mathbf{8}[110]}\{5\}$   & $D^*_{\bar{\mathbf{3}}[110]}\omega_{\mathbf{8}[110]}\{5\}$   & $D^*_{\bar{\mathbf{3}}[100]}\omega_{\mathbf{8}[100]}\{3\}$   \\
                                                                     & $D'_{1,\mathbf{\bar{3}}[100]}\eta_{\mathbf{8}[100]}$         & $D_{1,\mathbf{\bar{3}}[100]}\eta_{\mathbf{8}[100]}$          & $D_{\bar{\mathbf{3}}[111]}\omega_{\mathbf{8}[111]}$          & $D_{\bar{\mathbf{3}}[111]}\omega_{\mathbf{8}[111]}$          \\
                                                                     &                                                              & $D'_{1,\mathbf{\bar{3}}[100]}\eta_{\mathbf{8}[100]}$         & $D_{\bar{\mathbf{3}}[210]}\eta_{\mathbf{8}[210]}\{2\}$       & $D_{\bar{\mathbf{3}}[211]}\eta_{\mathbf{8}[211]}$            \\
                                                                     &                                                              & $D_{\bar{\mathbf{3}}[111]}\omega_{\mathbf{8}[111]}$          & $D_{1,\mathbf{\bar{3}}[100]}\eta_{\mathbf{8}[100]}$          & $D^*_{\bar{\mathbf{3}}[110]}\omega_{\mathbf{8}[110]}\{5\}$   \\
                                                                     &                                                              &                                                              & $D'_{1,\mathbf{\bar{3}}[100]}\eta_{\mathbf{8}[100]}$         & $D^*_{\bar{\mathbf{3}}[210]}\eta_{\mathbf{8}[210]}\{2\}$     \\
                                                                     &                                                              &                                                              & $D^*_{2,\mathbf{\bar{3}}[100]}\eta_{\mathbf{8}[100]}$          & $\grey{D_{1,\mathbf{\bar{3}}[100]}\eta_{\mathbf{8}[100]}}$          \\
                                                                     &                                                              &                                                              &                                                              & $\grey{D'_{1,\mathbf{\bar{3}}[100]}\eta_{\mathbf{8}[100]}}$         \\
                                                                     &                                                              &                                                              &                                                              & $D^*_{\bar{\mathbf{3}}[111]}\omega_{\mathbf{8}[111]}\{3\}$   \\
                                                                     &                                                              &                                                              &                                                              & $D^*_{\bar{\mathbf{3}}[211]}\eta_{\mathbf{8}[211]}\{1, \grey{2}\}$     \\[3pt]
        \hline
        \end{tabular}
        }  
		\caption{As Table~\ref{table:A1p_interpolating_list} but for the  $[000]E^+$ irrep.
             Degeneracies split in the form $\{m, \grey{n}\}$ indicate number of the operators that were included $(m)$ and not included $(n)$ in the basis of operators.}
		\label{table:Ep_interpolating_list} 
\end{table}

\begin{table}
		\centering
		\resizebox{1\columnwidth}{!}{%
        \begin{tabular}{ccccc}
        \hline
        $L/a_s=14$ & $L/a_s=16$ & $L/a_s=18$ & $L/a_s=20$ & $L/a_s=24$\\
        \hline
        $D^*_{\bar{\mathbf{3}}[100]}\omega_{\mathbf{8}[100]}$        & $D^*_{\bar{\mathbf{3}}[110]}\eta_{\mathbf{8}[110]}$          & $D^*_{\bar{\mathbf{3}}[110]}\eta_{\mathbf{8}[110]}$          & $D^*_{\bar{\mathbf{3}}[110]}\eta_{\mathbf{8}[110]}$          & $D^*_{\bar{\mathbf{3}}[110]}\eta_{\mathbf{8}[110]}$          \\
        $D^*_{\bar{\mathbf{3}}[110]}\eta_{\mathbf{8}[110]}$          & $D^*_{\bar{\mathbf{3}}[100]}\omega_{\mathbf{8}[100]}$        & $D^*_{\bar{\mathbf{3}}[100]}\omega_{\mathbf{8}[100]}$        & $D^*_{\bar{\mathbf{3}}[111]}\eta_{\mathbf{8}[111]}$          & $D^*_{\bar{\mathbf{3}}[111]}\eta_{\mathbf{8}[111]}$          \\
        $D_{\bar{\mathbf{3}}[110]}\omega_{\mathbf{8}[110]}$          & $D_{\bar{\mathbf{3}}[110]}\omega_{\mathbf{8}[110]}$          & $D_{\bar{\mathbf{3}}[110]}\omega_{\mathbf{8}[110]}$          & $D^*_{\bar{\mathbf{3}}[100]}\omega_{\mathbf{8}[100]}$        & $D_{\bar{\mathbf{3}}[110]}\omega_{\mathbf{8}[110]}$          \\
                                                                     & $D^*_{\bar{\mathbf{3}}[111]}\eta_{\mathbf{8}[111]}$          & $D^*_{\bar{\mathbf{3}}[111]}\eta_{\mathbf{8}[111]}$          & $D_{\bar{\mathbf{3}}[110]}\omega_{\mathbf{8}[110]}$          & $D_{\bar{\mathbf{3}}[210]}\eta_{\mathbf{8}[210]}$            \\
                                                                     & $D^*_{\bar{\mathbf{3}}[110]}\omega_{\mathbf{8}[110]}\{2\}$   & $D^*_{\bar{\mathbf{3}}[110]}\omega_{\mathbf{8}[110]}\{2\}$   & $D^*_{\bar{\mathbf{3}}[110]}\omega_{\mathbf{8}[110]}\{2\}$   & $D^*_{\bar{\mathbf{3}}[100]}\omega_{\mathbf{8}[100]}$        \\
                                                                     &                                                              & $D_{\bar{\mathbf{3}}[111]}\omega_{\mathbf{8}[111]}$          & $D_{\bar{\mathbf{3}}[111]}\omega_{\mathbf{8}[111]}$          & $D_{\bar{\mathbf{3}}[111]}\omega_{\mathbf{8}[111]}$          \\
                                                                     &                                                              &                                                              & $D_{\bar{\mathbf{3}}[210]}\eta_{\mathbf{8}[210]}$            & $D^*_{\bar{\mathbf{3}}[110]}\omega_{\mathbf{8}[110]}\{2\}$   \\
                                                                     &                                                              &                                                              & $D^*_{2,\mathbf{\bar{3}}[100]}\eta_{\mathbf{8}[100]}$          & $D^*_{\bar{\mathbf{3}}[210]}\eta_{\mathbf{8}[210]}$          \\
                                                                     &                                                              &                                                              & $D^*_{\bar{\mathbf{3}}[111]}\omega_{\mathbf{8}[111]}$        & $D^*_{\bar{\mathbf{3}}[111]}\omega_{\mathbf{8}[111]}$        \\
                                                                     &                                                              &                                                              & $D^*_{\bar{\mathbf{3}}[210]}\eta_{\mathbf{8}[210]}$          & $D^*_{\bar{\mathbf{3}}[211]}\eta_{\mathbf{8}[211]}\{2\}$     \\
                                                                     &                                                              &                                                              &                                                              & $D^*_{2,\mathbf{\bar{3}}[100]}\eta_{\mathbf{8}[100]}$          \\
                                                                     &                                                              &                                                              &                                                              & $D_{\bar{\mathbf{3}}[210]}\omega_{\mathbf{8}[210]}$          \\
                                                                     &                                                              &                                                              &                                                              & $D^*_{\bar{\mathbf{3}}[200]}\omega_{\mathbf{8}[200]}$        \\[3pt]
        \hline
        \end{tabular}
		}
		\caption{As Table~\ref{table:A1p_interpolating_list} but for the  $[000]A_2^+$ irrep.
                }
		\label{table:A2p_interpolating_list} 
\end{table}

\begin{table}
		\centering
		\resizebox{1\columnwidth}{!}{%
        \begin{tabular}{ccccc}
        \hline
        $L/a_s=14$ & $L/a_s=16$ & $L/a_s=18$ & $L/a_s=20$ & $L/a_s=24$\\
        \hline
        $D^*_{\bar{\mathbf{3}}[000]}\eta_{\mathbf{8}[000]}$          & $D^*_{\bar{\mathbf{3}}[000]}\eta_{\mathbf{8}[000]}$          & $D^*_{\bar{\mathbf{3}}[000]}\eta_{\mathbf{8}[000]}$          & $D^*_{\bar{\mathbf{3}}[000]}\eta_{\mathbf{8}[000]}$          & $D^*_{\bar{\mathbf{3}}[000]}\eta_{\mathbf{8}[000]}$          \\
        $D_{\bar{\mathbf{3}}[000]}\omega_{\mathbf{8}[000]}$          & $D_{\bar{\mathbf{3}}[000]}\omega_{\mathbf{8}[000]}$          & $D^*_{\bar{\mathbf{3}}[100]}\eta_{\mathbf{8}[100]}\{2\}$     & $D^*_{\bar{\mathbf{3}}[100]}\eta_{\mathbf{8}[100]}\{2\}$     & $D^*_{\bar{\mathbf{3}}[100]}\eta_{\mathbf{8}[100]}\{2\}$     \\
        $D^*_{\bar{\mathbf{3}}[100]}\eta_{\mathbf{8}[100]}\{2\}$     & $D^*_{\bar{\mathbf{3}}[100]}\eta_{\mathbf{8}[100]}\{2\}$     & $D_{\bar{\mathbf{3}}[000]}\omega_{\mathbf{8}[000]}$          & $D_{\bar{\mathbf{3}}[000]}\omega_{\mathbf{8}[000]}$          & $D_{\bar{\mathbf{3}}[000]}\omega_{\mathbf{8}[000]}$          \\
        $D^*_{\bar{\mathbf{3}}[000]}\omega_{\mathbf{8}[000]}$        & $D^*_{\bar{\mathbf{3}}[000]}\omega_{\mathbf{8}[000]}$        & $D^*_{\bar{\mathbf{3}}[000]}\omega_{\mathbf{8}[000]}$        & $D^*_{\bar{\mathbf{3}}[110]}\eta_{\mathbf{8}[110]}\{3\}$     & $D^*_{\bar{\mathbf{3}}[110]}\eta_{\mathbf{8}[110]}\{3\}$     \\
        $D_{\bar{\mathbf{3}}[100]}\omega_{\mathbf{8}[100]}\{2\}$     & $D_{\bar{\mathbf{3}}[100]}\omega_{\mathbf{8}[100]}\{2\}$     & $D_{\bar{\mathbf{3}}[100]}\omega_{\mathbf{8}[100]}\{2\}$     & $D^*_{\bar{\mathbf{3}}[000]}\omega_{\mathbf{8}[000]}$        & $D_{\bar{\mathbf{3}}[100]}\omega_{\mathbf{8}[100]}\{2\}$     \\
        $D^*_{\bar{\mathbf{3}}[100]}\omega_{\mathbf{8}[100]}\{3\}$   & $D^*_{\bar{\mathbf{3}}[110]}\eta_{\mathbf{8}[110]}\{3\}$     & $D^*_{\bar{\mathbf{3}}[110]}\eta_{\mathbf{8}[110]}\{3\}$     & $D_{\bar{\mathbf{3}}[100]}\omega_{\mathbf{8}[100]}\{2\}$     & $D^*_{\bar{\mathbf{3}}[000]}\omega_{\mathbf{8}[000]}$        \\
        $D^*_{\bar{\mathbf{3}}[110]}\eta_{\mathbf{8}[110]}\{3\}$     & $D^*_{\bar{\mathbf{3}}[100]}\omega_{\mathbf{8}[100]}\{3\}$   & $D^*_{\bar{\mathbf{3}}[100]}\omega_{\mathbf{8}[100]}\{3\}$   & $D^*_{\bar{\mathbf{3}}[111]}\eta_{\mathbf{8}[111]}\{2\}$     & $D^*_{\bar{\mathbf{3}}[111]}\eta_{\mathbf{8}[111]}\{2\}$     \\
        $D^*_{0,\mathbf{\bar{3}}[100]}\eta_{\mathbf{8}[100]}$        & $D_{\bar{\mathbf{3}}[110]}\omega_{\mathbf{8}[110]}\{3\}$     & $D_{\bar{\mathbf{3}}[110]}\omega_{\mathbf{8}[110]}\{3\}$     & $D^*_{\bar{\mathbf{3}}[100]}\omega_{\mathbf{8}[100]}\{3\}$   & $D_{\bar{\mathbf{3}}[110]}\omega_{\mathbf{8}[110]}\{3\}$     \\
        $D_{\bar{\mathbf{3}}[110]}\omega_{\mathbf{8}[110]}\{3\}$     & $D^*_{0,\mathbf{\bar{3}}[100]}\eta_{\mathbf{8}[100]}$        & $D^*_{\bar{\mathbf{3}}[111]}\eta_{\mathbf{8}[111]}\{2\}$     & $D_{\bar{\mathbf{3}}[110]}\omega_{\mathbf{8}[110]}\{3\}$     & $D_{\bar{\mathbf{3}}[210]}\eta_{\mathbf{8}[210]}$            \\
                                                                     & $D^*_{\bar{\mathbf{3}}[111]}\eta_{\mathbf{8}[111]}\{2\}$     & $D^*_{0,\mathbf{\bar{3}}[100]}\eta_{\mathbf{8}[100]}$        & $D^*_{0,\mathbf{\bar{3}}[100]}\eta_{\mathbf{8}[100]}$        & $D^*_{\bar{\mathbf{3}}[100]}\omega_{\mathbf{8}[100]}\{3\}$   \\
                                                                     & $D_{1,\mathbf{\bar{3}}[100]}\eta_{\mathbf{8}[100]}$          & $D^*_{\bar{\mathbf{3}}[110]}\omega_{\mathbf{8}[110]}\{6\}$   & $D^*_{\bar{\mathbf{3}}[110]}\omega_{\mathbf{8}[110]}\{6\}$   & $D^*_{\bar{\mathbf{3}}[200]}\eta_{\mathbf{8}[200]}\{2\}$     \\
                                                                     & $D^*_{\bar{\mathbf{3}}[110]}\omega_{\mathbf{8}[110]}\{6\}$   & $D_{1,\mathbf{\bar{3}}[100]}\eta_{\mathbf{8}[100]}$          & $D^*_{\bar{\mathbf{3}}[200]}\eta_{\mathbf{8}[200]}\{2\}$     & $D_{\bar{\mathbf{3}}[111]}\omega_{\mathbf{8}[111]}\{2\}$     \\
                                                                     & $D'_{1,\mathbf{\bar{3}}[100]}\eta_{\mathbf{8}[100]}$         & $D'_{1,\mathbf{\bar{3}}[100]}\eta_{\mathbf{8}[100]}$         & $D_{\bar{\mathbf{3}}[111]}\omega_{\mathbf{8}[111]}\{2\}$     & $D_{\bar{\mathbf{3}}[211]}\eta_{\mathbf{8}[211]}$            \\
                                                                     &                                                              & $D_{\bar{\mathbf{3}}[111]}\omega_{\mathbf{8}[111]}\{2\}$     & $D_{\bar{\mathbf{3}}[210]}\eta_{\mathbf{8}[210]}$            & $D^*_{0,\mathbf{\bar{3}}[100]}\eta_{\mathbf{8}[100]}$        \\
                                                                     &                                                              &                                                              & $D_{1,\mathbf{\bar{3}}[100]}\eta_{\mathbf{8}[100]}$          & $D^*_{\bar{\mathbf{3}}[110]}\omega_{\mathbf{8}[110]}\{6\}$   \\
                                                                     &                                                              &                                                              & $D'_{1,\mathbf{\bar{3}}[100]}\eta_{\mathbf{8}[100]}$         & $D^*_{\bar{\mathbf{3}}[210]}\eta_{\mathbf{8}[210]}\{5\}$     \\
                                                                     &                                                              &                                                              & $D^*_{2,\mathbf{\bar{3}}[100]}\eta_{\mathbf{8}[100]}\{2\}$     & $D_{\bar{\mathbf{3}}[200]}\omega_{\mathbf{8}[200]}\{2\}$     \\
                                                                     &                                                              &                                                              &                                                              & $D_{1,\mathbf{\bar{3}}[100]}\eta_{\mathbf{8}[100]}$          \\
                                                                     &                                                              &                                                              &                                                              & $D'_{1,\mathbf{\bar{3}}[100]}\eta_{\mathbf{8}[100]}$         \\
                                                                     &                                                              &                                                              &                                                              & $D^*_{\bar{\mathbf{3}}[111]}\omega_{\mathbf{8}[111]}\{4\}$   \\
                                                                     &                                                              &                                                              &                                                              & $D^*_{\bar{\mathbf{3}}[211]}\eta_{\mathbf{8}[211]}\{5\}$     \\[3pt]
        \hline
        \end{tabular}
		}
		\caption{As Table~\ref{table:A1p_interpolating_list} but for the  $[000]T_1^+$ irrep.}
		\label{table:T1p_interpolating_list} 
\end{table}

\clearpage
\section{Inclusion of \texorpdfstring{$G$-waves}{} } 
\label{appendix:wGwave}
As discussed in the main text, finite-volume energy levels appear in the energy region of interest (up to roughly the three-hadron threshold) that require the inclusion of $G$-wave or higher partial waves in the scattering analysis. 
More specifically, for the $[000]A^+_1$ irrep, there is a $\Dst \etaoct$ dominated level on the $L/a_s = 24$ volume at $a_tE_{\sf cm} \approx 0.70$ (faded purple in Fig.~\ref{fig:6f_spectra}) which requires a $\Dst \etaoct$ partial wave for it to be reproduced in the quantisation condition.
The lowest partial-wave contribution from this hadron-hadron channel is $\Dst \etaoct(\SLJ{3}{G}{4})$. 
Note that this level appears on top of the nearby $\Dst \etaoct$  non-interacting energy curve.
There is also a $G$-wave contribution from the lower channel, $\D \etaoct(\SLJ{1}{G}{4})$.

\begin{figure}
  \centering
	\includegraphics[width=1\linewidth]{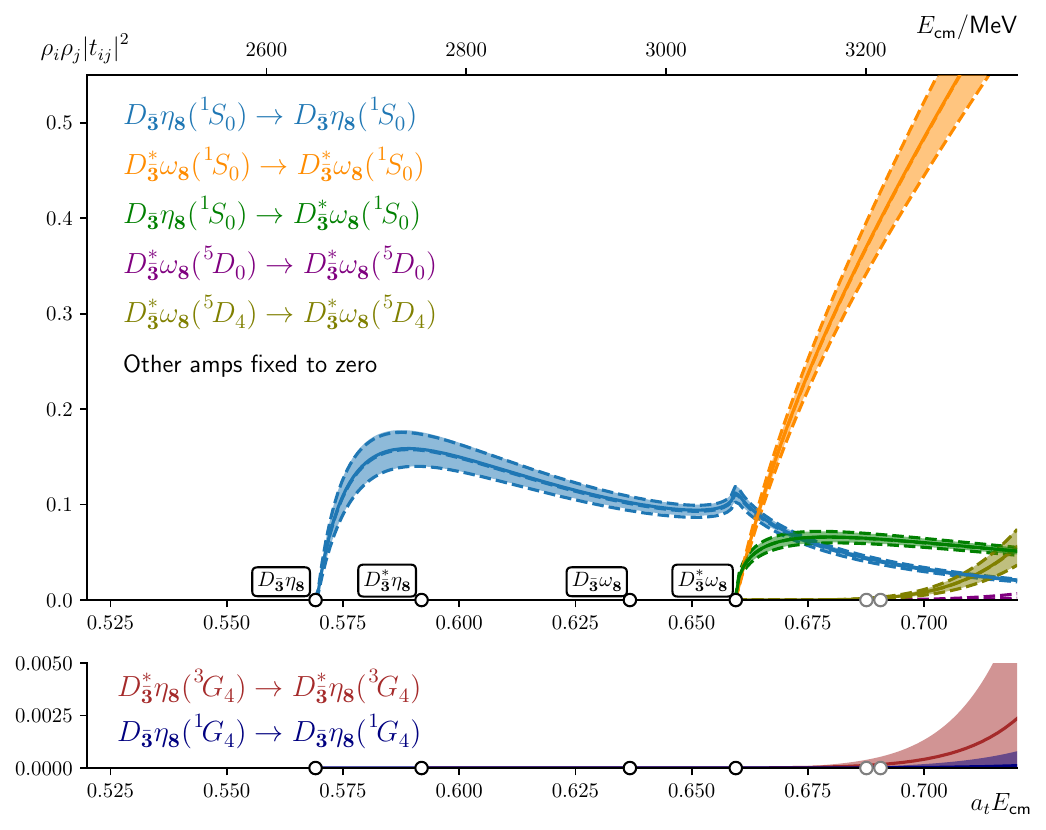}
	\caption{As for Fig.~\ref{fig:A1p6f_all_fits} but for the central values (middle dashed lines) and uncertainties (upper and lower dashed lines) of reference parameterisation fit to the $[000]A^+_1$ spectrum  and the central values (curves) and uncertainties (bands) of the analogous parameterisation with diagonal $\D \etaoct (\SLJ{1}{G}{4})$ and $\Dst \etaoct (\SLJ{3}{G}{4})$ amplitudes Eq.~\ref{eq:A1p6f_wGwave}.
            The additional diagonal $G$-wave amplitudes are displayed in the bottom panel. }
    \label{fig:A1p_wGwaves}
\end{figure}

To test if these $G$-wave contributions were important, 
we compared the reference parameterisation (Eq.~\ref{eq:A1p6f_ref})  to a parameterisation with the same form but with constant terms for the diagonal 
$\D \etaoct(\SLJ{1}{G}{4})$ and $\Dst \etaoct(\SLJ{3}{G}{4})$ amplitudes.
The latter parameterisation was fit to the same spectrum plus the faded purple level on the $L/a_s = 24$ volume at $a_tE_{\sf cm} \approx 0.70$.
The resultant fit parameters are
\begin{equation}
  \centering
  \resizebox{\textwidth}{!}{%
  \begin{tabular}{r@{ }ll}    
$c^{(0)} \{ \D\etaoct(\SLJ{1}{S}{0}) | \D\etaoct(\SLJ{1}{S}{0}) \}  \, = $ & $(2.00 \pm 0.10) $ & \multirow{10}{*}{%
\resizebox{0.66\textwidth}{!}{%
    $\begin{bmatrix*}[r]   1.00 &   0.17 &   0.44 &   0.68 &   0.49 &   0.52 &   0.14 &  -0.07 &  -0.05 &  -0.26\\
&  1.00 &   0.42 &   0.70 &   0.49 &   0.50 &   0.15 &  -0.09 &  -0.11 &  -0.23\\
&&  1.00 &   0.85 &   0.58 &   0.58 &   0.17 &  -0.12 &  -0.16 &  -0.26\\
&&&  1.00 &   0.72 &   0.70 &   0.21 &  -0.11 &  -0.14 &  -0.34\\
&&&&  1.00 &   0.05 &   0.18 &  -0.08 &  -0.10 &  -0.29\\
&&&&&  1.00 &   0.18 &  -0.10 &  -0.09 &  -0.29\\
&&&&&&  1.00 &  -0.10 &  -0.08 &   0.05\\
&&&&&&&  1.00 &   0.08 &   0.11\\
&&&&&&&&  1.00 &   0.04\\
&&&&&&&&&  1.00\end{bmatrix*}$ }} \\ 
$c^{(1)} \{ \D\etaoct(\SLJ{1}{S}{0}) | \D\etaoct(\SLJ{1}{S}{0})  \} = $& $(-16.5 \pm 0.3) \cdot a_t^2$ & \\
$c^{(2)} \{ \D\etaoct(\SLJ{1}{S}{0}) | \D\etaoct(\SLJ{1}{S}{0})  \} =$ & $(43.1 \pm 1.1) \cdot a_t^4$ & \\
$c^{(0)} \{ \D\etaoct(\SLJ{1}{S}{0}) | \Dst\omoct(\SLJ{1}{S}{0})  \} =$ & $(1.21 \pm 0.18)$ & \\
$c^{(0)} \{ \Dst\omoct(\SLJ{1}{S}{0}) |  \Dst\omoct(\SLJ{1}{S}{0})  \} =$ & $(2.79 \pm 0.06)$ & \\
$c^{(1)} \{ \Dst\omoct(\SLJ{1}{S}{0}) |  \Dst\omoct(\SLJ{1}{S}{0})  \} =$  & $(-4.57 \pm 0.12) \cdot a_t^2$ & \\
$c^{(0)} \{ \Dst\omoct(\SLJ{5}{D}{0}) |  \Dst\omoct(\SLJ{5}{D}{0})  \} =$ & $(0.05 \pm 0.05) \cdot a_t^{-4}$ & \\
$\gamma^{(0)} \{ \D\etaoct(\SLJ{1}{G}{4}) |  \D\etaoct(\SLJ{1}{G}{4})  \} =$  & $(20 \pm 80)  \cdot a_t^8$ & \\
$\gamma^{(0)} \{ \Dst\etaoct(\SLJ{3}{G}{4}) |  \Dst\etaoct(\SLJ{3}{G}{4})  \} =$ & $(-200 \pm 300)  \cdot a_t^8$ & \\
$\gamma^{(0)} \{ \Dst\omoct(\SLJ{5}{D}{4}) |  \Dst\omoct(\SLJ{5}{D}{4})  \} =$ & $(120 \pm 20)  \cdot a_t^{4}$ & \\[1.3ex]
&\multicolumn{2}{l}{ $\chisq= \frac{50.72}{52-10} = 1.21$\,,}
\end{tabular}
  }
\label{eq:A1p6f_wGwave}
\end{equation}
Comparing these fit parameters to Eq.~\ref{eq:A1p6f_ref}, we see that the values of the parameters for the non $G$-wave amplitudes are the same within one or two sigma or deviate by only a small amount in absolute terms.
This parameterisation is plotted in Fig.~\ref{fig:A1p_wGwaves} with its central values (curves) and statistical uncertainty (bands), alongside the central values (middle dashed lines) and statistical uncertainty (upper and lower dashed lines) of the reference parameterisation Eq.~\ref{eq:A1p6f_ref}.
It can be seen that the central values and statistical uncertainties match between the reference parameterisation and Eq.~\ref{eq:A1p6f_wGwave}.
The resultant diagonal $\D \etaoct(\SLJ{1}{G}{4})$ and $\Dst \etaoct(\SLJ{3}{G}{4})$ amplitudes are found to be very small over the energy region of interest. 
The $\chi^2$ in both cases are found to be very similar. 

We therefore conclude that the inclusion of the diagonal $G$-wave amplitudes had negligible impact on the determination of the other amplitudes.
Furthermore, the  $\Dst \etaoct(\SLJ{3}{G}{4})$ amplitude
was only needed to describe the $\Dst \etaoct$ dominated level on the $L/a_s = 24$ volume at $a_tE_{\sf cm} \approx 0.70$.
Throughout the rest of this study, we ignore $G$-wave contributions and ignore energy levels that require these waves if they are to be reproduced in the quantisation condition.
This leaves the $J^P = \{0, 1,2,3,4\}^+$ sectors to be constrained from the $[000]A^+_1, [000]A^+_2, [000]E^+$ and $[000]T^+_1$ finite-volume irreps.

\clearpage
\section{Indistinguishable vector-vector \texorpdfstring{$D$}{}-waves amplitudes} 
\label{appendix:indist_Dwave}
In this appendix we discuss how the $\Dst \omoct (\SLJ{3}{D}{1})$ and $\Dst \omoct (\SLJ{5}{D}{1})$ amplitudes cannot be uniquely determined from the quantisation condition when constraining them only with data from the $[000]T_1^+$ irrep.
In an analogous situation to what is seen in Ref.~\cite{Woss:2020ayi}, this stems from the fact that the $\SLJ{3}{D}{1}$ and $\SLJ{5}{D}{1}$ components (for the same vector-vector channel) of the matrix of kinematic functions, $\overline{\mathcal{M}}(s, L)$, are identical in the rest frame.

To see this explicitly, we recall that the matrix of kinematic functions is diagonal in hadron-hadron channel space and its form for a single channel is
\begin{equation}
\begin{split}
\overline{\mathcal{M}}_{\ell S Jm,  \ell' \! S'\! J' \! m'} = \delta_{SS'}\sum_{\ell''\!, m''}\frac{(4\pi)^{3/2}}{(k)^{\ell'' +1}} \frac{\sqrt{4\pi}}{\gamma L^3} c^{\vec{d}}_{\ell''\! m''}(k^{2}; L) \times \\
\sum_{m_\ell, m_{\ell'}, m_S}\langle \ell m_{\ell}, Sm_S| Jm \rangle \langle \ell' m_{\ell'}, Sm_S | J'm' \rangle \int d\Omega\ Y^*_{\ell m_{\ell}}Y^*_{\ell''\! m''}Y_{\ell' \! m_{\ell'}},
\end{split}
\end{equation}
where $k$ is the magnitude of relative momentum in the centre of mass frame for the channel in question, and 
\begin{equation}
    c^{\vec{d}}_{\ell m}(k^{2}; L) = \frac{\sqrt{4\pi}}{\gamma L^3} \Big{(}\frac{2\pi}{L} \Big{)}^{\ell -2}\mathcal{Z}^{\vec{d}}_{\ell m}(1;q^{2}),
\end{equation}
for $q = \frac{L}{2 \pi}k$, and $\mathcal{Z}^{\vec{d}}_{\ell m}$ the Lüscher zeta function whose definition can be found in Ref.~\cite{Woss:2020cmp}.
From the symmetries of the Lüscher zeta function in the rest frame, one can show  $\mathcal{Z}^{\vec{0}}_{\ell m} (\propto  c^{\vec{0}}_{\ell m})$ is only nonzero when  $\ell \in 2\mathds{Z}$ and $m \in 4\mathds{Z}$~\cite{Luscher:1990ux}. 
Furthermore, one can show $\mathcal{Z}^{\vec{0}}_{20}= 0$~\cite{Luscher:1990ux}.
With these properties, it follows that  
\begin{equation}
    \overline{\mathcal{M}}(\SLJ{2S+1}{D}{1}, m | \SLJ{2S'+1}{D}{1}, m') = \frac{4\pi c^{\vec{0}}_{00}}{k} \delta_{SS'}\delta_{mm'}.
\end{equation}
As a consequence, for any given reasonable parameterisation,  switching the $\SLJ{3}{D}{1}$ and $\SLJ{5}{D}{1}$ partial waves also yields a reasonable parameterisation (with the same \chisq).
Therefore, when presenting the results we refer to these partial waves collectively as $\Dst \omoct (\SLJ{X}{D}{1})$.
In principle these two waves can be distinguished using data from finite-volume irreps at nonzero total momentum, however, we do not consider such irreps here. 

The scenario of multiple  $\Dst \omoct$ $D$-wave partial waves contributing to the same $J^P$ also occurs in $J^P =2^+$ and $J^P =3^+$ (see Table~\ref{table:pw_sub}).
We note, however, that the presence of the $\smash{\mathcal{Z}^{\vec{0}}_{4m}}$ terms in these cases means these waves can be distinguished when considering only at-rest finite-volume irreps.
For example, in $J^P = 2^+$, the $m =m'= 2$ term for each  $S$ yields 
\begin{equation}
    \begin{split}
            \overline{\mathcal{M}}( \SLJ{1}{D}{2}, 2 | \SLJ{1}{D}{2}, 2) = \frac{4\pi c^{\vec{0}}_{00}}{k} + \frac{4\pi c^{\vec{0}}_{40}}{7k^{5}},\\
                \overline{\mathcal{M}}(\SLJ{3}{D}{2}, 2 | \SLJ{3}{D}{2}, 2) = \frac{4\pi c^{\vec{0}}_{00}}{k} - \frac{8\pi c^{\vec{0}}_{40}}{21k^{5}}, \\
    \overline{\mathcal{M}}(\SLJ{5}{D}{2}, 2| \SLJ{5}{D}{2}, 2) = \frac{4\pi c^{\vec{0}}_{00}}{k} + \frac{8\pi c^{\vec{0}}_{40}}{49k^{5}},
    \end{split}
\end{equation} 
and in $J^P = 3^+$, the $m =m'= 3$ term for each  $S$ yields 
\begin{equation}
    \begin{split}
                \overline{\mathcal{M}}(\SLJ{3}{D}{3}, 3 | \SLJ{3}{D}{3}, 3) = \frac{4\pi c^{\vec{0}}_{00}}{k} + \frac{4\pi c^{\vec{0}}_{40}}{7k^{5}}, \\
    \overline{\mathcal{M}}(\SLJ{5}{D}{3}, 3| \SLJ{5}{D}{3}, 3) = \frac{4\pi c^{\vec{0}}_{00}}{k} - \frac{6\pi c^{\vec{0}}_{40}}{7k^{5}},
    \end{split}
\end{equation} 
where in both cases the degeneracy is split by the $c^{\vec{0}}_{40}$ term.

\clearpage
\section{\texorpdfstring{$J^P = \{1,2,3\}^+$}{} parameterisations in the flavour \texorpdfstring{$\overline{\mathbf{15}}$ sector}{}}
\label{appendix:flavour_15bar_parameterisations}
In this appendix, we discuss in more detail the parameterisations considered for the $J^P = \{1,2,3\}^+$ amplitudes in the flavour $\overline{\mathbf{15}}$ sector presented in Section~\ref{section:other_15bar}.

We start with the $J^P =2^+$ amplitudes, constrained from the  $[000]E^+$ spectrum.
This finite-volume irrep also gets contributions from the $J^P =4^+$ amplitude $\Dst \omoct (\SLJ{5}{D}{4})$, which we described by a constant term with its value fixed from the reference parameterisation in the $[000]A^+_1$ analysis, i.e. from Eq.~\ref{eq:A1p15bar_ref}.
The $J^P =2^+$ amplitudes were parameterised using $K$-matrix polynomial parameterisations with Chew-Mandelstam phase space.
We take a diagonal constant $K$-matrix as the reference parameterisation, whose fit parameters are  
\begin{equation}
    \centering
    \resizebox{\textwidth}{!}{%
    \begin{tabular}{r@{ }ll}    
     $\gamma^{(0)} \{\D \etaoct (\SLJ{1}{D}{2}) | \D \etaoct (\SLJ{1}{D}{2})\} =$ & $(9 \pm 3 \pm 17)\cdot a_t^{4} $ &
     \resizebox{0.47\textwidth}{!}{
     \multirow{7}{*}{  $\left[ \begin{array}{rrrrrrr}    
      1.00 &  0.15 &  0.20 &  0.17 &  0.21 &  0.24 &  0.23 \\
           &  1.00 &  0.25 &  0.15 &  0.08 &  0.10 &  0.09 \\
           &       &  1.00 &  0.25 &  0.23 &  0.28 &  0.19 \\
           &       &       &  1.00 &  0.25 &  0.38 &  0.38 \\
           &       &       &       &  1.00 &  0.36 &  0.45 \\
           &       &       &       &       &  1.00 &  0.34 \\
           &       &       &       &       &       &  1.00 \\
     \end{array} \right]$ }
     } 
     \\ 
     $\gamma^{(0)} \{\D \omoct (\SLJ{3}{D}{2}) | \D \omoct (\SLJ{3}{D}{2})\} =$ & $(14 \pm 20 \pm 9)\cdot a_t^{4} $ & \\
     $\gamma^{(0)} \{\Dst \etaoct (\SLJ{3}{D}{2}) | \Dst \etaoct (\SLJ{3}{D}{2})\} =$ & $(-1 \pm 4 \pm 14)\cdot a_t^{4} $ & \\
     $\gamma^{(0)} \{\Dst \omoct (\SLJ{1}{D}{2}) | \Dst \omoct (\SLJ{1}{D}{2})\} =$ & $(20 \pm 19 \pm 18)\cdot a_t^{4} $ & \\
     $\gamma^{(0)} \{\Dst \omoct (\SLJ{3}{D}{2}) | \Dst \omoct (\SLJ{3}{D}{2})\} =$ & $(80 \pm 30 \pm 70)\cdot a_t^{4} $ & \\
     $\gamma^{(0)} \{\Dst \omoct (\SLJ{5}{D}{2}) | \Dst \omoct (\SLJ{5}{D}{2})\} =$ & $(110 \pm 30 \pm 30)\cdot a_t^{4} $ & \\
     $\gamma^{(0)} \{\Dst \omoct (\SLJ{5}{S}{2}) | \Dst \omoct (\SLJ{5}{S}{2})\} =$ & $(0.59 \pm 0.10 \pm 0.06) $ & \\[10pt]
     &\multicolumn{2}{l}{$\chisq = \frac{41.64}{41-7} = 1.22$\,.}
    \end{tabular}
    }
\label{eq:Ep_15bar_ref}
\end{equation}
This parameterisation is displayed in Fig.~\ref{fig:15bar_other_all_fits} (middle panel), alongside all other reasonable parameterisations to the data -- these are listed in Table~\ref{table:Ep15bar_parameterisation_table}.
Also shown are parameterisations with the same form  as Eq.~\ref{eq:Ep_15bar_ref} but the fixed $J^P =4^+$ amplitude parameter varied by $\pm \sigma$.
 It was found the resultant $J^P =2^+$ amplitudes were insensitive to the change of this parameter.
For all reasonable parameterisations the amplitudes were found to be small.
Only a constant term in each diagonal amplitude was needed to obtain reasonable descriptions to the data. 
Considering higher-order polynomials for the diagonal $\Dst\omoct(\SLJ{5}{S}{2})$ amplitude and nonzero constant terms for the $\Dst\omoct(\SLJ{5}{S}{2})$ to $D$-wave amplitudes resulted in small improvements in $\chisq$, but these were not needed for a reasonable description.
We considered parameterisations with nonzero constant terms for all off-diagonal $D$-wave to $D$-wave amplitudes, but these were found to be small and statistically consistent with zero.

Next we discuss the $J^P =3^+$ amplitudes, constrained from  $[000]A_2^+$. 
We use $K$-matrix parameterisations with constant and linear terms, with both phase-space prescriptions.
For our reference parameterisation, we take a diagonal constant $K$-matrix with Chew-Mandelstam phase space. 
Its fit parameters are 
\begin{equation}
    \centering
    \resizebox{0.9\textwidth}{!}{%
    \begin{tabular}{r@{ }ll}    
     $\gamma^{(0)} \{\D \omoct (\SLJ{3}{D}{3}) | \D \omoct (\SLJ{3}{D}{3})\} =$ & $(-10 \pm 20 \pm 30)\cdot a_t^{4} $ &
     \multirow{4}{*}{  $\left[ \begin{array}{rrrr}    
      1.00 &  0.41 &  0.31 &  0.43 \\
           &  1.00 &  0.51 &  0.69 \\
           &       &  1.00 &  0.59 \\
           &       &       &  1.00 \\
     \end{array} \right]$ }
     \\ 
     $\gamma^{(0)} \{\Dst \etaoct (\SLJ{3}{D}{3}) | \Dst \etaoct (\SLJ{3}{D}{3})\} =$ & $(-7 \pm 6 \pm 19)\cdot a_t^{4} $ & \\
     $\gamma^{(0)} \{\Dst \omoct (\SLJ{3}{D}{3}) | \Dst \omoct (\SLJ{3}{D}{3})\} =$ & $(60 \pm 20 \pm 30)\cdot a_t^{4} $ & \\
     $\gamma^{(0)} \{\Dst \omoct (\SLJ{5}{D}{3}) | \Dst \omoct (\SLJ{5}{D}{3})\} =$ & $(-20 \pm 30 \pm 50)\cdot a_t^{4} $ & \\[10pt]
     &\multicolumn{2}{l}{$\chisq = \frac{9.12}{16-4} = 0.76$\,.}
    \end{tabular}
    }
\label{eq:A2p15bar_ref}
\end{equation}
This parameterisation is displayed in Fig.~\ref{fig:15bar_other_all_fits} (bottom panel), along with its statistical uncertainty and uncertainty from mass and anisotropy variations.
The other reasonable parameterisations are also plotted in Fig.~\ref{fig:15bar_other_all_fits} (bottom panel) and a summary of these can be found in Table~\ref{table:A2p15bar_parameterisation_table}.
For all reasonable parameterisations, the amplitudes were found to be small.
Both diagonal parameterisations and parameterisations with nonzero off-diagonal terms were found to yield reasonable descriptions of the data. 

Finally, we discuss the $J^P =1^+$ amplitudes.
For parameterising the $J^P = 1^+$ amplitudes we use $K$-matrix polynomials with Chew-Mandelstam phase space, whilst 
for the $J^P = \{3, 4\}^+$ amplitudes we use constant $K$-matrices with Chew-Mandelstam phase space, with values fixed  to the reference parameterisations of Eq.~\ref{eq:A2p15bar_ref} and Eq.~\ref{eq:A1p15bar_ref}, respectively.
The fit parameters of the reference parameterisation are
\begin{equation}
    \centering
    \resizebox{\textwidth}{!}{%
    \begin{tabular}{r@{ }ll}  
     $\gamma^{(0)} \{\D \omoct (\SLJ{3}{D}{1}) | \D \omoct (\SLJ{3}{D}{1})\} =$ & $(30 \pm 20 \pm 70)\cdot a_t^{4} $ &
     \resizebox{.99\textwidth}{!}{
     \multirow{12}{*}{  $\left[ \begin{array}{rrrrrrrrrrrr}    
      1.00 & -0.01 &  0.05 & -0.13 & -0.03 &  0.21 & -0.02 &  0.06 &  0.13 &  0.16 &  0.30 &  0.18 \\
           &  1.00 & -0.98 & -0.04 & -0.02 &  0.02 &  0.00 &  0.01 &  0.01 &  0.02 & -0.01 &  0.02 \\
           &       &  1.00 & -0.09 &  0.02 &  0.05 &  0.03 & -0.03 &  0.04 &  0.00 &  0.02 & -0.01 \\
           &       &       &  1.00 &  0.18 & -0.46 & -0.17 &  0.02 & -0.36 & -0.10 & -0.01 & -0.05 \\
           &       &       &       &  1.00 &  0.01 &  0.15 & -0.17 &  0.15 & -0.07 &  0.00 & -0.08 \\
           &       &       &       &       &  1.00 & -0.04 &  0.16 &  0.13 &  0.19 &  0.11 &  0.19 \\
           &       &       &       &       &       &  1.00 & -0.98 &  0.22 &  0.03 &  0.16 & -0.04 \\
           &       &       &       &       &       &       &  1.00 & -0.14 &  0.00 & -0.14 &  0.06 \\
           &       &       &       &       &       &       &       &  1.00 & -0.05 &  0.14 & -0.01 \\
           &       &       &       &       &       &       &       &       &  1.00 &  0.37 &  0.54 \\
           &       &       &       &       &       &       &       &       &       &  1.00 &  0.57 \\
           &       &       &       &       &       &       &       &       &       &       &  1.00 \\
     \end{array} \right]$ }
     } 
     \\ 
     $\gamma^{(0)} \{\D \omoct (\SLJ{3}{S}{1}) | \D \omoct (\SLJ{3}{S}{1})\} =$ & $(-3.7 \pm 0.4 \pm 0.6) $ & \\
     $\gamma^{(1)} \{\D \omoct (\SLJ{3}{S}{1}) | \D \omoct (\SLJ{3}{S}{1})\} =$ & $(7.5 \pm 0.9 \pm 1.3)\cdot a_t^{2} $ & \\
     $\gamma^{(0)} \{\D \omoct (\SLJ{3}{S}{1}) | \Dst \etaoct (\SLJ{3}{S}{1})\} =$ & $(-0.42 \pm 0.14 \pm 0.23) $ & \\
     $\gamma^{(0)} \{\D \omoct (\SLJ{3}{S}{1}) | \Dst \omoct (\SLJ{3}{S}{1})\} =$ & $(-0.06 \pm 0.21 \pm 0.17) $ & \\
     $\gamma^{(0)} \{\Dst \etaoct (\SLJ{3}{D}{1}) | \Dst \etaoct (\SLJ{3}{D}{1})\} =$ & $(-18 \pm 7 \pm 16)\cdot a_t^{4} $ & \\
     $\gamma^{(0)} \{\Dst \etaoct (\SLJ{3}{S}{1}) | \Dst \etaoct (\SLJ{3}{S}{1})\} =$ & $(-1.8 \pm 0.3 \pm 0.6) $ & \\
     $\gamma^{(1)} \{\Dst \etaoct (\SLJ{3}{S}{1}) | \Dst \etaoct (\SLJ{3}{S}{1})\} =$ & $(3.2 \pm 0.7 \pm 1.5)\cdot a_t^{2} $ & \\
     $\gamma^{(0)} \{\Dst \etaoct (\SLJ{3}{S}{1}) | \Dst \omoct (\SLJ{3}{S}{1})\} =$ & $(0.36 \pm 0.08 \pm 0.62) $ & \\
     $\gamma^{(0)} \{\Dst \omoct (\SLJ{3}{D}{1}) | \Dst \omoct (\SLJ{3}{D}{1})\} =$ & $(-10 \pm 60 \pm 20)\cdot a_t^{4} $ & \\
     $\gamma^{(0)} \{\Dst \omoct (\SLJ{3}{S}{1}) | \Dst \omoct (\SLJ{3}{S}{1})\} =$ & $(0.68 \pm 0.13 \pm 0.09) $ & \\
     $\gamma^{(0)} \{\Dst \omoct (\SLJ{5}{D}{1}) | \Dst \omoct (\SLJ{5}{D}{1})\} =$ & $(40 \pm 50 \pm 90)\cdot a_t^{4} $ & \\[10pt]
     &\multicolumn{2}{l}{$\chisq = \frac{37.04}{51-12} = 0.95$\,.}
    \end{tabular}
    }
\label{eq:T1p15bar_ref}
\end{equation}
This parameterisation is plotted in Fig.~\ref{fig:15bar_other_all_fits} (top panel) with its statistical uncertainty and uncertainty from mass and anisotropy variations.
We also plot other reasonable parameterisations in Fig.~\ref{fig:15bar_other_all_fits}  (top panel),
including those with the same form as the reference parameterisation (Eq.~\ref{eq:T1p15bar_ref}) but with the fix parameters of the $J^P = \{3,4\}^+$ amplitudes varied by plus or minus one sigma from their central values,
and a summary of these can be found in Table~\ref{table:T1p15bar_parameterisation_table}.
It was found that the resulting $J^P =1^+$ amplitudes had a small dependence on the values of these fixed parameters.
Across the reasonable parameterisations all amplitudes were found to be small.

It was found that only constant terms for the diagonal amplitudes were needed to yield reasonable descriptions to the data, although considering 
constant plus first-order terms for the diagonal $\Dst\etaoct(\SLJ{3}{S}{1})$ and diagonal $\D\omoct(\SLJ{3}{S}{1})$ amplitudes significantly improved the $\chisq$.
We also considered parameterisations with constant terms for the off-diagonal $S$-wave to $S$-wave amplitudes and $S$-wave to $D$-wave amplitudes but these terms were found to be small.
Finally, we tested parameterisations with constant terms coupling  $D$-wave to $D$-wave amplitudes, however allowing for these terms did not lead to an improved description of the data. 

Overall, across the reasonable parameterisations in each of  $J^P = \{1,2,3\}^+$ in the flavour $\overline{\mathbf{15}}$ sector, all amplitudes were found to be small over the energy region of interest.

\clearpage
\section{Tables of parameterisations} 
\label{appendix:table_of_parameterisations}
In this appendix, we summarise parameterisations used for each irrep in each flavour sector.
Tables~\ref{table:A1p6f_parameterisation_table}-\ref{table:T1p6f_parameterisation_table} summarise the parameterisations for the flavour $\mathbf{6}$ sector, as presented in Section~\ref{section:6f_analysis},
whilst Tables~\ref{table:A1p15bar_parameterisation_table}-\ref{table:T1p15bar_parameterisation_table} summarise the parameterisations for the flavour $\overline{\mathbf{15}}$ sector, presented in Section~\ref{section:15bar_analysis}
and Appendix~\ref{appendix:flavour_15bar_parameterisations}.

As mentioned in the main text, some parameterisations utilised a pole-exclusion procedure when spurious bound states arose close to the constrained energy region for the fit parameters that minimised the $\chi^2$.
These are bound-state poles in the amplitude that arise despite no corresponding finite-volume energy level found in the lattice QCD computed spectrum.
As first introduced in our elastic study of $\D \etaoct$ $S$-wave scattering~\cite{Yeo:2024chk}, the procedure entails modifying the chi-squared,
$\chi^2 \rightarrow \chi^2 + \chi_{\text{add}}^2 $,
where this additional term penalises fits with spurious bound states close to the constrained energy region. 
We use 
\begin{equation}
\chi_{\text{add}}^2(E_p)= \alpha \Bigg{(} \frac{f(E_p)}{1+e^{-f(E_p)} }\Bigg{)}^2,
\end{equation}
where $f(E_p) = \frac{E^2_p-E^2_{l}}{E^2_{h}-E^2_{l}}$ for $E_p$ the mass of the spurious bound state for a given parameterisation, and we take  $(E_l,  E_h, \alpha) = (0.45, 0.57, 100)$.
This choice of constants is found to sufficiently push spurious bound states far enough below threshold such that they do not affect the behaviour of the physical scattering amplitude, whilst not over constraining the parameter space such that reasonable descriptions of the data cannot be achieved. 
We label parameterisations which were obtained with this procedure with ``$\dagger$'', and the chi-squared quoted is the total chi-squared, i.e. $\chi^2+  \chi^2_{\text{add}}$.

\begin{table}
\centering
    \resizebox{1\columnwidth}{!}{%
    \begin{tabular}{ccccc}
    \hline
    \hline
    \multicolumn{5}{c}{$K$-matrix polynomial}  \\
    \hline
    \hline
    \rule{0pt}{12pt}$\gamma^{(n)}\{\Dst\omoct({\SLJ{1}{S}{0}})| \Dst\omoct({\SLJ{1}{S}{0}})\}$ & $\gamma^{(n)}\{\D\etaoct({\SLJ{1}{S}{0}}) | \Dst\omoct({\SLJ{1}{S}{0}})\}$ & $\gamma^{(n)}\{\D\etaoct({\SLJ{1}{S}{0}}) | \Dst\omoct({\SLJ{1}{D}{0}})\}$ & $\gamma^{(n)}\{\Dst\omoct({\SLJ{1}{S}{0}}) | \Dst\omoct({\SLJ{1}{D}{0}})\}$ & $\chisq$  \\
    \hline
    \multicolumn{5}{c}{\rule{0pt}{12pt} Chew-Mandelstam phase space}  \\
    \hline
     \rule{0pt}{12pt}0   & 0    &  $-$  &  $-$  & $\frac{67.51}{51-7}=  1.53$            \\[3pt]
     0, 1                & 0    &  $-$  &  $-$  & $\frac{63.12}{51-8}=  1.47$            \\[3pt] 
     0, 1                & 0    &  0    &  $-$  & $\frac{63.12}{51-9}=  1.50$            \\[3pt]     
     0, 1                & 0    &  $-$  &  0    & $\frac{63.12}{51-9}=  1.50$            \\[3pt]     
     0, 2                & 0    &  $-$  &  $-$  & $\frac{62.74}{51-8}=   1.46$           \\[3pt] 
     0, 1, 2             & 0    &  $-$  &  $-$  & $\frac{61.90}{51-9}=   1.47$           \\[3pt]   
    \hline
    \multicolumn{5}{c}{\rule{0pt}{12pt} Simple phase space}  \\
    \hline
     \rule{0pt}{12pt}0   & 0    &  $-$  &  $-$  & $\frac{62.63}{51-7}=  1.42$          \\[3pt]  
     0, 1                & 0    &  $-$  &  $-$  & $\frac{61.64}{51-8}=  1.43$          \\[3pt]  
     0, 1                & 0    &  0    &  $-$  & $\frac{61.64}{51-9}=  1.47$          \\[3pt]  
     0, 1                & 0    &  $-$  &  0    & $\frac{61.64}{51-9}=  1.47$          \\[3pt]  
     0, 2                & 0    &  $-$  &  $-$  & $\frac{61.57}{51-8}=  1.43$          \\[3pt] 
     0, 1, 2             & 0    &  $-$  &  $-$  & $\frac{61.45}{51-9}=  1.46$          \\[3pt]
    \hline
    \hline
    \multicolumn{5}{c}{Inverse $K$-matrix polynomial}  \\
    \hline
    \hline
    \rule{0pt}{12pt}$c^{(n)}\{\Dst\omoct({\SLJ{1}{S}{0}})| \Dst\omoct({\SLJ{1}{S}{0}})\}$ & $c^{(n)}\{\D\etaoct({\SLJ{1}{S}{0}}) | \Dst\omoct({\SLJ{1}{S}{0}})\}$ & $c^{(n)}\{\D\etaoct({\SLJ{1}{S}{0}}) | \Dst\omoct({\SLJ{1}{D}{0}})\}$ & $c^{(n)}\{\Dst\omoct({\SLJ{1}{S}{0}}) | \Dst\omoct({\SLJ{1}{D}{0}})\}$ & \\
    \hline
    \multicolumn{5}{c}{\rule{0pt}{12pt} Chew-Mandelstam phase space}  \\
    \hline
     \rule{0pt}{12pt}0   & 0         &  $-$            & $-$            & $\frac{80.59}{51-7}=  1.83^\dagger $    \\[3pt]  
     $\bm{0, 1}$         & $\bm{0}$  & $\bm{-}$        & $\bm{-}$       & $\bm{\frac{50.80}{51-8}=  1.18}$        \\[3pt]  
     0, 1                & 0         &  0              & $-$            & $\frac{50.80}{51-9}=  1.21$             \\[3pt]  
     0, 1                & 0         &  $-$            & 0              & $\frac{50.80}{51-9}=  1.21$             \\[3pt] 
     0, 2                & 0         &  $-$            & $-$            & $\frac{51.46}{51-8}=  1.20$             \\[3pt]  
     0, 1, 2             & 0         &  $-$            & $-$            & $\frac{50.91}{51-9}=  1.21^\dagger$     \\[3pt]  
    \hline
    \multicolumn{5}{c}{\rule{0pt}{12pt} Simple phase space}  \\
    \hline
     \rule{0pt}{12pt}0   & 0    &  $-$  & $-$  & $\frac{65.72}{51-7}=  1.49^\dagger $    \\[3pt]  
     0, 1                & 0    &  $-$  & $-$  & $\frac{50.62}{51-8}=  1.18$             \\[3pt]  
     0, 1                & 0    &  $-$  & 0    & $\frac{50.61}{51-9}=  1.21$             \\[3pt]  
     0, 2                & 0    &  $-$  & $-$  & $\frac{51.10}{51-8}=  1.19$             \\[3pt] 
    \hline
    \end{tabular}
    }
    \caption{Reasonable parameterisations of the $J^P = 0^+$ amplitudes constrained by the $[000]A_1^+$ spectrum in the flavour $\mathbf{6}$ sector, as described in the main text. 
             All parameterisations have a second order polynomial for the diagonal $\D\etaoct(\SLJ{1}{S}{0})$ amplitude and a constant term with Chew-Mandelstam phase space for the diagonal $\Dst\omoct({\SLJ{5}{D}{4}})$ amplitude.
             For all other amplitudes, table entries indicate the order of each term in the polynomial not fixed to zero and ``$-$'' indicates all terms were fixed to zero. 
             Parameterisations utilising the pole-exclusion procedure are labelled with ``$\dagger$'' by the minimised \chisq.  
             The reference parameterisation is in bold.          
             }
    \label{table:A1p6f_parameterisation_table} 
\end{table}

\begin{table}
\centering
    \resizebox{1\columnwidth}{!}{%
    \begin{tabular}{cccc}
    \hline
    \hline
    \multicolumn{4}{c}{Inverse $K$-matrix ratio-of-polynomials}  \\
    \hline
    \hline
    \rule{0pt}{12pt}$d^{(n)}\{\Dst\omoct({\SLJ{5}{S}{2}})| \Dst\omoct({\SLJ{5}{S}{2}})\}$ & $c^{(n)}\{\Dst\omoct({\SLJ{5}{S}{2}})| \Dst\omoct({\SLJ{5}{S}{2}})\}$ &  $D$-wave to $\Dst\omoct({\SLJ{5}{S}{2}})$ &  \chisq \\
    \hline
    \multicolumn{4}{c}{\rule{0pt}{12pt} Chew-Mandelstam phase space}  \\
     \hline
     \rule{0pt}{12pt}1 & 0, 1         & $-$           & $\frac{46.35}{58-9}=  0.95^\dagger$                 \\[3pt]    
     $\bm{1}$          & $\bm{0, 1}$  & \checkmark    & $\bm{\frac{38.03}{58-15}=  0.95}$                   \\[3pt]    
     1                 & 0, 1         & \checkmark    & $\frac{36.74}{58-15}=  0.85$ (\textparagraph)       \\[3pt]    
     1                 & 0, 1         & \checkmark    & $\frac{40.24}{58-15}=  0.94$ (\S)                   \\[3pt]    
     1                 & 0, 2         & $-$           & $\frac{47.65}{58-9}=  0.97^\dagger$                 \\[3pt]   
     1                 & 0, 2         & \checkmark    & $\frac{38.02}{58-15}=  0.88$                        \\[3pt]   

     2                 & 0, 1 & $-$         & $\frac{46.22}{58-9}=  0.94$           \\[3pt]    
     2                 & 0, 1 & \checkmark  & $\frac{38.03}{58-15}=  0.95$          \\[3pt]    
     2                 & 0, 2 & $-$         & $\frac{56.30}{58-9}=  1.15$           \\[3pt]   
     2                 & 0, 2 & \checkmark  & $\frac{38.03}{58-15}=  0.88$          \\[3pt]    

    \hline
    \multicolumn{4}{c}{\rule{0pt}{12pt} Simple phase space}  \\
    \hline
     \rule{0pt}{12pt}1 & 0, 1    & $-$         & $\frac{46.19}{58-9}=  0.94$   \\[3pt]   
     1                 & 0, 1    & \checkmark  & $\frac{37.91}{58-15}=  0.88$  \\[3pt]    
     1                 & 0, 2    & $-$         & $\frac{46.28}{58-9}=  0.94$   \\[3pt]    
     1                 & 0, 2    & \checkmark  & $\frac{37.90}{58-15}=  0.88$  \\[3pt]   
     2                 & 0, 1    & $-$         & $\frac{46.10}{58-9}=  0.94$   \\[3pt]    
     2                 & 0, 1    & \checkmark  & $\frac{37.91}{58-15}=  0.88$  \\[3pt]    
     2                 & 0, 2    & $-$         & $\frac{53.67}{58-9}=  1.10$   \\[3pt]    
     2                 & 0, 2    & \checkmark  & $\frac{37.91}{58-15}=  0.88$  \\[3pt]    
     \hline
    \end{tabular}
    }
    \caption{As Table~\ref{table:A1p6f_parameterisation_table} but for $J^P =2^+$ amplitudes and $[000]E^+$. 
             All diagonal $D$-wave amplitudes had a constant term not fixed to zero. 
             Third column indicates parameterisations that had constant terms not fixed to zero for all off-diagonal $D$-wave to $\Dst\omoct({\SLJ{5}{S}{2}})$  amplitudes. 
             The parameterisation labelled ``\textparagraph'' (``\S'') has the $J^P= 4^+$ amplitude parameter fixed to the reference parameterisation's central value minus (plus) one sigma (Eq.~\ref{eq:A1p6f_ref}).
             For all other parameterisations the $J^P= 4^+$ amplitude parameter is fixed to the central value.}
    \label{table:Ep6f_parameterisation_table} 
\end{table}

\begin{table}
\centering
    \begin{tabular}{ccc}
    \hline
    \hline
    \multicolumn{3}{c}{$K$-matrix polynomial}  \\
    \hline
    \hline
    \rule{0pt}{12pt} Order for diagonal amplitudes &  Order for off-diagonal amplitudes & \chisq  \\
    \hline
    \multicolumn{3}{c}{\rule{0pt}{12pt} Chew-Mandelstam phase space}  \\
    \hline
     \rule{0pt}{12pt}$\bm{0}$   & $\bm{-}$      & $\bm{\frac{17.88}{22-4}=  0.99}$  \\[3pt] 
     0                   & 0             & $\frac{12.85}{22-10}=  1.07^\dagger$     \\[3pt]
     1                   & $-$           & $\frac{18.20}{22-4}= 1.01$               \\[3pt] 
     1                   & 0             & $\frac{13.52}{22-10}=  1.13^\dagger$     \\[3pt] 

    \hline
    \multicolumn{3}{c}{\rule{0pt}{12pt} Simple phase space}  \\
    \hline
     \rule{0pt}{12pt}0   & $-$          & $\frac{17.96}{22-4}=  1.00$             \\[3pt]  
     0                   & 0            & $\frac{14.47}{22-10}=  1.21^\dagger$    \\[3pt] 
     1                   & $-$             & $\frac{18.28}{22-10}= 1.02 $         \\[3pt] 
    \hline
    \end{tabular}
	\caption{As Table~\ref{table:A1p6f_parameterisation_table} but for $J^P = 3^+$ amplitudes and $[000]A_2^+$.
             First column indicates the order of each term in the polynomial for the diagonal amplitudes.
             Second column indicates the order of each term in the polynomial  for the off-diagonal amplitudes.}
	\label{table:A2p6f_parameterisation_table} 
\end{table}

\begin{table}
\centering
    \resizebox{1\columnwidth}{!}{%
    \begin{tabular}{cccccc}
    \hline
    \hline
    \multicolumn{5}{c}{$K$-matrix polynomial}  \\
    \hline
    \hline
    \rule{0pt}{12pt}Label & $\gamma^{(n)}\{\Dst\etaoct({\SLJ{3}{S}{1}})| \Dst\etaoct({\SLJ{3}{S}{1}})\}$ & $\gamma^{(n)}\{\Dst\etaoct({\SLJ{3}{D}{1}})| \Dst\etaoct({\SLJ{3}{D}{1}})\}$ & $\gamma^{(n)}\{\Dst\etaoct({\SLJ{3}{S}{1}})| \Dst\etaoct({\SLJ{3}{D}{1}})\}$ & $\chisq$  \\
    \hline 
    \multicolumn{5}{c}{Chew-Mandelstam phase space}  \\
    \hline
    \rule{0pt}{12pt}$(a)$      &  0, 1             & 0           & $-$        & $\frac{3.46}{10-3}=  0.49$                    \\[3pt]       
                    $\bm{(b)}$ &  $\bm{0, 1}$      & $\bm{0}$    & $\bm{0}$   & $\bm{\frac{2.80}{10-4}=  0.47}$               \\[3pt]       
                    $(c)$      &  0, 1             & 0           & 0          & $\frac{2.80}{10-4}=  0.47$ (\textparagraph)   \\[3pt]       
                    $(d)$      &  0, 1             & 0           & 0          & $\frac{2.79}{10-4}=  0.47$ (\S)               \\[3pt]       
                    $(e)$      &  0, 2             & 0           & $-$        & $\frac{3.60}{10-3}=  0.51$                    \\[3pt]       
                    $(f)$      &  0, 2             & 0           & 0          & $\frac{2.85}{10-4}=  0.47$                    \\[3pt]       
                    $(g)$      &  0, 1, 2          & 0           & $-$        & $\frac{2.65}{10-4}=  0.44$                    \\[3pt]       
                    $(h)$      &  $0, 1, 2$        & 0           & 0          & $\frac{2.58}{10-5}=  0.52$                    \\[3pt]

    \hline 
    \multicolumn{5}{c}{Chew-Mandelstam phase space}  \\
    \hline
    \rule{0pt}{12pt}$(i)$ &  0, 1      & 0    & $-$  & $\frac{3.80}{10-3}=  0.54$            \\[3pt]       
                    $(j)$ &  0, 1      & 0    & 0    & $\frac{2.88}{10-4}=  0.48$            \\[3pt]       
                    $(k)$ &  0, 2      & 0    & 0    & $\frac{3.96}{10-3}=  0.57$            \\[3pt]       
                    $(l)$ &  0, 2      & 0    & 0    & $\frac{2.95}{10-4}=  0.49$            \\[3pt]       
                    $(m)$ &  0, 1, 2   & 0    &  $-$  & $\frac{2.76}{10-4}=  0.46$           \\[3pt]      
                    $(n)$ &  0, 1, 2   & 0    &   0   & $\frac{2.54}{10-5}=  0.51$           \\[3pt]      
    \hline
    \end{tabular}
    }
    \caption{As Table~\ref{table:A1p6f_parameterisation_table} but for $J^P =1^+$ single-channel fits to $[000]T_1^+$.
             The parameterisation labelled ``\textparagraph'' (``\S'') has the $J^P= 3^+$ amplitude parameter fixed to the reference parameterisation's central value minus (plus) one sigma (Eq.~\ref{eq:A2p6f_ref}).
             For all other parameterisations the $J^P= 3^+$ amplitude parameter is fixed to the central value.
    }
    \label{table:T1p6f_single_channel_parameterisation_table} 
\end{table}

\begin{table}
\centering
    \resizebox{1\columnwidth}{!}{%
    \begin{tabular}{ccccc}
    \hline
    \hline
    \multicolumn{5}{c}{$K$-matrix polynomial}  \\
    \hline
    \hline
    \rule{0pt}{13pt}$\gamma^{(n)}\{\D\omoct({\SLJ{3}{S}{1}})| \D\omoct({\SLJ{3}{S}{1}})\}$ &  $\gamma^{(n)}\{\Dst\omoct({\SLJ{3}{S}{1}})| \Dst\omoct({\SLJ{3}{S}{1}})\}$ & $S$-wave to $S$-wave  & $S$-wave to $D$-wave & \chisq \\
    \hline
    \multicolumn{5}{c}{Chew-Mandelstam phase space}  \\
    \hline
     \rule{0pt}{12pt}0, 1 & 0         & $-$            & $-$           & $\frac{89.99}{82-10}=  1.25$                    \\[3pt]           
     $\bm{0, 1}$          & $\bm{0}$  & \checkmark     & $\bm{-}$      & $\bm{\frac{69.31}{82-13}=  1.00}$               \\[3pt]       
     0, 1                 & 0         & \checkmark     & $-$           & $\frac{70.05}{82-13}=  1.02$ (\textparagraph)   \\[3pt]            
     0, 1                 & 0         & \checkmark     & $-$           & $\frac{70.06}{82-13}=  1.02$ (\S)               \\[3pt]            
     0, 1                 & 0         & \checkmark     & \checkmark    & $\frac{47.75}{82-25}=  0.84$                    \\[3pt]           
     0, 1                 & 0, 1      & $-$            & $-$           & $\frac{89.75}{82-11}=  1.26$                    \\[3pt]            
     0, 1                 & 0, 1      & \checkmark     & $-$           & $\frac{69.24}{82-14}=  1.02$                    \\[3pt]            
     0, 1                 & 0, 1      & \checkmark     & \checkmark    & $\frac{47.41}{82-26}=  0.84$                    \\[3pt]           
     0, 1, 2              & 0         & $-$            & $-$           & $\frac{84.53}{82-11}=  1.19$                    \\[3pt]           
     0, 1, 2              & 0         & \checkmark     & $-$           & $\frac{62.57}{82-14}=  0.92$                    \\[3pt]           
     0, 1, 2              & 0         & \checkmark     & \checkmark    & $\frac{47.62}{82-26}=  0.85$                    \\[3pt]           
     0, 1, 2              & 0, 1      & $-$            & $-$           & $\frac{84.53}{82-12}=  1.21$                    \\[3pt]            
     0, 1, 2              & 0, 1      & \checkmark     & $-$           & $\frac{69.72}{82-15}=  0.94$                    \\[3pt]           
     0, 1, 2              & 0, 1      & \checkmark     & \checkmark    & $\frac{47.17}{82-27}=  0.86$                    \\[3pt]    

    \hline
    \multicolumn{5}{c}{Simple phase space}  \\
    \hline
     \rule{0pt}{12pt}0, 1 & 0         & $-$            & $-$           & $\frac{98.88}{82-10}=  1.37^\dagger $   \\[3pt]     
     0, 1                 & 0         & \checkmark     & $-$           & $\frac{77.79}{82-13}=  1.13^\dagger $   \\[3pt]     
     0, 1                 & 0         & \checkmark     & \checkmark    & $\frac{60.14}{82-25}=  1.06^\dagger $   \\[3pt]    

     0, 1                 & 0, 1      & $-$            & $-$           & $\frac{98.88}{82-11}=  1.39^\dagger $   \\[3pt]    
     0, 1                 & 0, 1      & \checkmark     & $-$           & $\frac{77.72}{82-14}=  1.14^\dagger $   \\[3pt]       
     0, 1                 & 0, 1      & \checkmark     & \checkmark    & $\frac{57.41}{82-26}=  1.03^\dagger $   \\[3pt]     

     0, 1, 2              & 0         & $-$            & $-$           & $\frac{93.10}{82-11}=  1.31^\dagger $   \\[3pt]    
     0, 1, 2              & 0         & \checkmark     & $-$           & $\frac{71.13}{82-14}=  1.05^\dagger $   \\[3pt]   
     0, 1, 2              & 0         & \checkmark     & \checkmark    & $\frac{60.16}{82-26}=  1.07^\dagger $   \\[3pt]   

     0, 1, 2              & 0, 1      & $-$            & $-$           & $\frac{92.81}{82-12}=  1.33^\dagger $   \\[3pt]    
     0, 1, 2              & 0, 1      & \checkmark     & $-$           & $\frac{70.43}{82-15}=  1.05^\dagger $   \\[3pt]      
     0, 1, 2              & 0, 1      & \checkmark     & \checkmark    & $\frac{57.13}{82-27}=  1.04^\dagger $   \\[3pt]     

    \hline
    \end{tabular}
    }
	\caption{As Table~\ref{table:A1p6f_parameterisation_table} but for $J^P =1^+$ amplitudes and  $[000]T_1^+$. 
          All parameterisations have a second order polynomial for the diagonal $\Dst\etaoct(\SLJ{3}{S}{1})$ amplitude and constant terms of the diagonal $D$-wave amplitudes.
          Third column indicates parameterisations that had constant terms not fixed to zero for all off-diagonal $S$-wave to $S$-wave  amplitudes.
          Fourth column indicates parameterisations that had constant terms not fixed to zero for all off-diagonal $S$-wave to $D$-wave amplitudes.
          The parameterisation labelled ``\textparagraph'' (``\S'') has the $J^P= \{3,4\}^+$ amplitudes' parameters fixed to the reference parameterisations' central values minus (plus) one sigma (Eq.~\ref{eq:A2p6f_ref} and Eq.~\ref{eq:A1p6f_ref}).
          For all other parameterisations the $J^P= \{3,4\}^+$ amplitudes' parameters are fixed to the central values.
        }
	\label{table:T1p6f_parameterisation_table} 
\end{table}

\begin{table}
\centering
    \resizebox{1\columnwidth}{!}{%
    \begin{tabular}{ccccc}
    \hline
    \hline
    \multicolumn{5}{c}{$K$-matrix polynomial}  \\
    \hline
    \hline
    \rule{0pt}{12pt}$\gamma^{(n)}\{\D\etaoct({\SLJ{1}{S}{0}})| \D\etaoct({\SLJ{1}{S}{0}})\}$ &  $\gamma^{(n)}\{\Dst\omoct({\SLJ{1}{S}{0}})| \Dst\omoct({\SLJ{1}{S}{0}})\}$ & $\gamma^{(n)}\{\D\etaoct({\SLJ{1}{S}{0}}) | \Dst\omoct({\SLJ{1}{S}{0}})\}$ & $S$-wave to $\Dst\omoct({\SLJ{1}{D}{0}})$ & $\chisq$  \\
    \hline
    \multicolumn{5}{c}{\rule{0pt}{12pt} Chew-Mandelstam phase space}  \\
    \hline
     \rule{0pt}{12pt}0, 1 & 0            & $-$        &  $-$              & $\frac{30.64}{31-5}=  1.18$          \\[3pt]  
     0, 1                 & 0            & 0          &  $-$              & $\frac{30.56}{31-6}=  1.22$          \\[3pt]  
     0, 1                 & 0            & 0          &  $\checkmark$     & $\frac{30.56}{31-8}=  1.33$          \\[3pt] 
     0, 1                 & 0, 1         & $-$        &  $-$              & $\frac{16.48}{31-6}=  0.66$          \\[3pt]  
     $\bm{0, 1}$          & $\bm{0, 1}$  & $\bm{0}$   &  $\bm{-}$         & $\bm{\frac{16.48}{31-7}=  0.69}$     \\[3pt]  
     0, 1                 & 0, 1         & 0          &  $\checkmark$     & $\frac{16.27}{31-9}=  0.74$          \\[3pt]  
     0, 1                 & 0, 1, 2      & $-$        &  $-$              & $\frac{16.07}{31-7}=  0.67$          \\[3pt]  
     0, 1                 & 0, 1, 2      & 0          &  $-$              & $\frac{16.07}{31-8}=  0.70$          \\[3pt]  
     0, 1                 & 0, 1, 2      & 0          &  $\checkmark$     & $\frac{15.92}{31-10}=  0.76$         \\[3pt]  
     0, 1, 2              & 0            & $-$        &  $-$              & $\frac{30.16}{31-6}=  1.21$          \\[3pt]  
     0, 1, 2              & 0            & 0          &  $-$              & $\frac{30.15}{31-7}=  1.26$          \\[3pt]  
     0, 1, 2              & 0            & 0          &  $\checkmark$     & $\frac{30.15}{31-9}=  1.37$          \\[3pt]  
     0, 1, 2              & 0, 1         & $-$        &  $-$              & $\frac{16.25}{31-7}=  0.68$          \\[3pt] 
     0, 1, 2              & 0, 1         & 0          &  $-$              & $\frac{16.25}{31-8}=  0.71$          \\[3pt]  
     0, 1, 2              & 0, 1         & 0          &  $\checkmark$     & $\frac{16.19}{31-10}=  0.77$         \\[3pt]  
     0, 1, 2              & 0, 1, 2      & $-$        &  $-$              & $\frac{15.85}{31-8}=  0.69$          \\[3pt]  
     0, 1, 2              & 0, 1, 2      & 0          &  $-$              & $\frac{15.85}{31-9}=  0.72$          \\[3pt]  
     0, 1, 2              & 0, 1, 2      & 0          &   $\checkmark$    & $\frac{15.83}{31-11}=  0.72$         \\[3pt]  
    \hline
    \end{tabular}
    }
    \caption{As for Table~\ref{table:A1p6f_parameterisation_table} but for the $J^P = 0^+$ amplitudes and $[000]A_1^+$ in the flavour $\overline{\mathbf{15}}$ sector.
             Fourth column indicates parameterisations that had constant terms not fixed to zero for off-diagonal $S$-wave to $\Dst\omoct({\SLJ{1}{D}{0}})$ amplitudes.
    }
    \label{table:A1p15bar_parameterisation_table} 
\end{table}

\begin{table}
\centering
    \resizebox{1\columnwidth}{!}{%
    \begin{tabular}{cccc}
    \hline
    \hline
    \multicolumn{4}{c}{$K$-matrix polynomial with Chew-Mandelstam phase space}  \\
    \hline
    \hline
    \rule{0pt}{12pt}$\gamma^{(n)}\{\Dst\omoct({\SLJ{5}{S}{2}})| \Dst\omoct({\SLJ{5}{S}{2}})\}$ &  $D$-wave to $\Dst\omoct({\SLJ{5}{S}{2}})$ & $D$-wave to $D$-wave& \chisq \\
    \hline
     \rule{0pt}{12pt}$\bm{0}$ & $\bm{-}$     & $\bm{-}$    & $\bm{\frac{41.64}{41-7}=  1.22}$                    \\[3pt] 
     0                        & $-$          & $-$         & $\frac{43.56}{41-7}=  1.28$ (\textparagraph)        \\[3pt] 
     0                        & $-$          & $-$         & $\frac{40.33}{41-7}= 1.19 $ (\S)                    \\[3pt] 
     0                        & \checkmark   & $-$         & $\frac{25.56}{41-13}=  0.91$                        \\[3pt] 
     0                        & \checkmark   & \checkmark  & $\frac{23.39}{41-28}=  1.80$                        \\[3pt] 
     0, 1                     & $-$          & $-$         & $\frac{38.92}{41-8}=  1.18$                         \\[3pt] 
     0, 1                     & \checkmark   & $-$         & $\frac{25.01}{41-14}=  0.93$                        \\[3pt] 
     0, 2                     & $-$          & $-$         & $\frac{36.61}{41-8}=  1.11$                         \\[3pt]
     0, 2                     & \checkmark   & $-$         & $\frac{25.08}{41-14}=  0.93$                        \\[3pt] 
    \hline
    \end{tabular}
    }
    \caption{As for Table~\ref{table:Ep6f_parameterisation_table} but for the $J^P =2^+$ amplitudes and $[000]E^+$ in the flavour $\overline{\mathbf{15}}$ sector.
             Fourth column indicates parameterisations that had constant terms not fixed to zero for all off-diagonal $D$-wave to $D$-wave amplitudes.
             }
                 \label{table:Ep15bar_parameterisation_table} 
\end{table}

\begin{table}
\centering
    \begin{tabular}{ccc}
    \hline
    \hline
    \multicolumn{3}{c}{$K$-matrix polynomial}  \\
    \hline
    \hline
    \rule{0pt}{12pt} Order for diagonal amplitudes &  Order for off-diagonal amplitudes & \chisq  \\
    \hline
    \multicolumn{3}{c}{\rule{0pt}{12pt} Chew-Mandelstam phase space}  \\
    \hline
     \rule{0pt}{12pt}$\bm{0}$   & $\bm{-}$      & $\bm{\frac{9.12}{16-4}=0.76 }$         \\[3pt] 
     0                          & 0             & $\frac{7.41}{16-10}= 1.23 ^\dagger$    \\[3pt] 
     0                          & 1             & $\frac{6.05}{16-10}= 1.01$             \\[3pt] 
     1                          & $-$           & $\frac{9.33}{16-4}=  0.78$             \\[3pt] 
     1                          & 1             & $\frac{6.11}{16-10}= 1.02$             \\[3pt]

    \hline
    \multicolumn{3}{c}{\rule{0pt}{12pt} Simple phase space}  \\
    \hline
     \rule{0pt}{12pt}0   & $-$          & $\frac{9.12}{16-4}= 0.76 $            \\[3pt] 
     0                   & 0            & $\frac{8.63}{16-10}= 1.44^\dagger$    \\[3pt]  
     1                   & $-$             & $\frac{9.33}{16-4}= 0.78 $         \\[3pt] 
    \hline
    \end{tabular}
	\caption{As Table~\ref{table:A2p6f_parameterisation_table}  but for the $J^P =3^+$ amplitudes and $[000]A^+_2$ in the flavour $\overline{\mathbf{15}}$ sector.
             }
	\label{table:A2p15bar_parameterisation_table} 
\end{table}

\begin{table}
    \resizebox{1\columnwidth}{!}{%
    \begin{tabular}{cccccc}
    \hline
    \hline
    \multicolumn{6}{c}{$K$-matrix polynomial with Chew-Mandelstam phase space}  \\
    \hline
    \hline
    \rule{0pt}{13pt} $\gamma^{(n)}\{\Dst\etaoct({\SLJ{3}{S}{1}})| \Dst\etaoct({\SLJ{3}{S}{1}})\}$ & $\gamma^{(n)}\{\D\omoct({\SLJ{3}{S}{1}})| \D\omoct({\SLJ{3}{S}{1}})\}$ &  $\gamma^{(n)}\{\Dst\omoct({\SLJ{3}{S}{1}})| \Dst\omoct({\SLJ{3}{S}{1}})\}$ & $S$-wave to $S$-wave  & $S$-wave to $D$-wave& \chisq \\
    \hline
     \rule{0pt}{12pt}0 & 0              & 0           & $-$            & $-$           & $\frac{82.36}{51-7}=  1.87$                \\[3pt] 
     0                 & 0              & 0           & \checkmark     & $-$           & $\frac{69.23}{51-10}=  1.69$               \\[3pt] 
     0                 & 0              & 0           & \checkmark     & \checkmark    & $\frac{44.17}{51-22}=  1.52$               \\[3pt] 
     0, 1              & 0              & 0           & $-$            & $-$           & $\frac{60.33}{51-8}=  1.40$                \\[3pt] 
     0, 1              & 0              & 0           & \checkmark     & $-$           & $\frac{56.43}{51-11}=  1.41$               \\[3pt] 
     0, 1              & 0              & 0           & \checkmark     & \checkmark    & $\frac{39.43}{51-23}=  1.41$               \\[3pt]
     0, 1              & 0, 1           & 0           & $-$            & $-$           & $\frac{42.27}{51-9}=  1.01^\dagger$        \\[3pt] 
     $\bm{0, 1}$       & $\bm{0, 1}$    & $\bm{0}$    & \checkmark     & $\bm{-}$      & $\bm{\frac{37.04}{51-12}=  0.95^\dagger}$  \\[3pt] 
     0, 1              & 0, 1           & 0           & \checkmark     & $-$           & $\frac{37.22}{51-12}=  0.95^\dagger$ (\textparagraph)       \\[3pt] 
     0, 1              & 0, 1           & 0           & \checkmark     & $-$           & $\frac{38.37}{51-12}=  0.98^\dagger$ (\S)        \\[3pt]
     0, 1              & 0, 1           & 0           & \checkmark     & \checkmark    & $\frac{22.73}{51-24}=  0.84^\dagger$          \\[3pt] 
     0, 1              & 0, 1           & 0, 1        & $-$            & $-$           & $\frac{41.96}{51-10}=  1.02^\dagger$       \\[3pt]
     0, 1              & 0, 1           & 0, 1        & \checkmark     & $-$           & $\frac{33.12}{51-13}=  0.87^\dagger$        \\[3pt] 
     0, 1              & 0, 1           & 0, 1        & \checkmark     & $-$           & $\frac{22.72}{51-25}=  0.87^\dagger$        \\[3pt] 
   \hline
    \end{tabular}
    }
	\caption{As Table~\ref{table:T1p6f_parameterisation_table}  but for the  $J^P =1^+$ amplitudes and $[000]T^+_1$ in the flavour $\overline{\mathbf{15}}$ sector.
              All diagonal $D$-wave amplitudes had a constant term not fixed to zero. 
             }
             		\label{table:T1p15bar_parameterisation_table} 
\end{table}

\newpage

\bibliographystyle{JHEP-2}
\bibliography{refs.bib}

\end{document}